\documentclass{aa} 
\usepackage{graphicx}
\usepackage{psfig}
\usepackage{epsfig}
\begin{document}
\title{The Geneva-Copenhagen survey of the Solar neighbourhood
\thanks{Based on observations made with the Danish 1.5-m telescope 
        at ESO, La Silla, Chile, and with the Swiss 1-m telescope at 
        Observatoire de Haute-Provence, France.}
}
\subtitle{Ages, metallicities, and kinematic properties of $\sim$14,000 F and
G dwarfs
}
\author{B. Nordstr{\"o}m \inst{1,4}
  \and
   M. Mayor \inst{3}
  \and
   J. Andersen \inst{2,5}
  \and
  J. Holmberg \inst{2,5}
  \and
  F. Pont \inst{3}
  \and
  B.R. J{\o}rgensen \inst{4}
  \and
  E. H. Olsen \inst{2}
  \and
  S. Udry \inst{3}
  \and
  N. Mowlavi \inst{3}
}
\offprints{B. Nordstr{\"o}m (birgitta@astro.ku.dk)}
\institute{
    Niels Bohr Institute for Astronomy, Physics \& Geophysics, 
    Blegdamsvej 17, 
    DK-2100 Copenhagen, Denmark
  \and
    Astronomical Observatory, NBIfAFG, Juliane Maries Vej 30,
    DK-2100 Copenhagen, Denmark
  \and
     Observatoire de Gen{\`e}ve, 51 Ch. des Maillettes,
     CH-1290 Sauverny, Switzerland
  \and
    Lund Observatory, Box 43, SE-221 00 Lund, Sweden
  \and
    Nordic Optical Telescope Scientific Association, Apartado 474,
        ES-38 700 Santa Cruz de La Palma, Spain.
}
\date{Received December 31, 2003; accepted January 23, 2004}
\abstract{We present and discuss new determinations of metallicity, rotation,
age, kinematics, and Galactic orbits for a complete, magnitude-limited, and
kinematically unbiased sample of 16,682 nearby F and G dwarf stars. Our
$\sim$63,000 new, accurate radial-velocity observations for nearly 13,500 stars
allow identification of most of the binary stars in the sample and, together
with published $uvby\beta$ photometry, Hipparcos parallaxes, Tycho-2 proper
motions, and a few earlier radial velocities, complete the kinematic
information for 14,139 stars. These high-quality velocity data are supplemented
by effective temperatures and metallicities newly derived from recent and/or
revised calibrations. The remaining stars either lack Hipparcos data or have
fast rotation. \\
A major effort has been devoted to the determination of new isochrone ages for
all stars for which this is possible. Particular attention has been given to a
realistic treatment of statistical biases and error estimates, as standard
techniques tend to underestimate these effects and introduce spurious features
in the age distributions. Our ages agree well with those by Edvardsson et al.
(\cite{edv93}), despite several astrophysical and computational improvements
since then. We demonstrate, however, how strong observational and theoretical
biases cause the distribution of the {\it observed} ages to be very different
from that of the {\it true} age distribution of the sample.\\
Among the many basic relations of the Galactic disk that can be reinvestigated
from the data presented here, we revisit the metallicity distribution of the G
dwarfs and the age-metallicity, age-velocity, and metallicity-velocity
relations of the Solar neighbourhood. Our first results confirm the lack of
metal-poor G dwarfs relative to closed-box model predictions (the ``G dwarf
problem''), the existence of radial metallicity gradients in the disk, the
small change in mean metallicity of the thin disk since its formation and the
substantial scatter in metallicity at all ages, and the continuing kinematic
heating of the thin disk with an efficiency consistent with that expected for a
combination of spiral arms and giant molecular clouds. Distinct features in the
distribution of the $V$ component of the space motion are extended in age and
metallicity, corresponding to the effects of stochastic spiral waves rather
than classical moving groups, and may complicate the identification of
thick-disk stars from kinematic criteria. More advanced analyses of this rich
material will require careful simulations of the selection criteria for the
sample and the distribution of observational errors.
}
\maketitle
%

\section{Background and motivation} \label{intro}

The Solar neighbourhood is the benchmark test for models of the Galactic disk:
The stars in a sample volume around the Sun provide a first estimate of the
mass density of the Galactic disk near the plane. Their distribution in age is
our record of the star formation history of the disk. Their overall and
detailed heavy-element abundances as functions of age are the fossil record of
the chemical evolution and enrichment history of the disk. Finally, the space
motions and Galactic orbits of the stars as functions of age are our clues to
the parallel dynamical evolution of the Galaxy and the degree of mixing of
stellar populations from different regions of the disk (see, e.g. the recent
review by Freeman \& Bland-Hawthorn \cite{freeman02}).

Providing the basic data for the stars of the Solar neighbourhood -- ages,
metallicities, velocities, and Galactic orbits -- may seem the simplest
observational task of all. Yet, the identification of the nearest stars and the
data needed for a proper characterisation of their main astrophysical
parameters remain seriously incomplete. Moreover, stars have often been
selected for observation from criteria which induce correlations between the
observed parameters that reflect the characteristics of the selection process
more strongly than those of the Galaxy. Intrinsic parameters such as the
frequency, age, and metallicity of groups of stars are intertwined with such
easily observable properties as brightness, colours, and proper motions in ways
that can make it difficult or impossible to retrieve the former from the
latter. Kinematic selection biases are particularly pernicious in this regard.

F- and G-type dwarf stars are favourite tracer populations of the history of
the disk. They are relatively numerous and sufficiently long-lived to survive
from the formation of the disk; their convective atmospheres reflect their
initial chemical composition; and ages can be estimated for at least the more
evolved stars by comparison with stellar evolution models. Photometry in the
Str{\"o}mgren $uvby\beta$ system is an efficient means to derive their
intrinsic properties from observation (Str{\"o}mgren \cite{stromgren63},
\cite{stromgren87}). Recent studies based mainly on $uvby\beta$ photometry are,
e.g. Feltzing et al. (\cite{feltz01}), Holmberg (\cite{holmb01}), and
J{\o}rgensen (\cite{bjarner00}), while more detailed spectroscopic studies of
the chemical history of selected subsets of stars have been made by, e.g.
Edvardsson et al. (\cite{edv93}), Fuhrmann (\cite{fuhrm98}), Reddy et al.
(\cite{reddy03}), and Bensby et al. (\cite{bensby03}).

Extensive $uvby\beta$ photometric surveys of the nearby F and G stars have been
performed by Olsen (\cite{eho83,eho93,eho94a,eho94b}). Accurate parallaxes and
proper motions have become available for large numbers of these stars from the
Hipparcos (ESA \cite{hipp97}) and Tycho-2 (H{\o}g et al. \cite{tycho2})
catalogues. The bottleneck so far has been the corresponding radial-velocity
data. These are needed, first, to complete the three-dimensional space motions
for all the stars and improve the statistical accuracy of the derived
age-velocity relations. They also serve to identify those stars which just
happen to pass near the Sun at present, but were formed elsewhere and have
witnessed another evolutionary history than that of the Solar circle. Finally,
and perhaps most importantly, repeated radial-velocity measurements allow
identification of the large fraction of stars in the photometric samples which
are binaries, and for which the derived astrophysical and kinematical data will
be inaccurate and potentially misleading.

The key contribution of this paper consists of new, accurate, radial velocities
for an all-sky, magnitude-limited and kinematically unbiased sample of
$\sim$13,500 nearby F and G stars, based on $\sim$63,000 individual
photoelectric observations. Because this data set is unlikely to be superseded
in essential respects until the results of the GAIA mission (Perryman et al.
\cite{gaia01}) and/or the RAVE project (Steinmetz \cite{steinmetz}) become
available, we have also recalibrated and redetermined the astrophysical
parameters ($T_{eff}$, $M_v$, and [Fe/H]) for all stars in our sample. Much
effort has been devoted to the fundamental issue of determining reliable
isochrone ages for as many stars as possible, and we believe that a rather more
realistic assessment of the errors of such ages has been obtained. Finally, we
have computed individual Galactic orbits for all stars with adequate data. 

\begin{table}[ht] 
\caption[]{Table 1 is only available in electronic form at the CDS. The first 
two pages of the table, listing the first 100 stars, are given at the end of 
this  paper as sample of its content and format. }
\end{table} 

The resulting data set should place a wide range of studies of the evolution of
the Galactic disk on a new and considerably improved footing. We caution,
however, that no such thing as a fully unbiased sample exists in Galactic
astronomy, and the reader is strongly urged to carefully study Sect.
\ref{completeness}, where we discuss the key issue of completeness of the data,
especially as regards the derived ages and masses. The catalogue with the
complete data set for 16,682 stars will be available
electronically\footnote{Tables 1 and 2 are available in electronic form by
anonymous ftp to the CDS at cdsarc.u-strasb.fr (130.79.128.5) or via
http://cdsweb.u-strasb.fr/cgi-bin/qcat?J/A+A/418/989} (sample pages are shown 
in Table \ref{catalogue.tab}).

The complementarity between the present study and the much-quoted paper by
Edvardsson et al. (\cite{edv93}) deserves clarification. The 189 stars studied
by Edvardsson et al. (\cite{edv93}) were drawn from the kinematically unbiased
sample presented here, selecting equal numbers of stars in 10 metallicity bins
through the range of [Fe/H] seen in the (thin and thick) disk. Evolved F dwarfs
were selected, excluding known binaries and fast-rotating stars, so that
interstellar reddening and isochrone ages could be determined. Thus, unevolved
stars were avoided, young metal-poor and old metal-rich stars were excluded
{\it a priori} by the colour cutoffs used, and the sample was strongly biased
in favour of metal-poor stars. The sample discussed here was explicitly
designed to alleviate these selection biases and further benefits from the
advent of the Hipparcos and Tycho-2 data but, for obvious reasons, lacks the
detailed, precise, and homogeneous spectroscopy that was the centrepiece of the
study by Edvardsson et al. (\cite{edv93}).

The present paper is organised as follows: Sect. \ref{hdsel} clarifies the
selection criteria used to define the sample, and Sect. \ref{observations}
describes the new radial velocities and other basic observational material
available for the stars. Sect. \ref{calibrations} explains the (partly new)
calibrations and analysis methods we have used to derive reliable astrophysical
parameters from the raw data, and Sect. \ref{completeness} discusses the
completeness and various biases in the resulting parameter sets. We briefly
rediscuss some of the classical diagnostic diagrams in Sect. \ref{discussion}
and finally summarise our conclusions and some of the prospects for the future
in Sect. \ref{conclusion}.

\section{Sample definition}\label{hdsel}

From the outset, it was clear that the definitive selection and
characterisation of the nearby F and G dwarfs should be made from a
comprehensive photometric survey in the Str{\"o}mgren $uvby\beta$ system. But
in order to undertake such a survey, an observing list is needed. Thus, the
first step in the project was to establish an apparent-magnitude limited,
kinematically unbiased all-sky sample of stars within which the F and G dwarfs
would also be volume-complete to a sufficiently large distance ($\sim$40 pc).

For this, we chose the HD catalogue (Cannon \& Pickering \cite{hdcat}), the 
only all-sky spectral catalogue then available. All A5-G0 stars brighter than
$m_{vis}$ = 8.3, G0 stars in the interval 8.30$\le m_{vis}\le$8.40, and all G5
or just G stars (no subtypes) brighter than $m_{vis}$ = 8.6 were selected. On
the one hand, the A5 spectral-type limit is early enough to include even quite
metal-poor F stars; on the other hand, most K0-type stars in the HD catalogue
are giant stars for which luminosities, distances, tangential velocities, and
ages  cannot be reliably determined.

It was desirable to also include any old, metal-rich dwarfs and thus improve
the observational basis for a reassessment of the ``G-dwarf problem''. However,
observing the vast numbers of HD K-type giants is an inefficient way to
identify such stars. We therefore added all 1,277 G0V-K2V stars south of
$\delta$ = -26$\degr$ from the Michigan Spectral Catalogues (Houk \& Cowley
1975, Houk 1978, 1982) that were previously unobserved, but estimated from the
spectral types and $m_{pg}$ to be within a distance of 50 pc, with generous
allowance for the uncertainties. This sample should be complete to about the
same distance limit (40 pc) as the rest of the G-dwarf sample, but covers only
part (28\%) of the entire sky.

The vast majority of all these stars ($\sim$30,000 in total) were then
observed in the $uvby\beta$ photometric surveys described below, and the
sample for the radial-velocity programme was defined from the measured
photometric indices. Subsequently, the Hipparcos and Tycho-2 catalogues have
provided accurate parallaxes and proper motions for nearly all the stars for
which radial velocities had been measured.

\subsection{Initial photometric survey}\label{survey}

From the observing lists assembled as described above, nearly all stars were
observed at least once in the Str{\"o}mgren $uvby\beta$ system by Olsen
(\cite{eho83,eho93,eho94a,eho94b}). The resulting catalogues were merged with
previous sources of $uvby\beta$ photometry (Str{\"o}mgren \& Perry
\cite{stromper65}; Crawford et al. \cite{crawbar66, crawbar70, crawbar71a,
crawbar71b, crawbar72, crawbar73}; Gr{\o}nbech \& Olsen \cite{gronbech76,
gronbech77}, and Olsen \& Perry \cite{ehoper84}). The combined FG photometric
catalogue thus constructed contains a total of 30,465 stars. 

This all-sky sample is complete to the magnitude limits described above and 
should be volume complete for the F and G dwarfs to a distance of $\sim$40 pc. 
In the southern cap where modern MK spectral types are available, the G0V-K2V 
stars within the same distance are also included. This homogeneous, complete, 
and kinematically unbiased database was used to select stars for the 
radial-velocity programme, using the criteria described below. 

\begin{figure}[thbp] 
\resizebox{\hsize}{!}{\includegraphics[angle=-90]{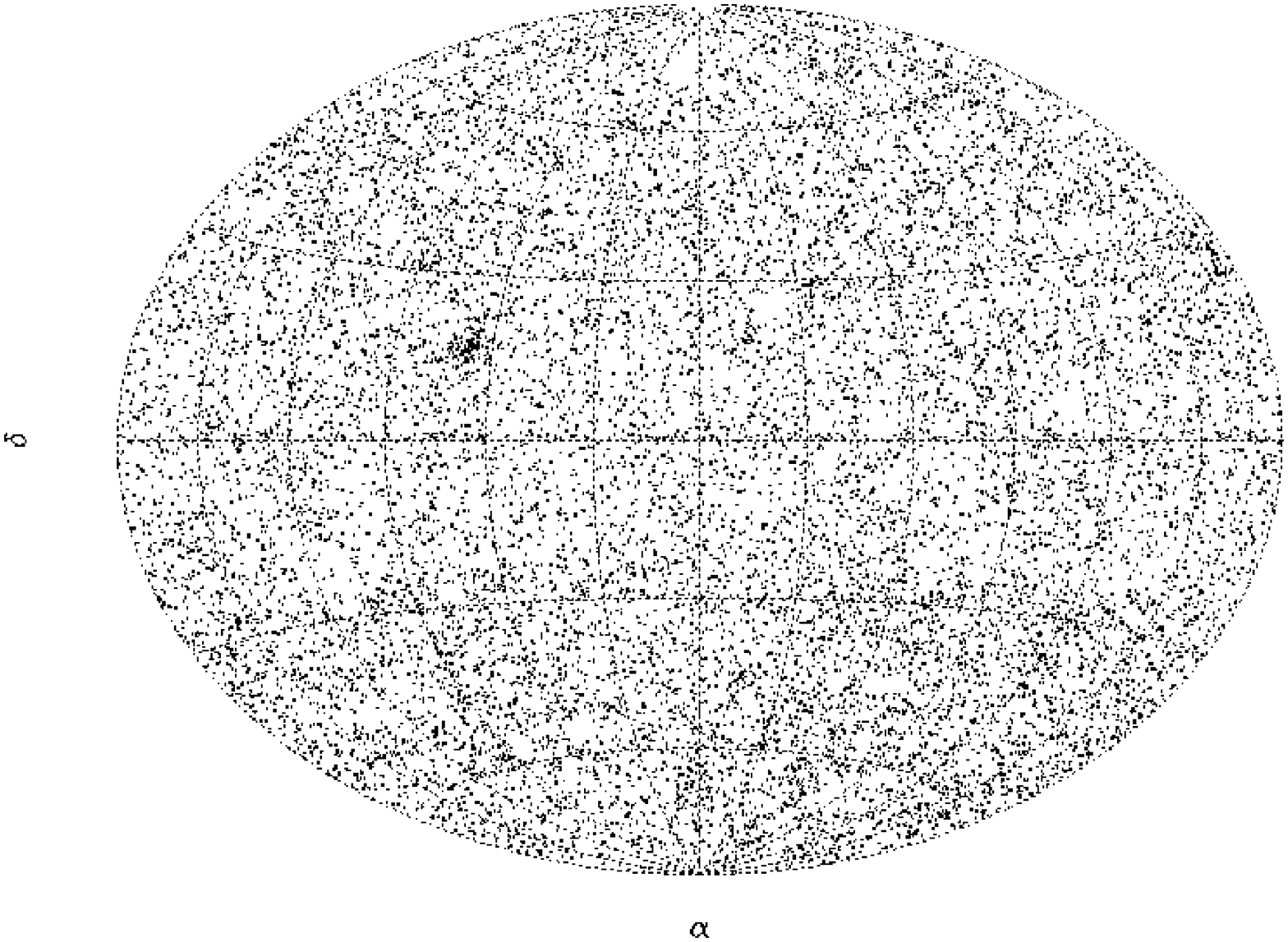}}
\resizebox{\hsize}{!}{\includegraphics[angle=-90]{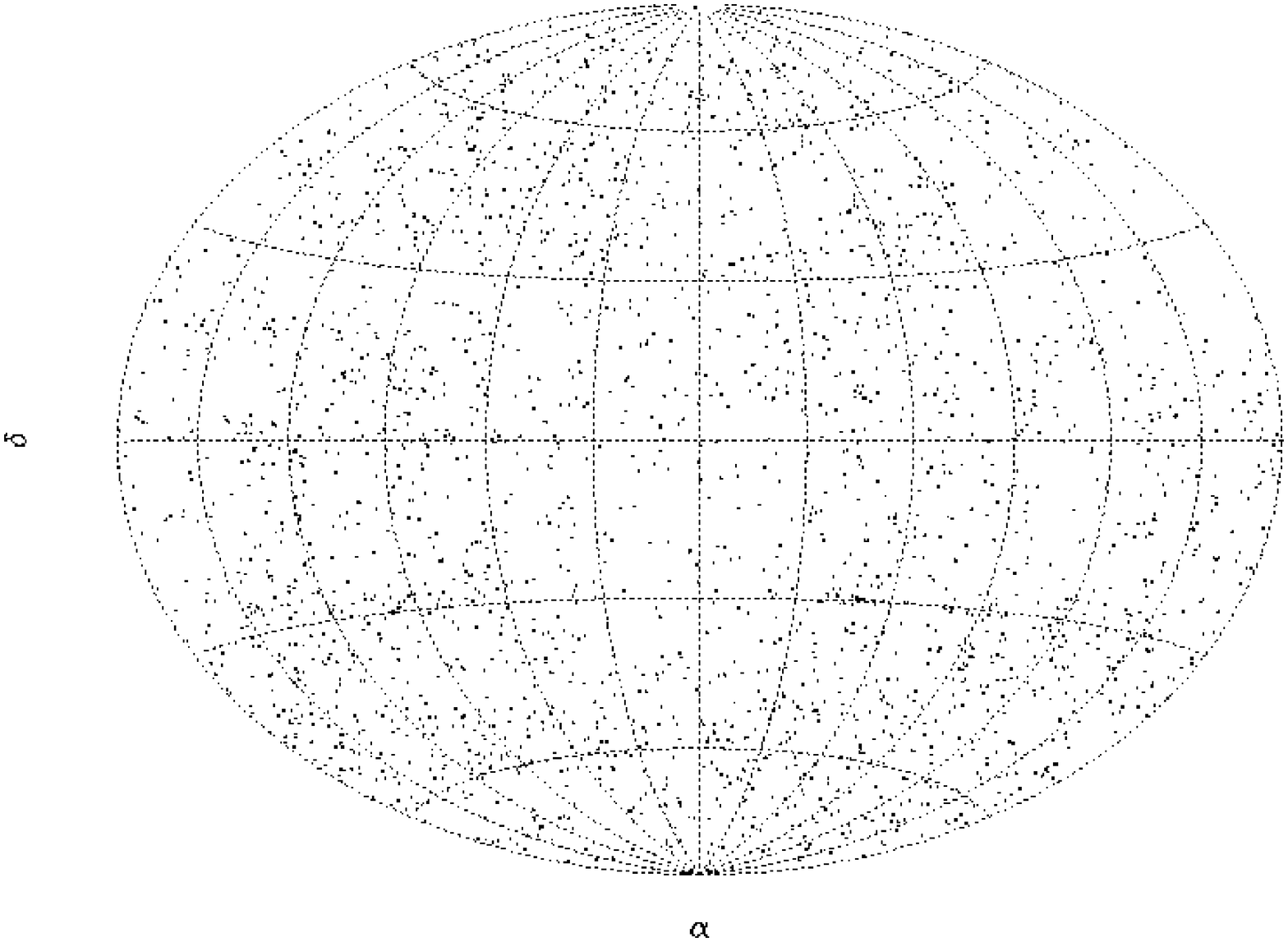}}
\caption{Distribution on the sky of the 16,682 programme stars. {\it Top:}
The 14,139 stars with radial velocity data (note the overdensity of stars in
the Hyades cluster and south of $\delta = -26\degr$). {\it Bottom:} The 2,543
stars with no radial velocity. 
}\label{vr}
\end{figure}

\subsection{Sample definition for the catalogue}\label{sample}

The sample of stars for which radial velocities have been obtained was
defined from the complete FG catalogue. Slightly generous limits in
photometry space have been adopted to allow a selection of FG dwarf stars
based on physical criteria like mass, temperature, metallicity, etc.

The following four samples were selected, merged, and cleaned of duplicate
entries:

\begin{enumerate}

\item All stars for which the F-star calibrations of Crawford (1975) and
Olsen (1988) are valid.

\item All stars with no $\beta$ value and 0.240 $\leq b-y \leq$ 0.460, $[m_1]
\geq$ 0.120, $\delta c_1 \leq$ 0.400, and $V \leq$ 9.600

\item All stars with 0.205 $\leq b-y \leq$ 0.240, $[m_1] \geq$ 0.120, $\delta
c_1 \leq$ 0.400, and $V \leq$ 9.600.

\item All stars for which the G- and K-star calibrations of Olsen (1984) are
valid.

\end{enumerate}

Criterion 3 ensures that no metal poor F-stars will be lost on the hot side  of
the F-type stars. Both criterion 2 and 3 compensate for some missing $\beta$
observations. Criteria 1 and 4 and the $V$-limit 9.6 ensure that a number of
fainter stars observed for calibration purposes are also included. After
removal of a few known supergiants and other irrelevant objects, the list
contains a total of 16,682 objects. 

The distribution of the full sample of 16,682 stars over the sky is shown in
Fig.~\ref{vr}, in equatorial coordinates and in an equi-area projection. Note
the concentration of stars in the Hyades -- observed with special care for
calibration purposes -- and the addition of the latest-type dwarfs south of
$\delta = -26\degr$. Apart from these features, the sample is very uniformly
distributed on the sky.

\section{Observational data}\label{observations}

For reference, we summarise in the following the basic observational data
underlying the astrophysical and kinematical parameters derived for the
programme stars. Our new radial-velocity observations are described in
detail; for other data, the relevant sources are given.

\subsection{Str{\"o}mgren $uvby\beta$ photometry}\label{phot}

\begin{figure}[htbp] 
\resizebox{\hsize}{!}{\includegraphics[angle=0]{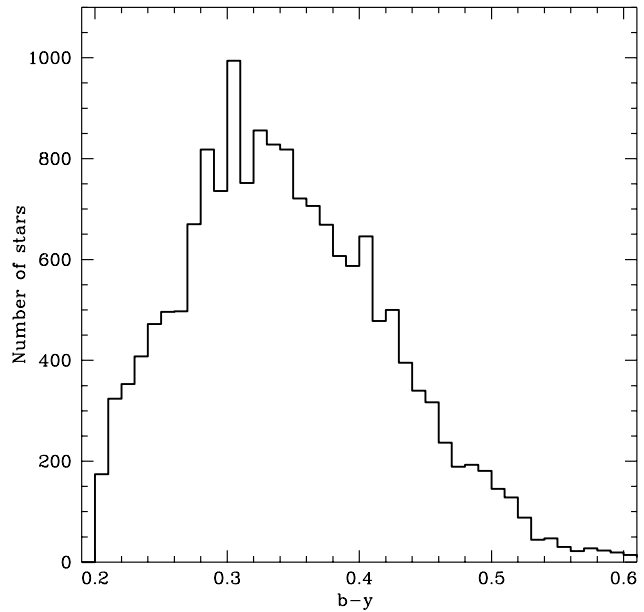}} 
\caption{Distribution of the whole sample in {\it b-y} colour.} 
\label{byhist}
\end{figure}

The merged catalogue of $uvby\beta$ photometry from which our programme stars
were selected was briefly described above (Sect. \ref{sample}); much additional
explanation and extensive notes are given in the original catalogues (Olsen
\cite{eho83,eho93,eho94a,eho94b}). With time, the $\beta$ observations were
extended to include all stars with a {\em b-y} value indicating that $\beta$
might be a useful reddening-free temperature indicator, except that $\beta$
observations for a small fraction of the northern sky are still pending. The
distribution of the sample in {\it b-y} colour is shown in Fig. \ref{byhist}.

\subsection{Radial velocities}\label{rv}

When the first phase of this project was initiated, not even the Bright Star
Catalog (Hoffleit \& Jaschek \cite{bsc82}) had complete radial-velocity
coverage. Of the $\sim$1,500 missing southern stars in that catalogue, the
early-type half were observed with conventional spectrographic techniques
(Andersen \& Nordstr{\"o}m \cite{an83a,an83b}; Nordstr{\"o}m \& Andersen
\cite{na85}), the late-type half with CORAVEL (Andersen et al. \cite{jaetal85};
see also Sect. \ref{coravel}).

For the fainter stars, new observations were obtained as detailed below --
altogether, a total of 62,993 new radial-velocity observations of 13,464
programme stars. Adding earlier literature data, complete kinematical
information is available for a total of 14,139 stars.

\subsubsection{CORAVEL observations}\label{coravel}

The bulk of the radial-velocity data presented here was obtained with the
photoelectric cross-correlation spectrometers CORAVEL (Baranne et al.
\cite{baranne79}; Mayor \cite{mayor85}). Operated at the Swiss 1-m telescope
at Observatoire de Haute-Provence, France, and the Danish 1.5-m telescope at
ESO, La Silla, the two CORAVELs cover the entire sky between them, and their
fixed, late-type cross-correlation template spectra efficiently match the
spectra of the large majority of our programme stars.

Initially, specific observing programmes were targeted to primarily cover the
thick-disk stars which were assumed to be very old, and the evolved thin-disk
stars for which ages can be determined. When a separate programme to observe
all the late-type southern stars of the Hipparcos survey was initiated (Udry
et al. \cite{udryetal97}), most of the remaining southern unevolved F stars
were included as well. Subsequently, a good fraction of the northern half of
the stars has been also observed in a separate Geneva programme on the
Hipparcos stars.

In all programmes, two or more observations were made for almost all stars over
a substantial time base. This allows to define more reliable mean velocities,
but also to identify most of the spectroscopic binaries which, if unrecognised,
yield misleading astrophysical parameters from the observed magnitudes and
colour indices. Between the two telescopes, a total of 60,476 CORAVEL
observations have been made of 12,941 of the programme stars discussed in this
paper - some 1,000 nights' worth of data.

The catalogue presented here contains the mean radial velocity for each star
together with the summary data on the observations as described in the
Appendix. Our computations of the observational errors, criterion for
detecting variable (i.e. binary) stars, and our treatment of double-lined
binaries are discussed in Sect. \ref{rverrors} below.

Many stars on the main programme are primaries of close double stars.
CORAVEL observations were made of the fainter companions to many of these
stars in order to ascertain whether they are physically bound or merely
optical pairs. These data will be made available separately and are not
discussed further here.

\subsubsection{CfA observations}

The fixed-resolution CORAVEL mask is optimised for sharp-lined spectra, and the
cross-correlation profile rapidly becomes too broad and shallow to yield
accurate radial velocities for stars rotating faster than 40-50 km~s$^{-1}$, as
is the case for a large fraction of stars with 0.20 $\leq b-y \leq$ 0.27. 

In order to recover as many as possible of the early F stars of the sample,
several hundred stars rotating too rapidly for CORAVEL were observed with the
digital spectrometers (Latham \cite{dwl85}) of the Harvard-Smithsonian Center
for Astrophysics (CfA). These instruments yield accurate results for single
stars with rotational velocities up to $\sim120$ km~s$^{-1}$ (Nordstr{\"o}m et
al. \cite{bnetal94}) and also perform well on double-lined spectra with
two-dimensional cross-correlation techniques (Latham et al. \cite{dmvir96}).

CfA radial velocities for 595 stars were published by Nordstr{\"o}m et al.
(\cite{bnetal97}) and have been used in the present catalogue when CORAVEL data
were missing or less accurate (2,517 observations of 523 stars). Rapidly
rotating stars south of declination -40$\degr$ cannot be reached by the CfA
instruments and thus have no new radial-velocity data.

\begin{figure}[hbtp] 
\resizebox{\hsize}{!}{\includegraphics{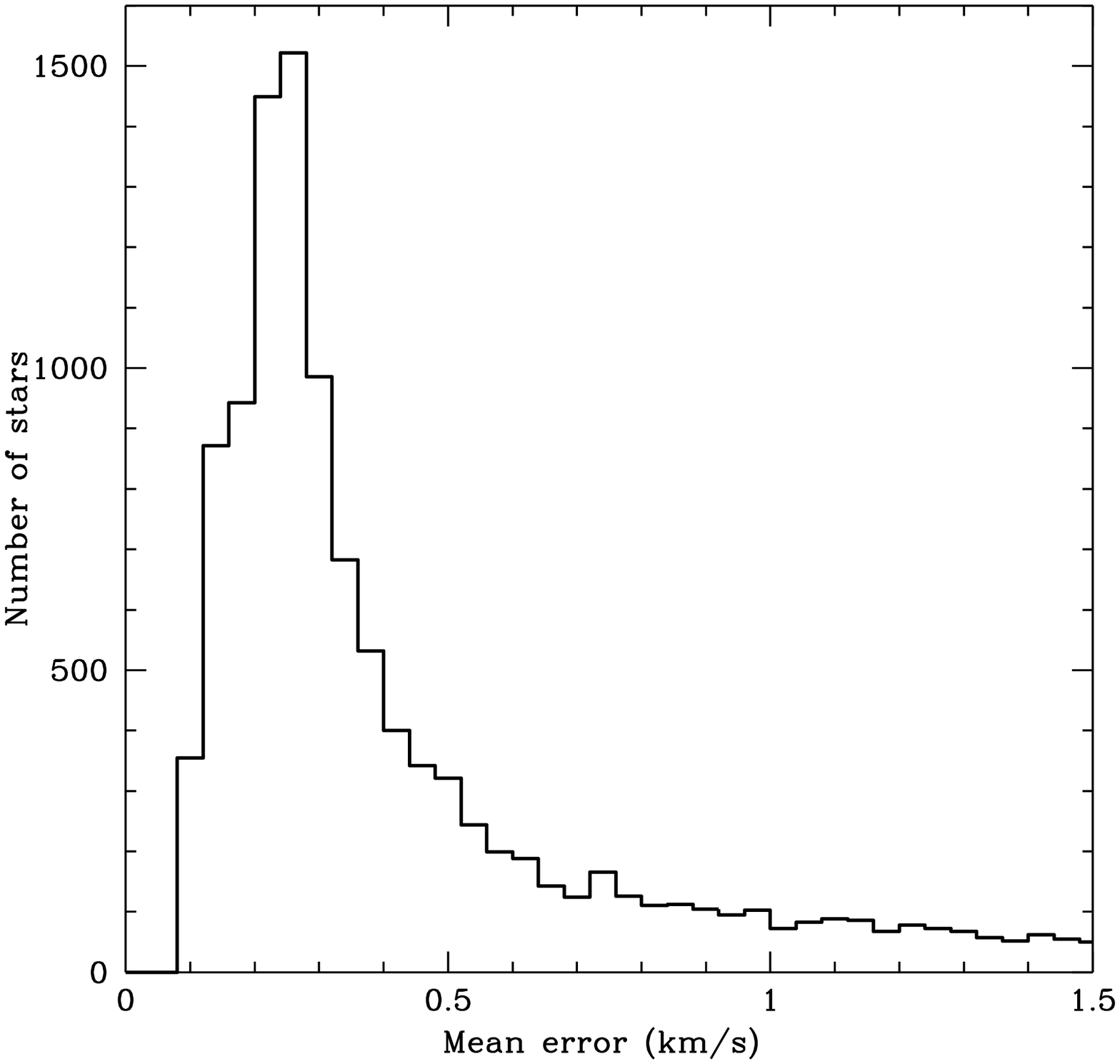}}
\resizebox{\hsize}{!}{\includegraphics{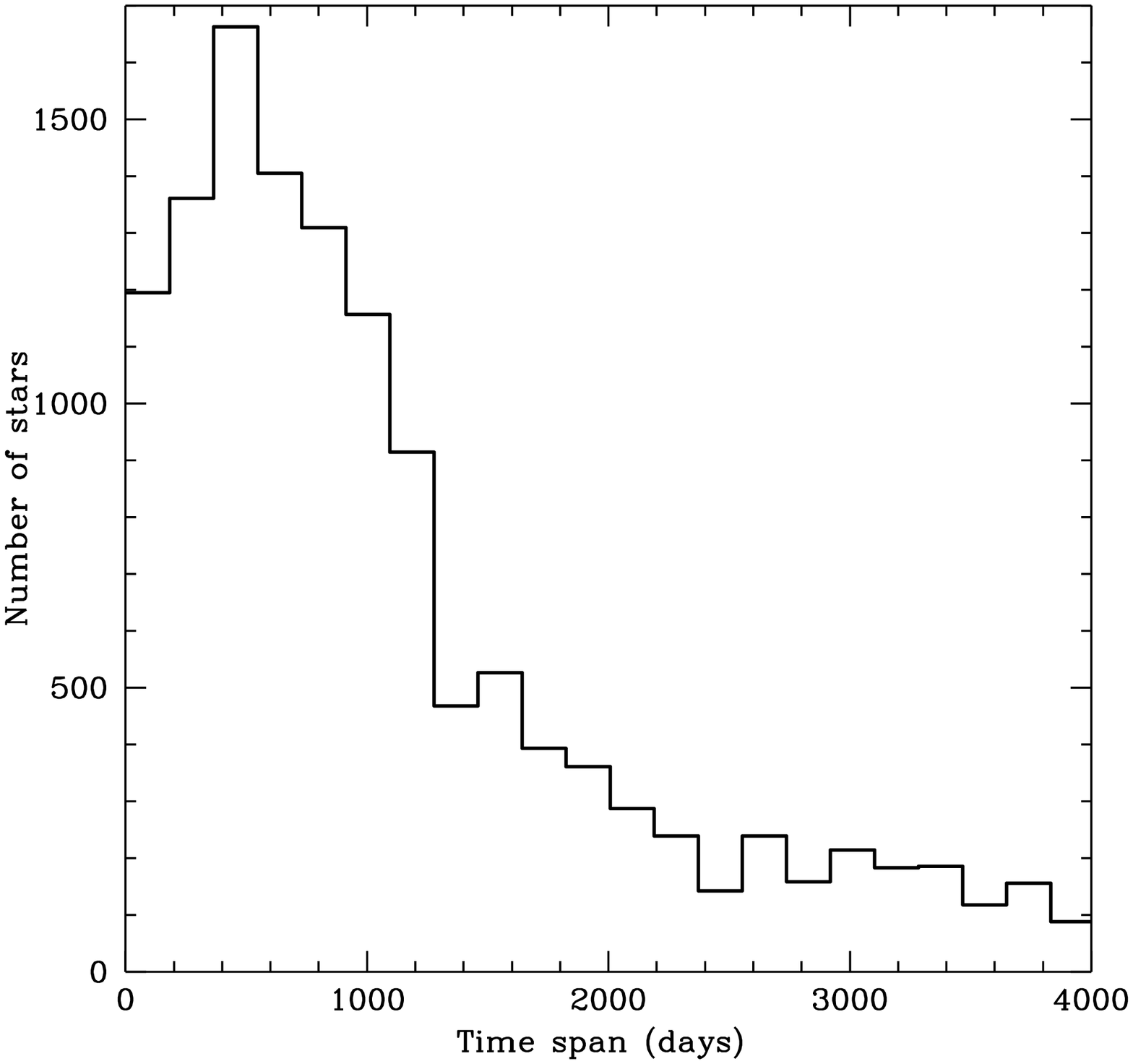}}
\caption{Distribution of the mean errors of the mean radial velocities in
the catalogue ($top$), and the time span covered for each star ($bottom$).} 
\label{vrstat}
\end{figure}

\subsubsection{Literature data}

When this programme was initiated, radial-velocity data existed in the
literature essentially only for stars in the Bright Star Catalog. As the
earlier data were mostly of reasonably good quality, these bright stars have
generally not been reobserved. Literature data for a total of 675 such stars
and others not covered by the new programmes have been taken, as far as
possible, from the compilation of Barbier-Brossat \& Figon (2000), and bring
the total number of stars with radial velocity data to 14,139.

\subsubsection{Variability criterion and binary detection}\label{rverrors}

Each radial-velocity observation is associated with an internal error
estimate, $\epsilon$, while an external error estimate is provided by the
standard deviation, $\sigma$, of repeated observations at different epochs.
From these and the number of observations, $n$, the probability $P(\chi^2)$
that the observed scatter is due to measuring errors alone may be computed as
described in greater detail by Andersen \& Nordstr{\"o}m (\cite{an83b}).
$P(\chi^2) < 0.01$ is adopted as our criterion for certain velocity
variability -- mostly due to binary orbital motion -- for both the CORAVEL
and CfA data.

Normally, the mean error of the mean radial velocity is computed as $\sigma *
n^{-1/2}$. However, if fortuitous good agreement between a few observations
results in $\sigma < \epsilon$, then $\epsilon * n^{-1/2}$ is given instead
as a more realistic estimate of the mean error of the average velocity.

Occasionally, a double correlation peak may identify a spectroscopic binary
from just a single observation, but normally two or more observations are
available. In such cases, the centre-of-mass velocity and the mass ratio of a
double-lined binary may be computed by the method of Wilson (\cite{olinw41})
without a full orbital solution, if the velocities can be properly assigned to
the two components. For the 510 systems for which this has been possible, the
systemic velocity is given instead of the raw average of the observations, and
the mass ratio is given in Table 2 (electronic form only). The binary
population of the sample is further discussed in Sect. \ref{binaries}.

Fig.~\ref{vrstat} shows the distribution of the mean errors of the mean radial
velocities in the sample (upper panel), and of the time span covered by the
observations of each star (lower panel). As will be seen, the mean error of a
mean radial velocity is typically $\sim$0.25 km~s$^{-1}$ and only rarely
exceeds 1 km~s$^{-1}$. The observations typically cover a time span of 1-3
years, but occasionally extend over more than a decade.

\subsection{Rotational velocities}\label{rotation}

For stars with significant rotation, the width of the cross-correlation profile
is a good indicator of $v$sin$i$. For the stars with CORAVEL observations,
$v$sin$i$ has been computed using the calibrations of Benz \& Mayor
(\cite{benzm80,benzm84}). For the CfA observations, the $v$sin$i$ of the
best-fitting template spectrum is a good measure of the rotation of the
programme star (Nordstr{\"o}m et al. \cite{bnetal94,bnetal97}).

\begin{figure}[htbp] 
\resizebox{\hsize}{!}{\includegraphics[angle=0]{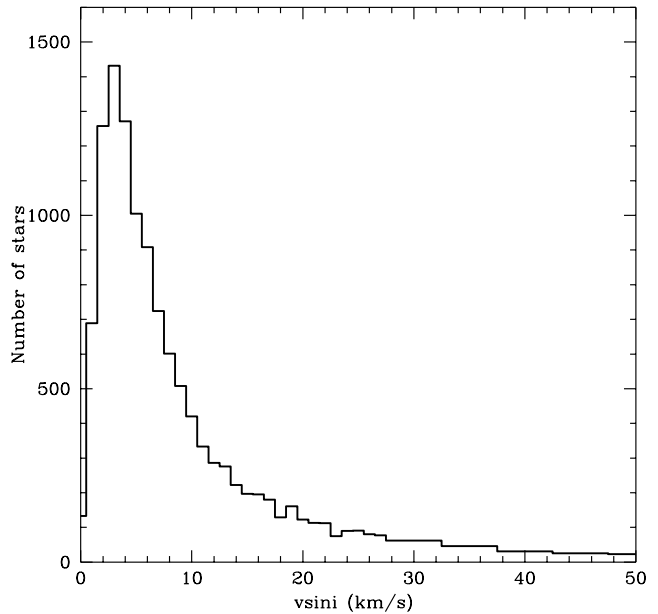}}
\caption{Distribution of rotational velocities in the sample.} 
\label{vsini} 
\end{figure}

For the slowest rotators, more elaborate procedures are needed to derive very
accurate rotational velocities; for the fastest rotators, the shallow
cross-correlation profiles yield results of low accuracy. Accordingly,
$v$sin$i$ as derived from the observations are only given to the nearest
km~s$^{-1}$, and from 30 km~s$^{-1}$ and upwards only to the nearest 5 or 10
km~s$^{-1}$. As seen in Fig. \ref{vsini}, the great majority of the programme
stars have rotations below 20 km~s$^{-1}$.

\subsection{Parallaxes}\label{pi}

\begin{figure}[htbp] 
\resizebox{\hsize}{!}{\includegraphics[angle=0]{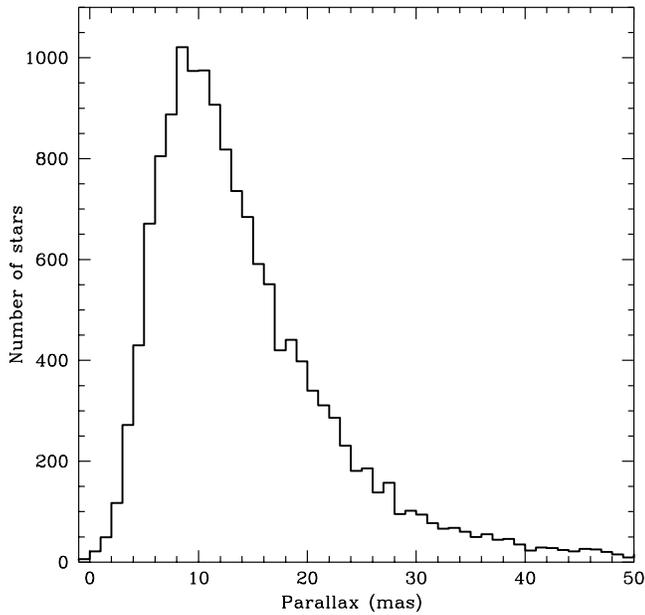}}
\resizebox{\hsize}{!}{\includegraphics[angle=0]{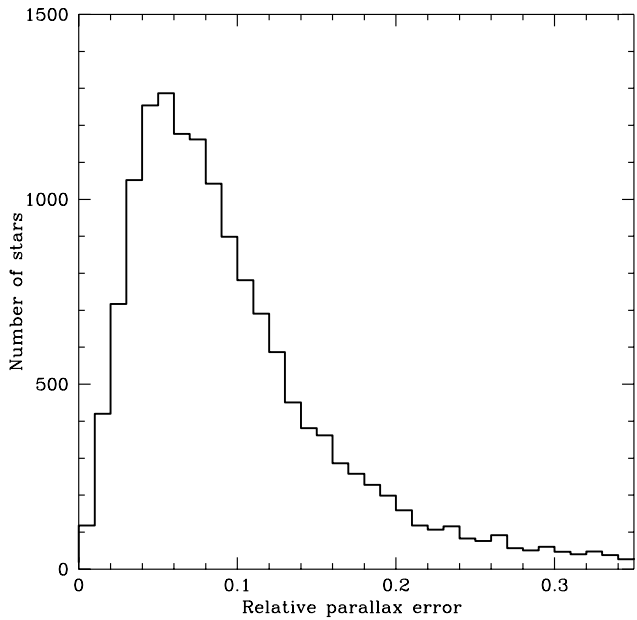}}
\caption{ Distribution of Hipparcos parallaxes ($top$) and their relative
errors ($bottom$) for the whole sample.} 
\label{parallax} 
\end{figure}

Good distances are crucial in order to compute accurate absolute magnitudes,
space motions, and parameters derived from them. Trigonometric parallaxes,
generally of very good accuracy, are available from Hipparcos for the majority
of our relatively nearby programme stars (ESA \cite{hipp97}).
Fig.~\ref{parallax} shows the distribution of the parallaxes ($\pi$) and their
relative errors ($\sigma_\pi/\pi$); most are better than 10\%, nearly all
better than 20\%. The computation of distances for all our stars is discussed
in Sect. \ref{distance}.

\subsection{Proper motions}\label{mu}

Accurate proper motions are available for the vast majority of the stars from
the Tycho-2 catalogue (H{\o}g et al. \cite{tycho2}). This catalogue is
constructed by combining the Tycho star-mapper measurements of the Hipparcos
satellite with the Astrographic Catalogue based on measurements in the Carte
du Ciel and other ground-based catalogues. By this procedure the baseline for
determining proper motions is extended up to nearly a century, against only
3.5 years for the Hipparcos mission itself.

For a few stars, mostly very bright stars or close binaries without a Tycho-2
proper motion, a Hipparcos or Tycho measurement has been used instead. The
typical mean error in the total proper motion vector is 1.8 milliarcsec/year,
corresponding to a mean propagated error in the space velocity of 0.7
km~s$^{-1}$ from the proper motions alone (i.e., neglecting parallax and 
radial-velocity errors).

\section{Derived astrophysical parameters}\label{calibrations}

In order for the stellar data to be useful in discussing the evolutionary
history of the Solar neighbourhood, a number of astrophysically interesting
parameters must be derived from the raw observational data. In most cases,
calibrations of the photometric indices in terms of intrinsic parameters are
found in the literature, except as noted below. We discuss each calibration
in turn in the following.

\subsection{Interstellar reddening}\label{excess}

\begin{figure}[htbp] 
\resizebox{\hsize}{!}{\includegraphics[angle=0]{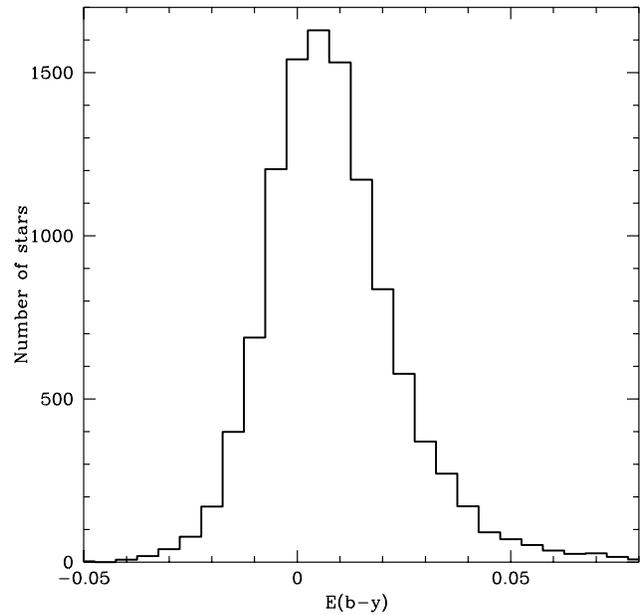}} 
\caption{Distribution of reddening values in the sample.} 
\label{ebyhist} 
\end{figure}

{\em E(b-y)} can be computed for F stars with $\beta$ observations from the
intrinsic colour calibration by Olsen (\cite{eho88}). It has been applied in
the photometric temperature and distance determinations if {\em E(b-y)}
$\ge$0.02 and the distance is above 40 pc; otherwise the stars are assumed to
be unreddened. Most stars with no value of {\em E(b-y)} are late-type dwarfs
within 40 pc, which will have negligible reddening anyway. As seen in Fig.
\ref{ebyhist}, very few of the stars have {\em E(b-y)} $>$ 0.05 mag. We note
that {\em E(b-y)} may be overestimated for the hottest, brightest, and most
distant early F stars (Burstein \cite{burst03}).

\subsection{Effective temperatures}\label{Teff}

\begin{figure}[htbp] 
\resizebox{\hsize}{!}{\includegraphics[angle=0]{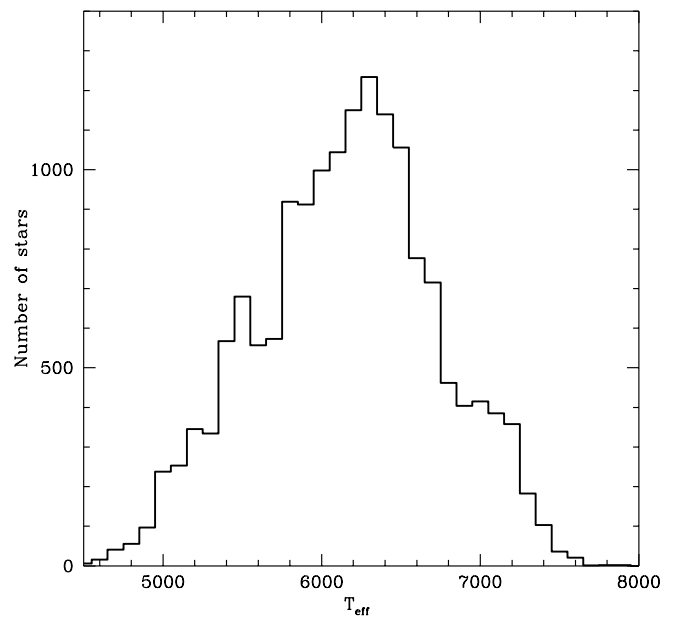}} 
\caption{Distribution of the sample in $T_{eff}$.}
\label{tehist}
\end{figure}

Effective temperatures for all the programme stars have been determined from
the reddening-corrected $b-y,$ $c_1$, and $m_1$ indices and the calibration of
Alonso et al. (\cite{alonso96}) which is based on the infrared flux method.
The resulting temperatures have been compared to the determinations by
Barklem et al. (\cite{barklem02}), based on a fit to the Balmer line wings
using the latest broadening theory. The agreement is excellent, with a mean
difference of only 3 K and a dispersion of 94 K. We have also compared our
results to the spectroscopic excitation temperatures determined by Bensby et
al. (\cite{bensby03}) for 63 of our stars; the latter are on average 93 K
higher than ours, with a dispersion around the mean of only 57 K. The
distribution of the effective temperatures in the sample is shown in Fig.
\ref{tehist}.

\subsection{Metal abundances}\label{FeH}

The accurate determination of metallicities for F and G stars is one of the
strengths of the Str{\"o}mgren $uvby\beta$ system. Among the available
calibrations, we have used that by Schuster \& Nissen (\cite{schuni89}) for the
majority of the stars. In our sample, $\sim$600 stars are covered by both the F
and G star calibrations of Schuster \& Nissen (\cite{schuni89}), and the mean
difference in [Fe/H] is 0.06 with a dispersion around the mean of only 0.07. We
have further compared these photometric metallicities with the homogeneous
spectroscopic values for F and G stars by Edvardsson et al. (\cite{edv93}) and
Chen et al. (\cite{chenyq00}). The agreement is excellent, with mean
differences of only 0.02 and 0.00 dex and dispersions around the mean of 0.08
and 0.11, respectively. A further comparison with the compilation by Taylor
(\cite{taylor03}) shows a mean difference of only 0.01 dex, but a larger
dispersion of 0.12 dex, as expected for a compilation from many sources of
varying quality. 

Within the range of validity of the Schuster \& Nissen (\cite{schuni89})
calibration, we thus find the photometric metallicities to have no
significant zero-point offset and remarkably small dispersion when compared
to high-quality spectroscopic values. However, as pointed out most recently
by Twarog et al. (\cite{twarog02}), the Schuster \& Nissen (\cite{schuni89})
calibration seems to give substantial systematic errors in the metallicity
computed for the very reddest G and K dwarfs ($b-y > 0.46$), where very few
spectroscopic calibrators were available at that time. Because our sample
contains an appreciable number of such red stars and more spectroscopic
metallicities in this range have become available, we decided to derive an
improved metallicity calibration for these stars, as follows:

\begin{figure}[thbp] 
\resizebox{\hsize}{!}{\includegraphics[angle=0]{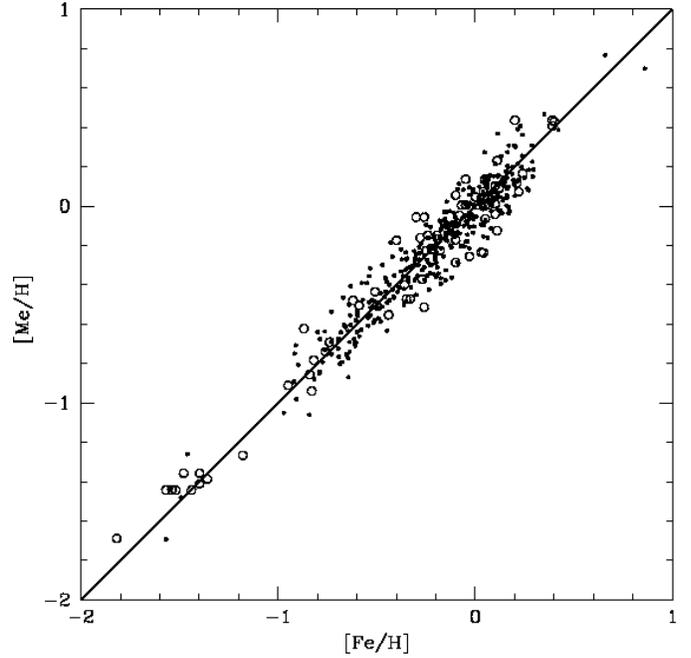}} 
\caption{Comparison between our final photometric metallicities ([Me/H]) and
the spectroscopic ([Fe/H]) values used to establish the calibrations. Open 
circles denote the cool (GK) stars, dots the hot (F) stars (see text).}
\label{fehcomp}
\end{figure}

From the high-resolution spectroscopic studies of Flynn \& Morell
(\cite{flynnmo97}), Tomkin \& Lambert (\cite{tomkin99}), Thor{\'e}n \&
Feltzing (\cite{thoren00}), and Santos et al. (\cite{santos01}), we have
extracted metallicities for 72 dwarf stars in the colour range 0.44 $\leq b-y
\leq$ 0.59 and performed a new fit of the $uvby$ indices to these values,
using the same terms as the Schuster \& Nissen (\cite{schuni89}) G-star
calibration. The resulting calibration equation is: \\

\noindent $[Fe/H] = -2.06+24.56m_{1}-31.61m_{1}^{2}-53.64m_{1}(b-y) \\ 
~~~+73.50m_{1}^{2}(b-y)+[26.34m_{1}-0.46c_{1}-17.76m_{1}^{2}]c_{1}$ \\

The fit of the photometric metallicities from this calibration to the
spectroscopic reference values is shown in Fig.~\ref{fehcomp} ({\it open
circles}). The dispersion around the (zero) mean is 0.12 dex.

Spectroscopic abundances for such cool dwarfs remain affected by both
observational and theoretical uncertainties (see, e.g. Thor{\'e}n \& Feltzing
\cite{thoren00}). However, the determinations selected here seem to represent
the current state of the art, and we have used our new calibration to compute
photometric metallicities for the $\sim$1500 stars in our sample with $b-y>$
0.46. For the $\sim$600 stars in the interval $0.44<b-y<0.46$, the new
calibration agrees with that by the Schuster \& Nissen (\cite{schuni89}) to
within 0.00 dex in the mean, with a dispersion of 0.12 dex.

About 2400 of our stars with high temperatures and low gravities are outside
the range covered by the Schuster \& Nissen (\cite{schuni89}) calibration.
For these stars we have adopted the calibration of $\beta$ and $m_1$ by
Edvardsson et al. (\cite{edv93}), when valid. For the stars in common, the
two calibrations agree very well (mean difference of 0.00 dex, dispersion
only 0.05). For stars outside the limits of both calibrations, we have
derived a new relation, using the same terms as  Schuster \& Nissen
(\cite{schuni89}) for F stars. In addition to the above new spectroscopic
sources, we used Burkhart \& Coupry (\cite{burkhart91}), Glaspey et al.
(\cite{glaspey94}) and Taylor (\cite{taylor03}) to extend the coverage in
{\em b-y}, $m_1$, $c_1$, and [Fe/H]. From 342 stars in the ranges: 0.18 $\leq 
b-y \leq$ 0.38, 0.07 $\leq m_{1} \leq$ 0.26, 0.21 $\leq c_{1} \leq$ 0.86 and 
-1.5 $\leq [Fe/H] \leq$ 0.8, we derive the following calibration equation:\\

\noindent $[Fe/H] = 9.60-61.16m_{1}+81.25m_{1}(b-y)\\
~~~~~~~~~+224.65m_{1}^{2}(b-y) -153.18m_{1}(b-y)^{2}\\
~~~~~~~~~+[12.23-90.23m_{1}+38.70(b-y)]\log(m_{1}-c_{3}),$ \\

\noindent where $c_3 = 0.45-3.98(b-y)+5.08(b-y)^{2}$. 

\begin{figure}[bhtp] 
\resizebox{\hsize}{!}{\includegraphics[angle=0]{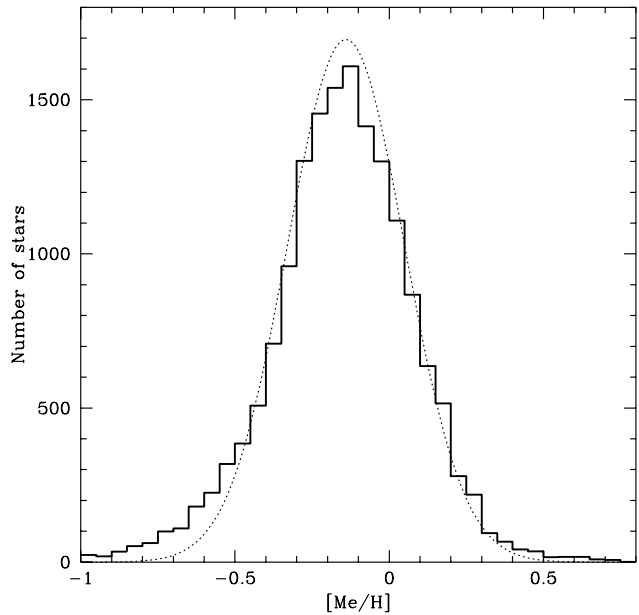}} 
\caption{Distribution of metallicities for the whole sample (full 
histogram). For comparison, the dotted curve shows a Gaussian distribution 
with mean of -0.14 and dispersion of 0.19 dex, covering the same area as the 
histogram.}
\label{mehhist} 
\end{figure} 

The fit of these photometric metallicities to the spectroscopic values is shown
in Fig.~\ref{fehcomp} ({\it dots}); the dispersion around the relation is 0.10
dex. For the stars in common, the new calibration and that by Schuster \&
Nissen (\cite{schuni89}) again agree very well (mean difference 0.02 dex,
dispersion only 0.04). More detail on the new calibration is given by Holmberg
(\cite{holmb04}). 

The distribution of the photometric metallicities derived as described above is
shown in Fig.~\ref{mehhist}. A Gaussian curve (with a mean of -0.14 and a
dispersion of 0.19 dex) has been plotted to highlight the tail of metal-poor
stars in the real distribution. This metallicity distribution for F- and G-type
dwarfs is almost identical to the one found for K-type giants by Girardi \&
Salaris (2001), with a mean of -0.12 and a dispersion of 0.18 dex.

\subsection{Distances and absolute magnitudes}\label{distance}

Most of our programme stars are nearby and have trigonometric parallaxes of
excellent quality from Hipparcos (see Sect. \ref{pi} and Fig.~\ref{parallax}).
We have therefore chosen to first determine distances for our stars based on
the Hipparcos parallaxes, either directly or indirectly. The distances are used
to compute tangential space motion components from the proper motions, and
absolute magnitudes used in the determination of ages and masses. 

\begin{figure}[htbp] 
\resizebox{\hsize}{!}{\includegraphics{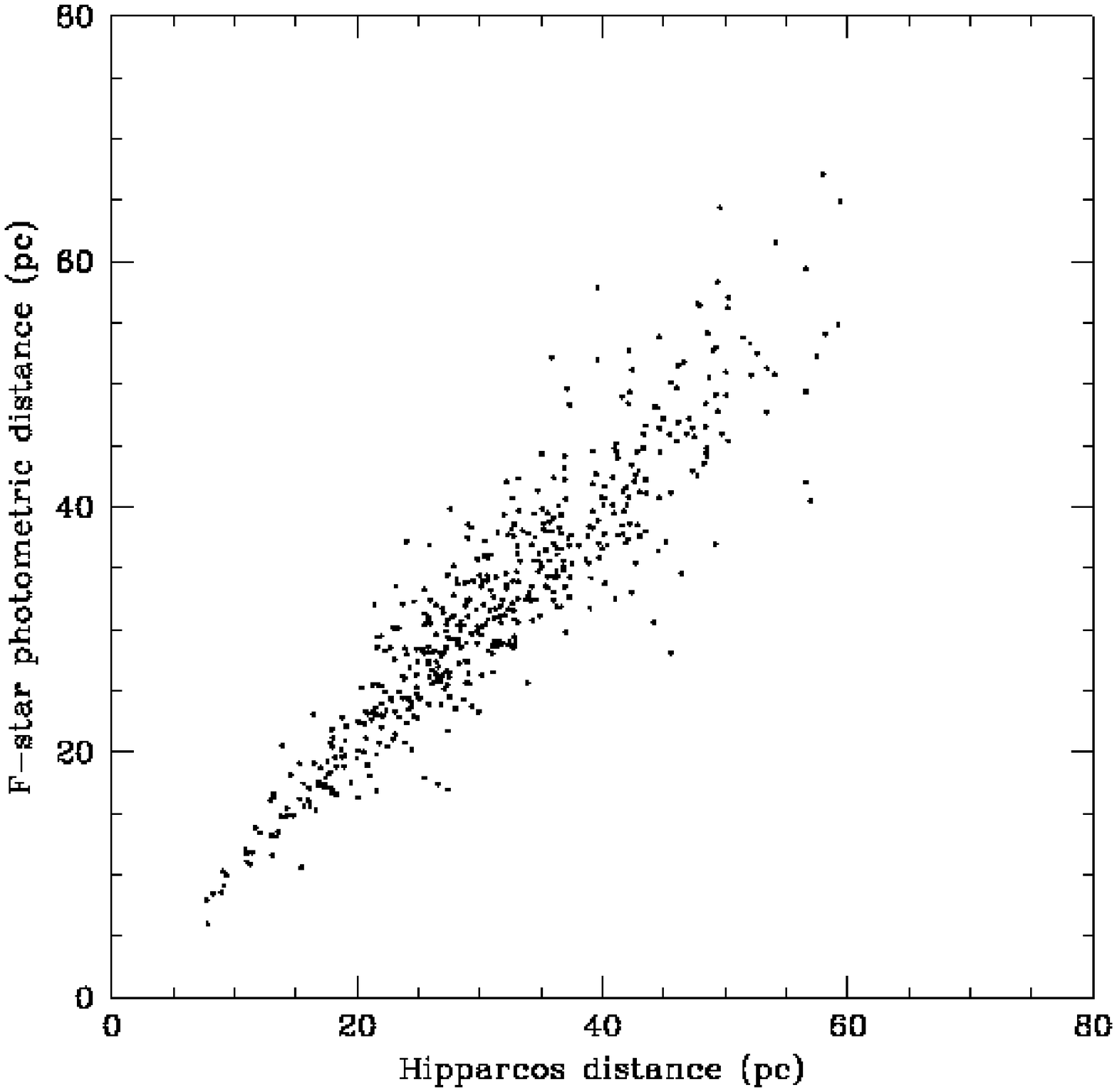}}
\resizebox{\hsize}{!}{\includegraphics{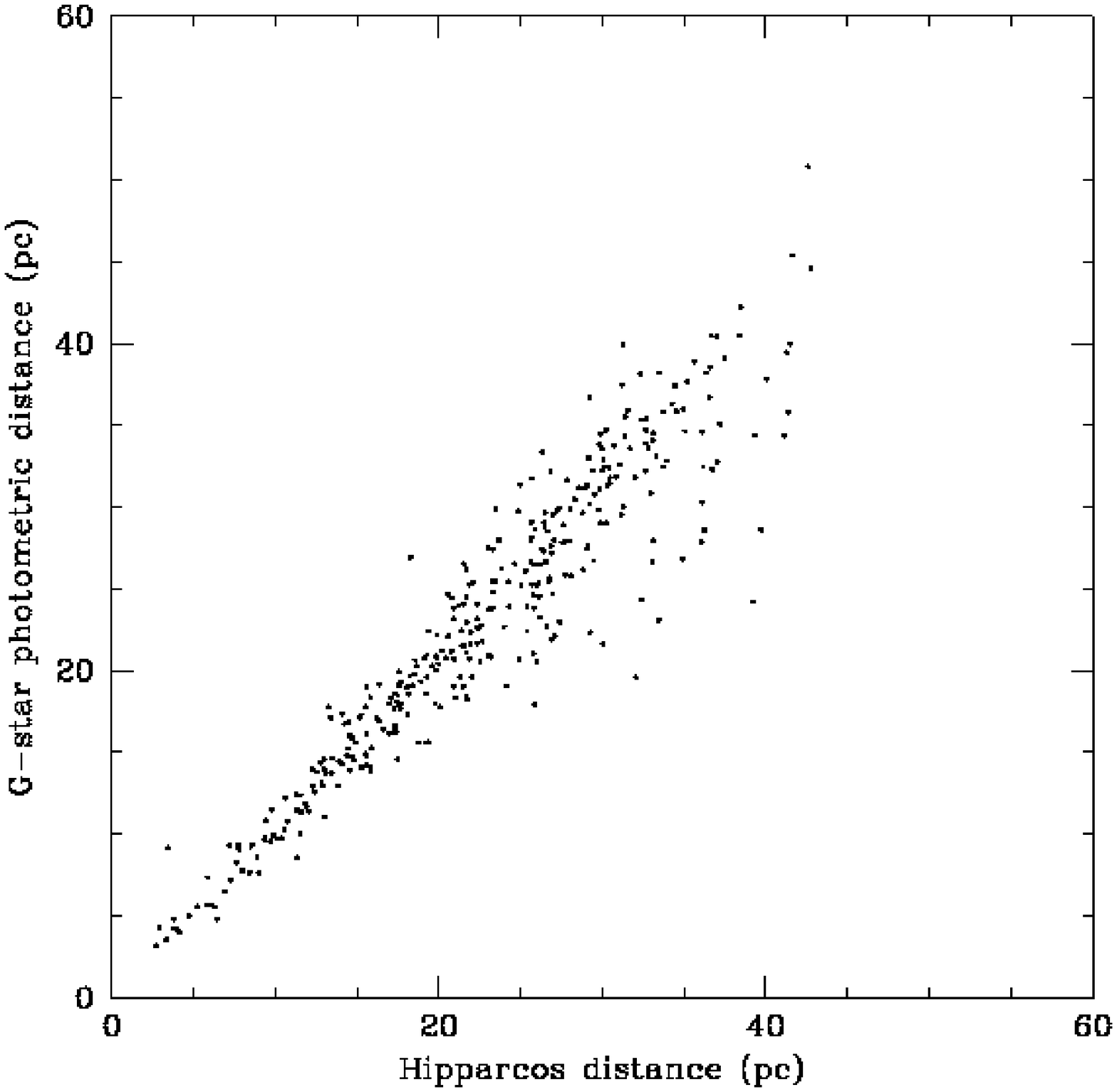}}
\caption{Photometric vs. Hipparcos distances for the single main sequence 
stars with parallax errors below 3\%. {\it Top:} F dwarfs; {\it bottom:} G
dwarfs.} 
\label{hipcomp}
\end{figure}

When the Hipparcos parallax is either unavailable or less accurate, a
photometric parallax is used. We have adopted the distance calibrations for F
and G dwarfs by Crawford (\cite{craw75}) and Olsen (\cite{eho84}); if both are
valid for the same star, the F star calibration is preferred (Note that this
calibration requires a $\beta$ value). We have checked the photometric
distances against the subset of Hipparcos parallaxes with relative errors below
3\% (Fig.~\ref{hipcomp}). The trigonometric and (distance independent)
photometric parallaxes agree very well, with no significant colour-dependent
bias: The photometric distances have an uncertainty of only 13\%. 

Accordingly, the Hipparcos distance is adopted if the parallax is accurate to
13\% or better; otherwise we adopt the photometric distance. However, the
photometric distance calibrations are not valid for binaries, giants, and many
kinds of peculiar stars. Such stars reveal themselves by large discrepancies
between the trigonometric and photometric distance estimates. The (few) stars
with photometric distances deviating more than $3\sigma$ from the Hipparcos
distances are flagged in the catalogue as suspected binaries or giants, and no
photometric distance is given if the Hipparcos parallax is too inaccurate as a
distance indicator on its own (235 stars). Similarly, no distance is given for
stars which lack the necessary photometry (typically the $\beta$ index) and/or
reliable Hipparcos parallaxes or fall outside the photometric calibrations
(1214 stars). These stars are all quite distant and of marginal relevance to
the overall sample.

\begin{figure}[htbp] 
\resizebox{\hsize}{!}{\includegraphics[angle=0]{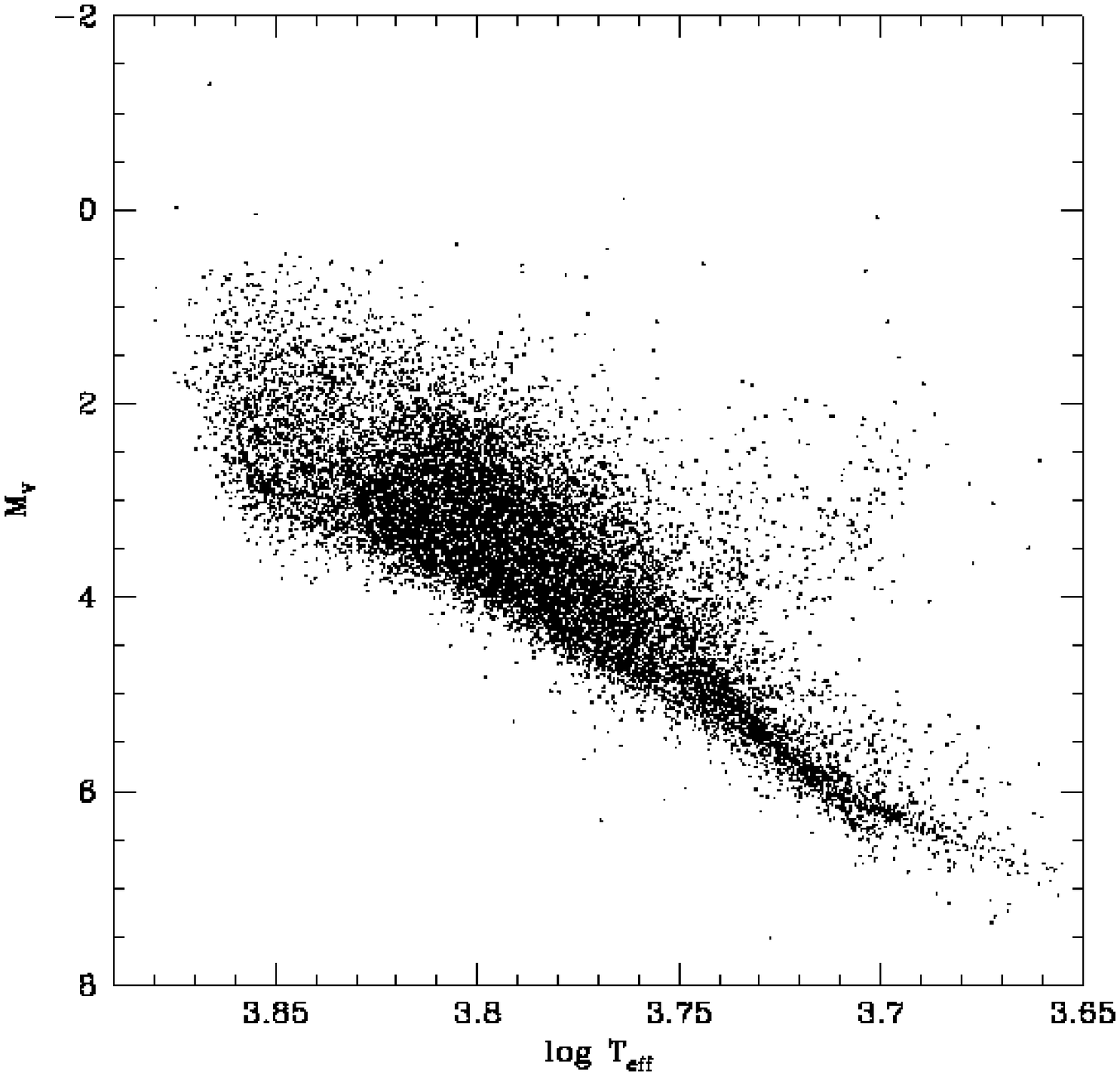}}
\resizebox{\hsize}{!}{\includegraphics[angle=0]{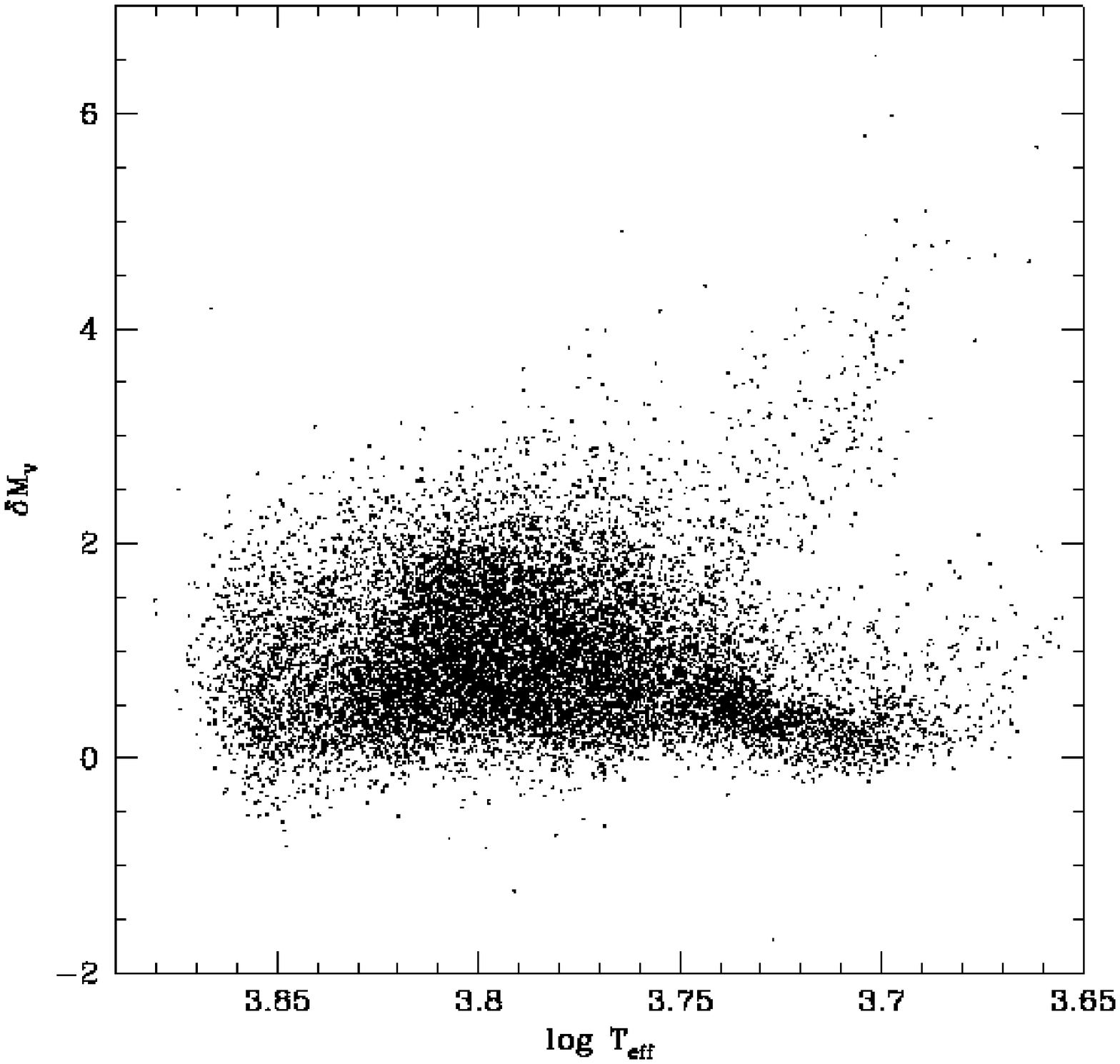}} 
\caption{$M_{V}$ ($top$) and $\delta M_{V}$ ($bottom$) vs. $\log T_{eff}$ 
for our sample, which by design consists of F and G dwarf stars.} 
\label{mvdmv}
\end{figure}

From the adopted distances and the observed $V$ magnitudes we have computed
the absolute magnitudes given in the catalogue, correcting for interstellar
extinction when known. Moreover, a $\delta M_{V}$ index has been calculated
as the magnitude difference between the star and the theoretical ZAMS at the
same colour and metallicity, as an indicator of the degree of evolution of
each star. Fig.~\ref{mvdmv} shows the distribution of $M_{V}$ and $\delta
M_{V}$ values for the sample.

\subsection{Ages}\label{ages}

Individual stellar ages are crucial in order to place the observed chemical
and kinematical properties of the stars in an evolutionary context. Because
of their importance, we have devoted a great deal of effort to finding the
most reliable way to determine ages for the stars in our sample and assessing
their individual and systematic errors. 

We start by noting that observable diagnostics of stellar ages are basically:
{\it (i):} chromospheric activity, and {\it (ii):} evolution away from the ZAMS
in the HR diagram. Both have strengths and limitations, as discussed recently
by Lachaume et al. (\cite{lachaume99}) and Feltzing et al. (\cite{feltz01}).

Chromospheric age determinations rely on the decline of stellar activity with
time (see Soderblom et al. \cite{soderb91} for an overview). The strength of
this technique is that it can be used for both F, G, and K-type dwarfs,
including very young stars. A drawback is that the chromospheric activity
indicators (e.g. X-ray and Ca II emission) decay into invisibility at about the
age of the Sun, so the method cannot be used for the older stars which are of
main interest for Galactic evolution. A more basic problem is the time
variability of stellar activity, similar to the activity cycles and more
dramatic phenomena of the Sun, such as the Maunder minimum. Further, stellar
activity is caused mainly by rotation, which decreases with age but can be
influenced by, e.g. tidal interaction in binary systems (Kawaler 1989). A
chromospheric age can therefore be completely wrong for reasons that cannot be
clarified without additional (substantial) observational data. 

Isochrone ages are determined by placing the stars in the theoretical HR
diagram (Fig. \ref{mvdmv}), using the observed $T_{eff}$, $M_v$, and [Fe/H] and
reading off the age (and mass) of the stars by interpolation between
theoretically computed isochrones. Edvardsson et al. (\cite{edv93}) exemplify
this technique in the present context. Given the presence of observational
errors, isochrone ages can only be determined for stars that have evolved
significantly away from the ZAMS, where all the isochrones converge. This
precludes the determination of reliable isochrone ages for unevolved (i.e.
relatively young) stars, and also for G and K dwarfs which evolve along the
ZAMS for the first long period of their life.

Many of our stars are considerably older than the Sun. Moreover, chromospheric
activity indicators exist only for a small fraction of them. Accordingly, we
have chosen to derive isochrone ages for our stars, recognising that meaningful
results will not be possible for all our stars.

\subsubsection{Selection of stellar models}\label{selectmod}

Selecting an appropriate set of theoretical evolution models and verifying its
correspondence with the observed stars is the first crucial step in any
determination of isochrone ages. The youngest stars in our sample are massive
enough that the stellar models must incorporate convective core overshooting
where appropriate. Several such models exist and are, in fact, very similar in
the theoretical plane, but employ rather different transformations to the
standard colour systems (see, e.g., the detailed comparison in Nordstr{\"o}m et
al. \cite{naa97}). We have preferred, therefore, to compute effective
temperatures and luminosities for the programme stars and compare with the
models directly in the $\log T_{eff} - M_v$ plane.

In preparation, we have compared the latest models from both the Geneva
(Mowlavi et al. \cite{nami98}; Lejeune \& Schaerer \cite{lejeune01}) and Padova
groups (Girardi et al. \cite{girardi00}, Salasnich et al. \cite{salasn00}). The
two sets of models yield essentially the same ages (to within 10\%), but have
significant and different limitations for our purposes: The Geneva models
extend to very large ages, but are computed for a relatively coarse grid of
masses $\geq 0.8 M_{\odot}$, which precludes a proper determination of masses
and ages for our coolest dwarf stars and leads to some numerical problems in
the detailed isochrone interpolations. The Padova models extend to stars well
below the lower mass limit of our sample, but the isochrones are terminated at
an age of 17.8 Gyr, complicating the proper computation of mean ages and age
errors for the oldest stars in our sample. Ages much in excess of 17.8 Gyr
exceed all recent estimates of the age of the Universe, however, so on balance,
we have chosen the Padova models for the final age determination.

\subsubsection{Choice of model compositions}

The next issue concerns the choice of chemical composition for the models. It
has long been known (e.g Edvardsson et al. \cite{edv93}, Reddy et al.
\cite{reddy03}) that disk stars with [Fe/H] $<$ 0 exhibit an average
enhancement of the $\alpha$-elements which rises approximately linearly to
[$\alpha$/Fe] $\simeq +0.25$ at [Fe/H] = -1 and remains constant at that or
perhaps a slightly higher level in even more metal-poor stars. The total
heavy-element content of metal-deficient disk stars is thus somewhat higher
than the heavy-element content of the Sun scaled by the observed [Fe/H].

Moreover, recent work (e.g. Fuhrmann \cite{fuhrm98}, Bensby et al.
\cite{bensby03}) has found that thick-disk stars appear somewhat more
$\alpha$-enhanced than thin-disk stars in the range $-1 <$ [Fe/H] $< 0$, which
spans the vast majority of our sample. There is, however, no consensus on a
precise criterion to distinguish between stars of the thin and thick disks, in
particular whether thick-disk stars are all extremely old and/or all moderately
metal-poor. This makes it impractical to identify the $\sim 5\%$ thick-disk
stars and estimate separate $\alpha$-enhancements for thin- and thick-disk
stars. Moreover, while Padova isochrones are available for the Solar mixture of
heavy elements as well as with an enhanced $\alpha$-element content for some
values of [Fe/H], the assumed $\alpha$-enhancement ([$\alpha$/Fe] $\simeq
+0.35$) is considerably greater than appropriate for most of our stars.

Fortunately, a simpler procedure appears sufficient. As demonstrated most
recently by VandenBerg (\cite{davb00}), isochrones computed with solar-scaled
and $\alpha$-enhanced compositions are almost indistinguishable, provided the
total heavy-element content $Z$ remains constant. For the metal-poor stars, we
therefore select scaled-Solar composition Padova isochrones with a somewhat
higher [Fe/H] than the directly observed value, as described below. 

Specifically, we assume an $\alpha$-enhancement that is zero for Fe/H] $\geq
0$, rises linearly with decreasing [Fe/H] to +0.25 dex at [Fe/H] = -1.0 and
+0.4 dex at [Fe/H] = -1.6, then remains flat at +0.4 dex at all lower
metallicities. Following VandenBerg (\cite{davb00}), we then increase the
observed [Fe/H] by 75\% of the corresponding value of [$\alpha$/Fe] as the best
estimate of the total heavy-element content of each star. We note that
VandenBerg (\cite{davb00}) found his procedure to be less satisfactory for the
most metal-rich compositions, but those results were derived for a constant
[$\alpha$/Fe] = +0.3 dex, even for [Fe/H] $> 0$. The above approximation should
remain fully satisfactory in the slightly metal-poor regime with the far
smaller $\alpha$-enhancements adopted here.

\begin{figure}[htbp] 
\resizebox{\hsize}{!}{\includegraphics[angle=0]{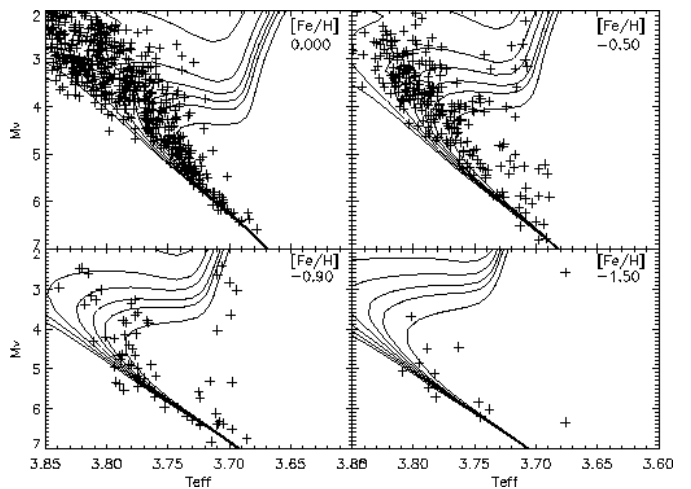}}
\caption{Comparison between the observed stars (known binaries excluded) and
Padova isochrones for 0, 2, 4, 6, 8, 10, and 15 Gyr at [Fe/H] = $0.00\pm0.02$,
$-0.50\pm0.05$, $-0.90\pm0.20$, and $-1.50\pm0.25$, after allowing for the
$\alpha$-enhancement and temperature corrections discussed in the text.} 
\label{isocomp}
\end{figure}

\subsubsection{Adjusting the temperature scales}\label{tempadj}

Having described our procedure for choosing models of appropriate heavy-element
content for stars of different metallicity, we ask whether these models provide
a satisfactory fit to the observed stars. This is especially important for the
effective temperatures, which are notoriously difficult to predict in an
absolute sense from stellar models as well as from observation (see, e.g.
Lebreton \cite{lebreton01}). A temperature mismatch between the models and
observed stars will enter directly into the derived ages. 

Because our sample is expected to include very old stars, the comparison must
be made on the unevolved main sequence, i.e. for $M_v >$ +5.5. Good agreement
is found for Solar and very metal-poor compositions, but for intermediate
values of [Fe/H] the models are too hot by small, but significant amounts, as
also found by Lebreton (\cite{lebreton01}). Ignoring this offset would drive
the low-mass stars to spuriously high ages, given the tight spacing of the
isochrones in this mass range.

Accordingly, we have applied temperature corrections to the models that amount
to a $\delta \log T_{eff}$ of -0.015 at [Fe/H] = -1.5, rising linearly to
$\delta \log T_{eff}$ = -0.022 at [Fe/H] = -1.0 and dropping linearly again to
zero at [Fe/H] = -0.3. With these corrections, we obtain the isochrone fits to
the lower main sequence shown in Fig. \ref{isocomp}, which we consider
satisfactory.

\subsubsection{Statistical biases in age determinations}\label{agebias}

The classical way to determine an isochrone age is to plot the observed stars
and computed isochrones together in the theoretical HR diagram, either the
$\log T_{eff} - M_V$ diagram (Fig. \ref{mvdmv}, top) or the $\log T_{eff}$ vs.
$\delta M_V$ variety (Fig. \ref{mvdmv}, bottom) used by Edvardsson et al.
(\cite{edv93}; see their Figs. 10-11). Errors are then derived by varying each
of the independent variables $\log T_{eff}$, $M_V$, and [Fe/H] by their
estimated observational errors and noting the changes in the resulting age.

However, age is a highly non-linear function of position in the HR diagram; the
distribution of the observed parameters is highly non-uniform as well; and the
observational errors are not always negligible compared to the ranges over
which these distributions and the derived ages vary considerably. An age
probability distribution function computed without regard to these effects will
therefore be biased; moreover, in the simple, ``classical'' approach it is also
incompletely sampled. Biased ages and misleading error estimates are the likely
result.

Statistical biases effecting the determination of isochrone ages include the
following: 

\begin{enumerate}

\item Stellar evolution accelerates strongly when stars leave the main
sequence; therefore, the density of stars in the HR diagram will be much higher
in the main-sequence region than away from it. This, in turn, causes more stars
to be scattered by observational errors from the main sequence into the
subgiant region than the reverse, and leads to a bias in favour of high ages.
Note that this effect is in fact exacerbated if stars with poorly determined
ages are eliminated, since unevolved stars necessarily have poorly-determined
ages.

\item Standard initial mass functions (IMF) rise towards lower stellar masses;
a given isochrone will therefore not have equal numbers of stars in equal mass
steps, but the density of stars will rise towards the ZAMS. Ignoring this
effect will also lead to a positive age bias.

\item Standard disk metallicity distributions (see Sect. \ref{gdwarf}) contain
many more metal-rich than metal-poor stars; observational errors will therefore
again scatter more stars from the metal-rich peak of the distribution into the
metal-poor tail than in the opposite direction. This will cause the
corresponding ages to be derived from isochrones that are too metal-poor, i.e.
too hot, and again a positive age bias results (the converse argument applies
to the tail of ``super metal-rich'' stars).

\item Apart from such ``intrinsic'' effects in the data, the distribution of
stars in the HR diagram will be non-uniform because, e.g. of a non-uniform age
distribution of the stars themselves, or as a result of the criteria used to
define the sample. Notably, the distribution in the HR diagram of our full
magnitude-limited sample will be quite different from that of the
volume-limited subsample, due to the inclusion of luminous, evolved stars from
much larger distances.

\end{enumerate}

\subsubsection{Age determination for the sample stars}\label{agedet}

The techniques we have developed to allow for these biases are superficially
similar to those discussed recently by Lachaume et al. (\cite{lachaume99}) and
Reddy et al. (\cite{reddy03}) as regards the treatment of the evolution bias
referred to above. There are, however, important differences, in that we treat
all three parameters $\log T_{eff}$, $M_V$, and [Fe/H] equally, include several
additional sources of bias, and consider the whole chain of astrophysical links
from data to age. Our method is outlined below and described in greater detail
by J{\o}rgensen \& Lindegren (\cite{bjarner04}).

Briefly, for every point in a dense grid of interpolated Padova isochrones we
compute the probability $P$ that the star could in reality be located there
(and thus have the corresponding age), given its nominal position in the
three-dimensional HR ``cube'' defined by $\log T_{eff}$, $M_V$, and [Fe/H]. To
do so, we assume that the associated observational errors have a Gaussian
distribution:

$P = \exp (-(\Delta T_{eff})^2 / 2 \sigma_{T_{eff}}^2) *
     \exp (-(\Delta M_v)^2 / 2 \sigma_{M_v}^2) *\\
     ~~~~~~~~~~~\exp (-(\Delta [Fe/H])^2 / 2 \sigma_{[Fe/H]}^2)$

\noindent Here, $\Delta T_{eff}$ etc. are the differences between the observed
parameters of the actual star and the isochrone points considered. We assume
constant errors ($\sigma$) of 0.01 dex in $\log T_{eff}$ and 0.1 dex in [Fe/H]
throughout; for $M_V$ we use the individual error estimate if an Hipparcos
parallax better than 13\% exists; otherwise, the standard photometric value of
0.28 mag in {\it (m-M)} is adopted.

\begin{figure}[htbp] 
\begin{center}
\resizebox{5.5cm}{!}{\includegraphics[angle=0]{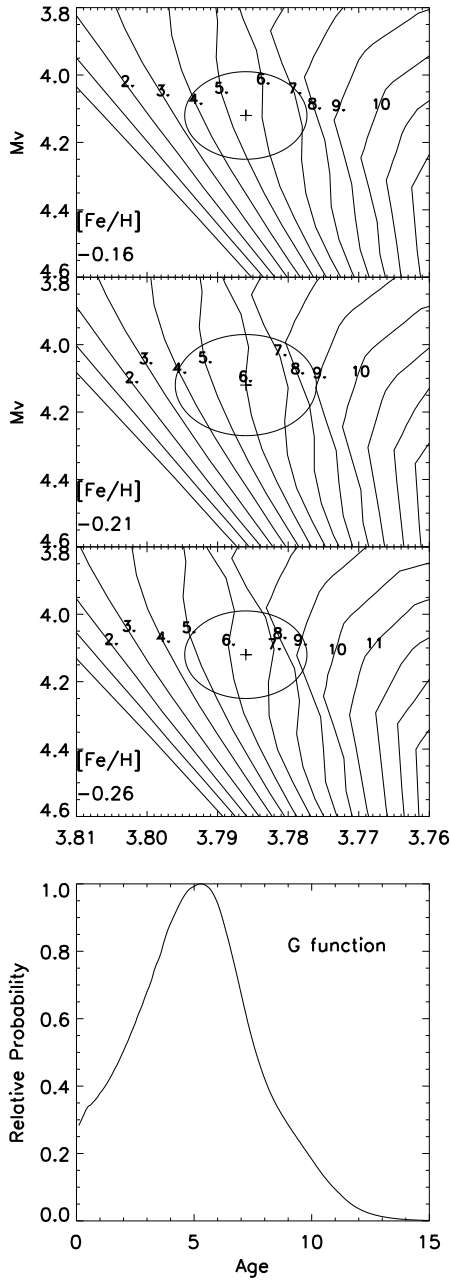}} 
\end{center}
\caption{{\it Top:} Three slices of the three-dimensional HR ``cube'' showing
the observed point (central panel) and three cuts through the 1-$\sigma$ error
ellipsoid. The probability distribution function (G-function, {\it bottom}) is
computed from $all$ points on the isochrones, not just those on or inside the
error ellipsoid.}
\label{gfunk_ok}
\end{figure}

Integrating over all points gives the global likelihood distribution for the
possible ages of the star, conditioned to account for observational biases as
described below. We call this the ``G-function'' and normalise it to unity at
maximum. The most probable age for the star is then determined as the value for
which the G-function has its maximum (see Fig. \ref{gfunk_ok}). 

The determination of the maximum value itself is a non-trivial task. Because of
numerical noise due to the finite sampling of the isochrones, a simple maximum
of the raw function results in spurious high-frequency features in the derived
age distributions which go undetected in small samples, but have dramatic
effects in densely-populated diagrams such as Figs. \ref{amrmag} and \ref{avr}.
The median of, say, the upper 50\% of the function yields a more stable
estimate, but if the corresponding age range includes one of the limits (0 or
17.8 Gyr), the estimate will be biased away from the limit, leading to
spuriously low ages for the oldest stars (cf. Fig. \ref{gfunk_bad}). A Gaussian
fit as used by Reddy et al. (\cite{reddy03}) is also more stable, but is a poor
approximation in the frequent cases when the G-function is distinctly non-
Gaussian (Fig. \ref{gfunk_bad}). 

After extensive tests with simulated and real data, smoothing the G-function
slightly with a kernel depending on the width of the unsmoothed function was
found to be the optimum procedure. The maximum of the smoothed function then
yields a stable age estimate without significant bias. Fig. \ref{gfunk_ok}
illustrates the procedure in the well-behaved case of a star located in a region
of the HR diagram where the isochrones are well separated, and the maximum of
the G-function yields a well-defined age.

Finally, the maximum value of the probability function found for any point on
the isochrones is a measure of the degree to which the star is covered by the
models. A small value signifies peculiar stars or large observational errors
which preclude any realistic age determination.

In computing the G-functions, we have accounted for the biases described in
Sect. \ref{agebias} as follows:
\begin{enumerate}

\item {\it Speed of evolution.} The isochrones are more widely separated in
phases of rapid evolution, so such phases automatically receive lower weight in
the integrations. We emphasize that not only points within a 1-$\sigma$ (or
3-$\sigma$, Reddy et al. \cite{reddy03}) error ellipse are included, but all
points on the isochrones in {\it all three} dimensions.

\item {\it Stellar mass function}. The varying density of stars on an isochrone
towards the main sequence is accounted for by weighting each point according to
the IMF (Kroupa et al. \cite{kroupa93}). It can be argued that the slope of the
actual distribution in the magnitude-limited sample will be lower than that of
the IMF due to the preferential inclusion of brighter, higher-mass stars, but
the effect is hard to quantify and in any case small, as the range in masses
covered by the sample is small.

\item {\it Metallicity bias.} The excess of apparently metal-poor stars caused
by observational scatter from the large peak of stars of near-Solar metallicity
is straightforward in concept. Allowing rigorously for it in practice is
another matter: A fully Bayesian approach requires an estimate of the a priori
distribution which is a priori unknown and, moreover, rather different for the
complete, magnitude-limited catalogue and for the volume-limited sample which
will no doubt be preferred in many applications (compare Figs. \ref{mehhist}
and \ref{gprob}); other subsamples would no doubt be different again. It
appears unreasonable that the catalogued age of a given star should depend on
the subsample of stars discussed together with it. 

Moreover, the first-order astrophysical effects considered earlier in the
procedure already seem to allow for these effects. First, if significant, the
metallicity bias should appear as an excess of positive residuals at low [Fe/H]
when photometric metallicities are compared with spectroscopic determinations;
this is not seen in our data (nor by Edvardsson et al. \cite{edv93}). Second,
our revised metallicity calibration (see Fig. \ref{fehcomp}), by design, yields
the correct mean spectroscopic [Fe/H] for a given mean photometric metallicity.

Finally, and probably most importantly in view of the sensitivity of the
derived ages to small temperature shifts, the samples of stars used to
``normalise'' the temperature scale of the models to that of the observed stars
(Fig. \ref{isocomp}) will already be affected by any residual metallicity bias.
Our temperature shifts will therefore allow for it to first order. In fact, it
could be argued that these corrections may, if anything, be too large because
the stars were drawn from the full, magnitude-limited sample which
preferentially includes young, metal-rich stars. 

In summary we believe that, for general use, little if any metallicity bias of
significance remains in the ages given in the catalogue. If particularly
precise ages are needed for certain types of stars, well-defined subsamples
should be extracted and all steps in the analysis reviewed and/or repeated,
including the temperature and metallicity calibrations, [$\alpha$/Fe] ratio(s),
model compositions and bolometric corrections, and the a priori distributions
of the relevant parameters. 

\item {\it Age bias etc.} A strongly peaked age distribution (e.g. due to a
starburst) could give biases analogous to those discussed above. No such peak
is expected, and its effects would again depend on the (sub)sample considered.
We have also not imposed any upper limit on the derived ages: While true ages
greater than $\sim$13 Gyr are implausible, observational error will cause some
determinations to exceed this limit, and imposing a cutoff will bias the mean
age of the oldest stars.

\end{enumerate}

In cases (1)-(2) we correct for the a priori information in a fully Bayesian
manner. Including also a priori metallicity and age distributions in the age
determination at this stage would build our prejudices concerning the
enrichment and star-formation history of the disk into our age estimates for
individual stars. We have therefore not made such corrections to the ages
listed in the catalogue (equivalent to assuming flat a priori distributions.) 

Using the above procedures, G-functions have been computed for all stars in our
sample with the necessary input data, including binary stars etc. An electronic
table of these functions, which illustrate the determinacy of each age
determination at a glance, will be made available to interested readers by
request to the corresponding author (B.R.J.; {\it bjarne@astro.lu.se}).

Bayesian probability theory offers an independent, alternative method to
compute unbiased age estimates under similar conditions, provided that the a
priori distributions of the relevant parameters can be estimated with
sufficient accuracy. An end-to-end Bayesian approach of this type is described
and applied to the data of Edvardsson et al. (\cite{edv93}) by Pont \& Eyer
(\cite{pont04}).

\subsubsection{Estimating errors for the ages}\label{ageerr}

Realistic error estimates are crucial in any applications of the ages. From a
well-behaved G-function such as that shown in Fig. \ref{gfunk_ok} we derive
(separate) 1-$\sigma$ lower and upper age limits as the points where the
G-function reaches a value of 0.6. Extensive Monte Carlo simulations using
artificial stars with typical observational errors have confirmed that indeed
68\% of the recovered ages then fall within $\pm 1\sigma$ of the correct age. 

When both the upper and lower 1-$\sigma$ age limits fall within the range of
the isochrones, 0-17.8 Gyr, the catalogue lists both the most likely age of the
star as well as its upper and lower limits. In the following, we refer to such
cases as {\it ``well-defined''} ages (note that this term by itself does not
imply a small error, only that the {\it error estimate} is reliable!). For more
demanding applications, the sample should no doubt be restricted to single
stars and perhaps also to stars with age errors below a specified limit; all
information needed to do so is readily available in the catalogue.

\begin{figure}[htbp] 
\resizebox{\hsize}{!}{\includegraphics[angle=0]{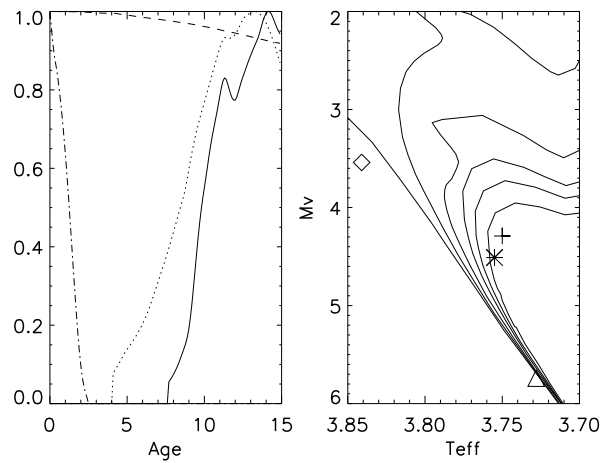}} 
\caption{{\it Right:} Examples of stars located in regions of the HR diagram
where reliable ages cannot be determined, and ({\it left}) the corresponding
G-functions. The stars plotted as plus, asterisk, triangle, and diamond symbols
in the HR diagram correspond to the solid, dotted, dashed, and dot-dashed
curves, respectively. Isochrones are for 0, 2, 4, 6, 8, and 10 Gyr.}
\label{gfunk_bad}
\end{figure}

If the G-function peaks within the valid age range but one of the limits is
outside it, that limit is not given in the table, indicating that the age is
uncertain and its error also poorly defined. For stars near or beyond the
limits of the isochrone set (very young or very old stars with large
observational errors, duplicity or other spectral peculiarities), the
G-function may peak at or even outside the age limits of the isochrones (see
Fig. \ref{gfunk_bad}). In such cases, no value is given for the age, only the
estimated upper or lower limits.

Finally, for the lowest-mass stars which have not evolved perceptibly, the
G-function will show no well-defined maximum (see Fig. \ref{gfunk_bad}). If the
G-function is too flat to reach the 1-$\sigma$ confidence level (0.6) anywhere
in the range 0-17.8 Gyr, or if the maximum probability value entering the
computation of the G-function indicates that the star falls significantly
outside the isochrone set, no age is given at all.

\begin{figure}[htbp] 
\resizebox{\hsize}{!}{\includegraphics[angle=0]{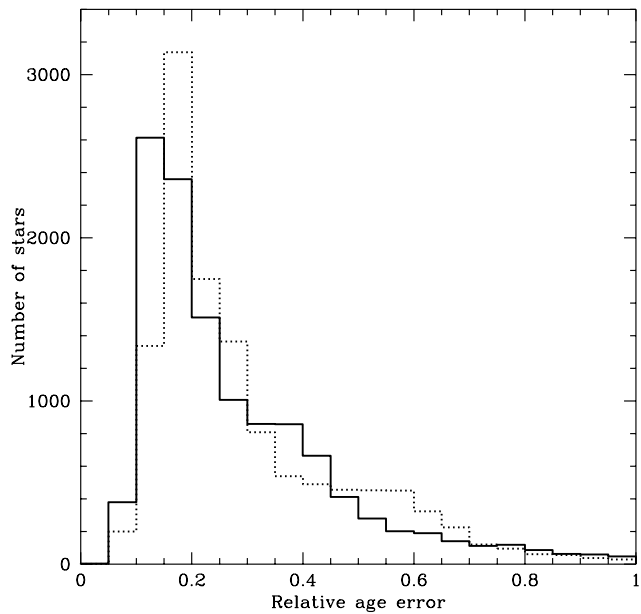}}
\caption{The distribution of upper (dotted line) and lower (solid 
line) 1-$\sigma$ relative errors for the ages in the catalogue.}
\label{relagedist}
\end{figure}

\begin{figure}[htbp] 
\resizebox{\hsize}{!}{\includegraphics[angle=0]{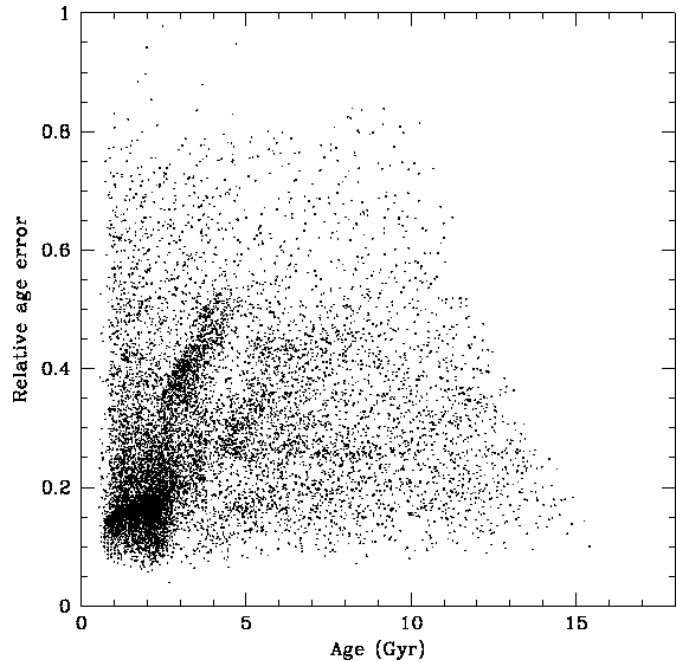}}
\caption{Relative age error (mean of lower and upper bounds) vs. age for the
11,445 stars with ``well-defined'' ages.}
\label{relagerror}
\end{figure}

The distribution of the relative age errors is shown in Fig. \ref{relagedist},
while Fig. \ref{relagerror} shows the mean relative age error as a function of
age for the stars with ``well-defined'' ages. Note that the condition that both
lower $and$ upper age limits should be determined removes old stars from the
upper right in Fig. \ref{relagerror}. The impression of increasing precision
with age that results when a numerical cutoff is mistaken for a physical upper
age limit (e.g. Fig. 4 of Feltzing et al. \cite{feltz01}) is, of course, an
illusion.

We have compared our error estimates with those derived in the classical
manner, i.e. by varying each input parameter by $\pm 1\sigma$ and adding the
age errors in quadrature. We find that the latter are often underestimated by
almost a factor 2. We attribute this to three causes: {\it (i):} Varying only
one parameter at a time significantly underestimates the true range of values
over which the age variations must be explored; {\it (ii):} the technique
effectively samples a total of only six points on the three-dimensional
probability function which we integrate in detail to compute the G-function;
and {\it (iii):} the standard way to add errors assumes implicitly that the
G-function is Gaussian, which is manifestly not the rule (cf. Fig.
\ref{gfunk_bad}). Some of the error estimation techniques for isochrone ages in
earlier literature have in reality estimated fitting errors rather than true
uncertainties, and the errors have likely been significantly underestimated in
several cases. 

In the total sample of 16,682 stars, these criteria yield age estimates for
13,636 stars (82\%), of which 11,445 stars (84\%) have ``well-defined'' ages by
the above definition. 9,428 (82\%) of these well-defined ages have estimated
errors below 50\%, 5,472 (47\%) even below 25\%. Eliminating known binaries of
all types leaves us with 9,158 presumably single stars (83\% of all 11,060 such
stars in the sample) with derived age values, 7,566 (83\%) of which have ages
that qualify as ``well-defined''. Of these in turn, 6,144 single stars (81\%)
have ages better than 50\% and 3,528 (46\%) better than 25\%, respectively. 

\begin{figure}[htbp] 
\resizebox{\hsize}{!}{\includegraphics[angle=0]{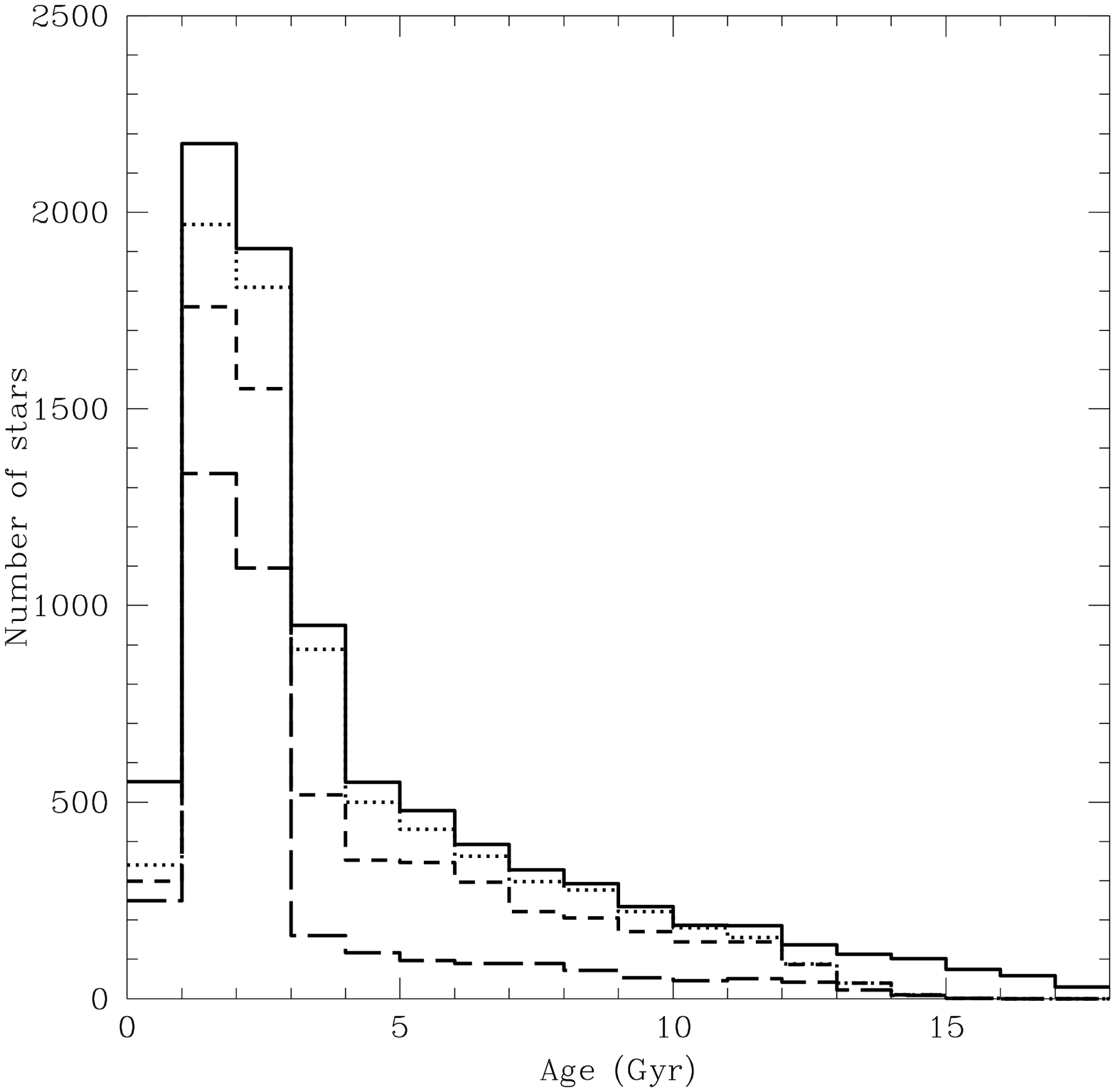}}
\resizebox{\hsize}{!}{\includegraphics[angle=0]{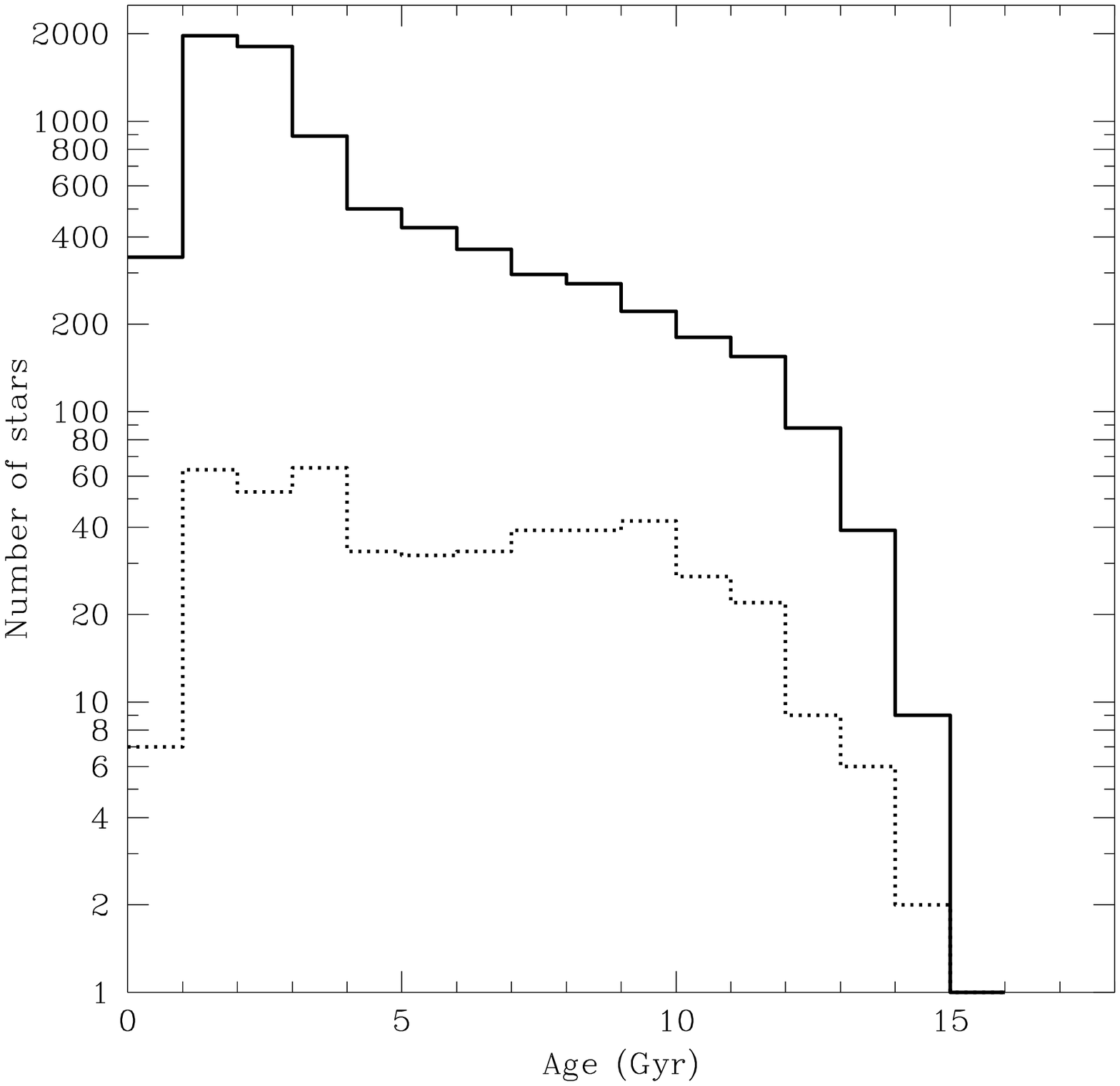}}
\caption{{\it Top:} Age distributions for single stars in the full sample. The
curves show (top to bottom): {\it (i)} all ages (solid); {\it (ii)}
``well-defined'' ages (dots); {\it (iii)} ages with errors $<50$\%
(short-dashed); and {\it (iv)} ages with errors $<25$\% (long-dashed). {\it
Bottom:} ``Well-defined'' ages in the full, magnitude-limited sample (solid)
and in the volume-limited subsample for $d < 40$ pc (dots). Note that
increasing the demand on accuracy progressively removes most of the oldest
stars.}
\label{agedist}
\end{figure}

Fig. \ref{agedist} shows the distributions of derived ages for the complete
(magnitude-limited) sample for increasingly strict limits on the accuracy of
the ages, and also compares the distributions of the magnitude-limited and
volume-limited samples. We stress that, due to the biases in the selection and
age computation procedures already discussed, {\it none} of these diagrams has
a simple interpretation in terms of the star formation history of the Solar
neighbourhood.

\subsubsection{Checking the results}\label{agecheck}

We have subjected our age derivation procedures to a wide range of numerical
and other checks. In analogy with the Monte Carlo simulations used to verify
the error estimates as described above, we have created artificial samples of
stars of specified ages and a range of masses from the isochrones, computed
$uvby\beta$ indices and $M_V$ values using the reverse transformations of those
applied to the observed data, and added realistic random errors to the
simulated observations. These artificial stars have then been subjected to our
age determination procedure in exactly the same manner as the observed stars.
We find that, within the limits of the computed errors, we recover the input
ages without significant systematic error (but of course with a loss of stars
in the parts of the HR diagram where ages cannot be determined).

Another reality check is to determine ages for stars in open clusters with good
$uvby$ photometry and known reddening (including, but not limited to such
favourable cases as NGC 3680; Nordstr{\"o}m et al. \cite{naa97}) as if they
were single stars. Again, good agreement is found with the results of detailed
isochrone fits to the entire cluster sequence, but of course the unevolved
lower main-sequence stars yield large error bars. Similar consistency checks
have been made in binary stars with good data for the individual components.
Further discussion of these tests is given by J{\o}rgensen \& Lindegren
(\cite{bjarner04}).

A particular concern was to ensure that our procedure does not introduce
metallicity-dependent systematic errors that could distort the resulting
age-metallicity relations (AMR). In order to do so, we have created
artificial AMRs of specified shape, with a distribution of metallicities at
each age corresponding to the observational scatter, and with a uniform
distribution of stellar masses from the ZAMS value to the maximum reached for
the assigned age and metallicity. For the experiment, we assumed both an AMR
with [Fe/H] increasing linearly with time and another (completely unphysical)
AMR in which [Fe/H] {\it decreased} linearly with time. Simulated observations
and realistic errors were computed and ages and metallicities rederived from
the artificial data. As before, many stars were lost for which reliable ages
could not be determined, but those with small calculated errors delineated the
input AMR without any systematic error -- a result which inspires confidence
in our method. Details on these simulations are given in Holmberg
(\cite{holmb04}).

\begin{figure}[htbp] 
\resizebox{\hsize}{!}{\includegraphics[angle=0]{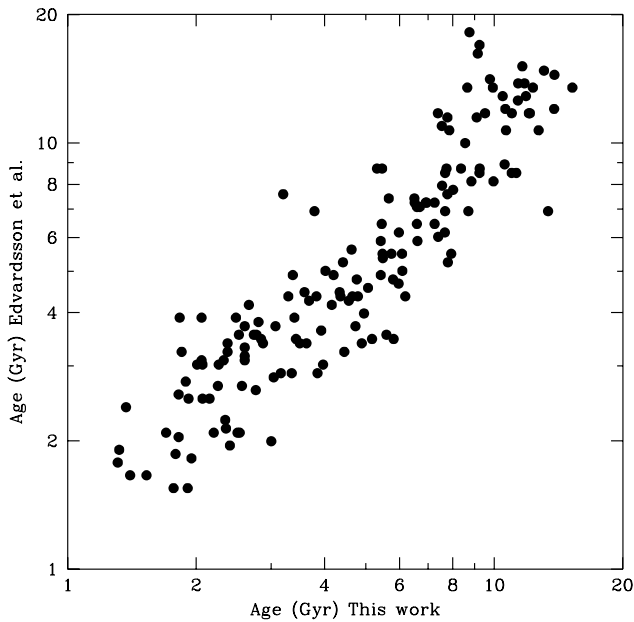}}
\caption{Ages from Edvardsson et al. (\cite{edv93}) vs. our results for the 160
stars in common with ``well-defined'' ages.}
\label{edvage}
\end{figure}

A further external check was made by comparing with the ages by Edvardsson et
al. (\cite{edv93}). Of their 182 stars with ages, our procedure yields
estimates of any quality for 179 stars and ``well-defined'' ages (see above)
for 160 stars. Fig. \ref{edvage} compares the two sets of ages for the latter
sample; a linear fit yields the relation\\
\noindent $\log Age_{Edv93} = (0.11\pm0.03)+(0.89\pm0.04)*\log Age_{our}$.\\
\noindent I.e., our ages are on average slightly smaller than those of
Edvardsson et al. (\cite{edv93}) for younger stars (where their $M_V$ derived
from $\delta c_1$ is biased towards brighter values), while our ages agree well
for the oldest stars. The scatter around the $45\degr$ line is 0.12 dex; the
fit only reduces it to 0.11 dex. Edvardsson et al. (\cite{edv93}) estimated
that their ages had errors of about $\pm0.1$ dex. Our estimated mean relative
error for the same sample is also 0.10 dex, somewhat larger for the youngest
and smaller for the oldest stars, but the sample excludes the very oldest stars
for which an upper age limit cannot be properly determined.

The agreement between our ages and those derived by Edvardsson et al.
(\cite{edv93}) is quite gratifying: One would have expected a larger dispersion
in Fig. \ref{edvage} if the errors in the two age determinations were
completely uncorrelated, which suggests that the errors are primarily
observational rather than systematic. We recall that even though the same
photometric data have been used, the stellar models (including opacities and
convection descriptions), temperature and metallicity calibrations, absolute
magnitude determinations (Hipparcos parallaxes vs. $\delta M_V$ derived from
the $\delta c_1$ index), treatment of the $\alpha$-enhancement in metal-poor
stars, and the method for computing the ages have all changed in the
intervening decade. Finding a relation such as that seen in Fig. \ref{edvage}
strengthens one's confidence in the whole procedure.

We have also compared our results with ages computed with the Bayesian method
of Pont \& Eyer (\cite{pont04}). Apart from a scale difference of $\sim25$\%,
due to a different metallicity scale and their choice of the median rather than
maximum of the probability distribution as the preferred age, there is no
significant difference between, e.g., age-metallicity relations derived with
the two sets of ages.

For completeness, we finally compared our new age determinations (with errors
$<$25\%) with the chromospheric ages derived by Rocha-Pinto et al.
(\cite{rocha00}) for the 85 stars in common in the range 1-20 Gyr. The plot is
quite similar to Fig. 8 of Feltzing et al. (\cite{feltz01}), with no trace of
any correlation between the two sets of ages. 

In conclusion, while noting the uncertainties, we consider our age scale to be
the best presently available for the whole sample. However, if the age of a
particular subgroup of stars, e.g. thick-disk stars, is needed to the highest
precision, then the detailed elemental composition of a carefully defined
sample of such stars should be determined and models of precisely the same
composition be computed to derive their ages. 

It must be emphasised that our reliability tests pertain to the ages derived
for {\it individual stars}. For any realistic distribution of true ages for a
{\it complete sample of stars}, one must take into account that many of the
least-evolved stars, especially the old low-mass stars, will have no derived
age in the catalogue, simply because the observations cannot measure their
evolution. We caution, therefore, that the true age distribution of the full
sample cannot be derived directly from the catalogue; careful simulation of the
biases operating on the selection of the stars and their age determination will
be needed to obtain meaningful results in investigations of this type.

Finally, we recall that many stars in our sample are binary or multiple
systems, for which the derived ages (and metallicities) will be unreliable. In
general, there is insufficient information available to recover the data for
the individual binary components from the combined photometry, but the great
majority of the visual and spectroscopic binaries in the sample are known and
identified in the catalogue (see \ref{binaries}), and can thus be excluded in
studies where absolute statistical completeness is not important.

\subsection{Masses}\label{mass}

Mass estimates are needed as the main clue to the evolutionary history of the
stars in the sample. Because each point on a model isochrone corresponds to a
specific mass value as well as an age, we can compute an M-function describing
the probability distribution of model masses for an observed star from the
Padova models, exactly analogous to the G-functions for the ages (see Sect.
\ref{agedet}).

\begin{figure}[htbp] 
\resizebox{\hsize}{!}{\includegraphics[angle=0]{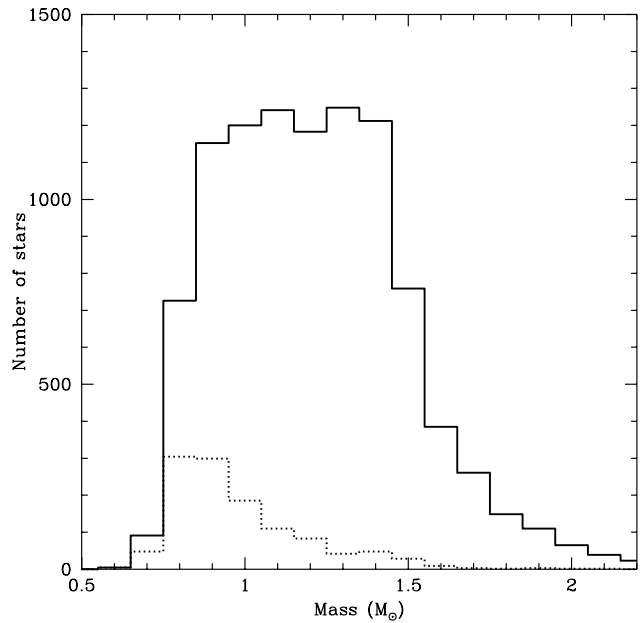}}
\caption{The distribution of derived masses for single stars in the full 
(solid line) and volume-limited sample for $d < 40$ pc (dotted line).}
\label{massdist}
\end{figure}

The M-functions are much better behaved than the G-functions and generally
yield good masses also for stars to which no meaningful age can be assigned.
Individual error estimates are also given for all masses in the catalogue; they
average about 0.05 $M_{\odot}$. Fig. \ref{massdist} shows the distribution of
the derived masses in both the magnitude-limited and volume-limited samples.
The low-mass limit at 0.65 $M_{\odot}$ reflects the red colour cutoff of our
sample.

\subsection{Space velocities}\label{UVW}

Space velocity components $(U,V,W)$ have been computed for all the stars from
their distances, proper motions, and mean radial velocities. $(U,V,W)$ are
defined in a right-handed Galactic system with $U$ pointing towards the
Galactic centre, $V$ in the direction of rotation, and $W$ towards the north
Galactic pole. No correction for the Solar motion has been made in the
tabulated velocities. Our radial velocities are of superior accuracy (Fig.
\ref{vrstat}) and the average error of the Tycho-2 proper motions corresponds
to only 0.7 km s$^{-1}$ in the tangential velocities, so the dominant source of
error in the space motions is the distance. Accounting for all these sources,
we find the average error of our space motions to be 1.5 km s$^{-1}$ in each
component ({\it U, V,} and {\it W}).

\begin{figure}[htbp] 
\resizebox{8cm}{!}{\includegraphics[angle=0]{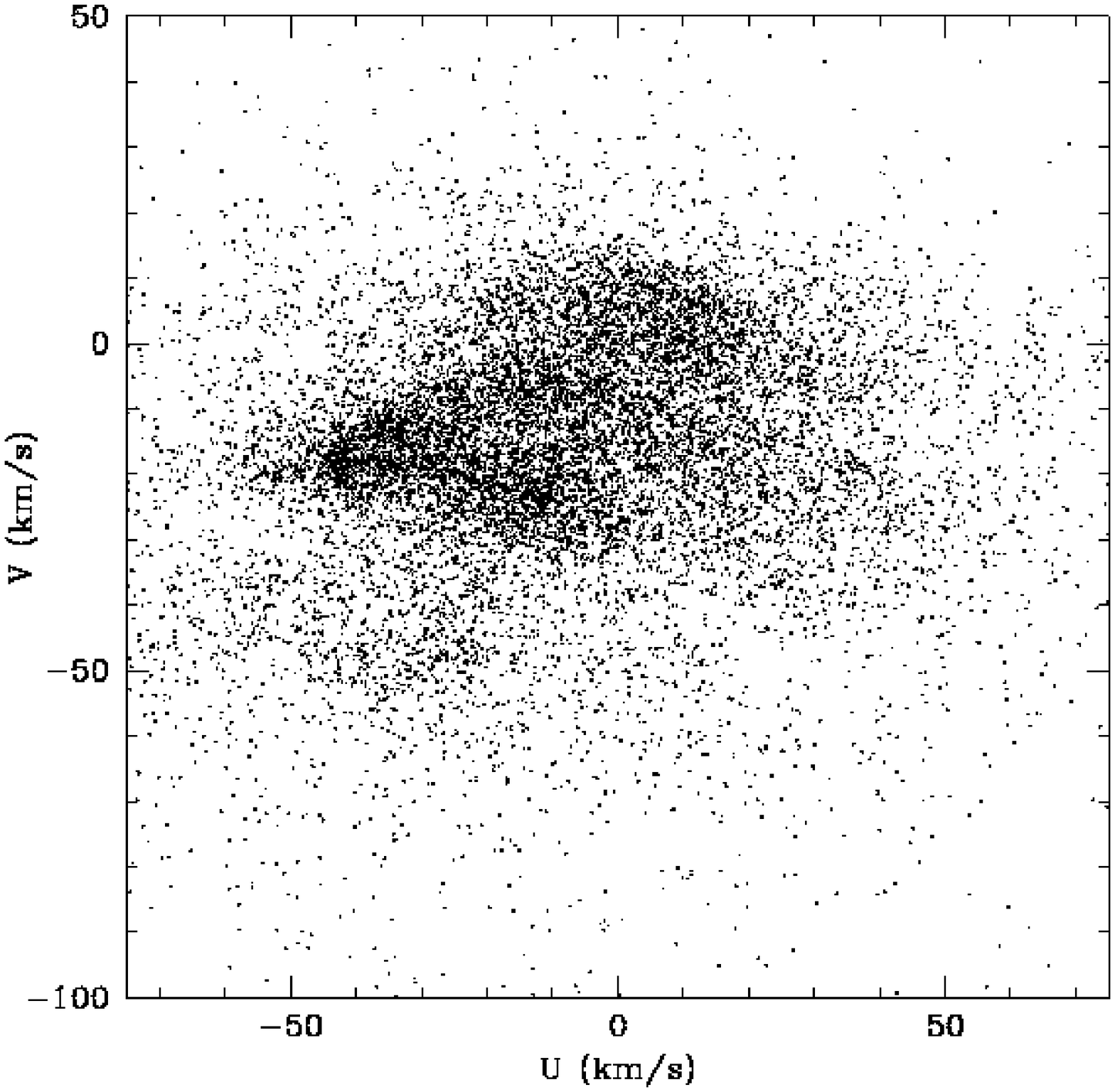}}
\resizebox{8cm}{!}{\includegraphics[angle=0]{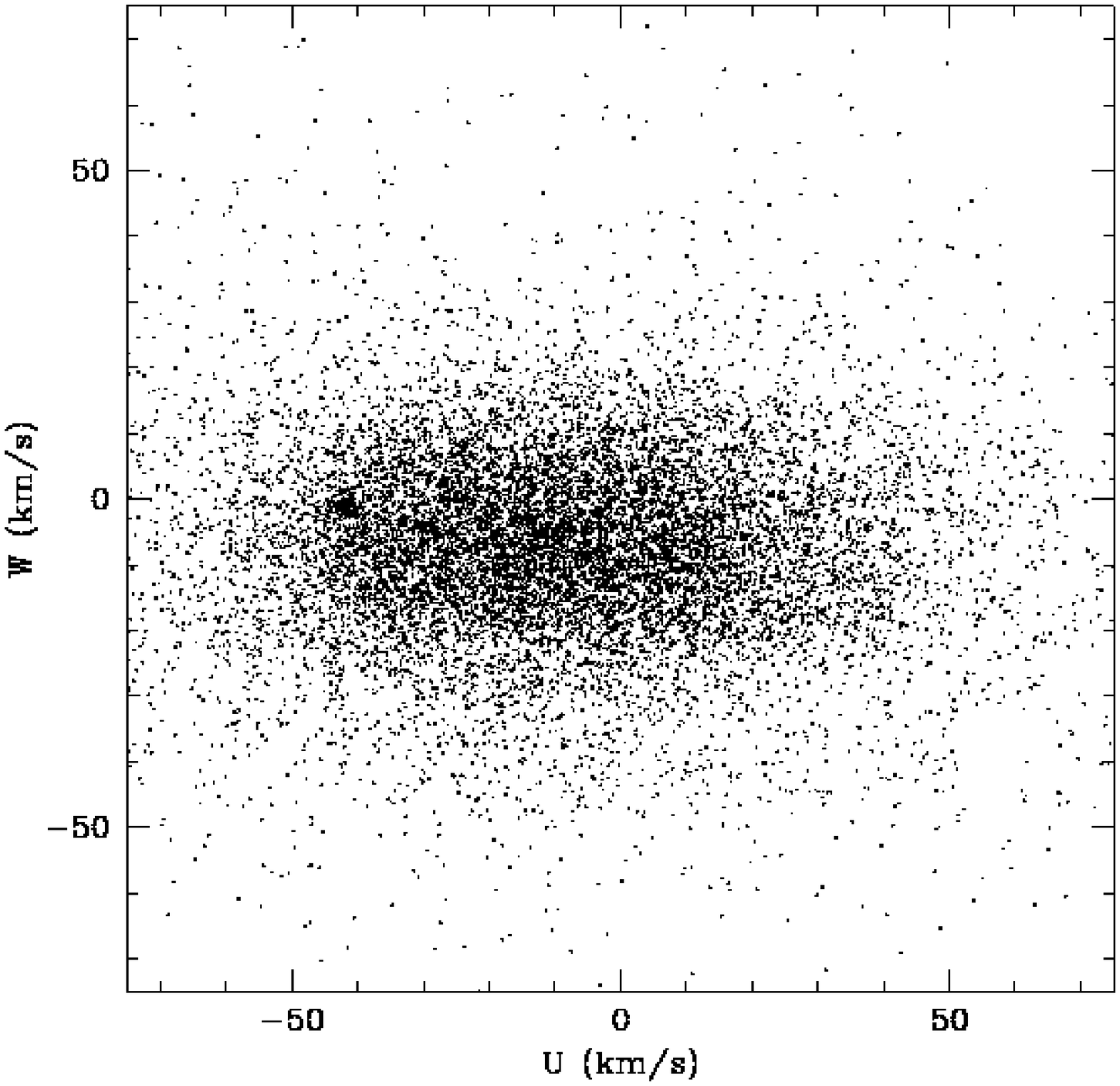}}
\resizebox{8cm}{!}{\includegraphics[angle=0]{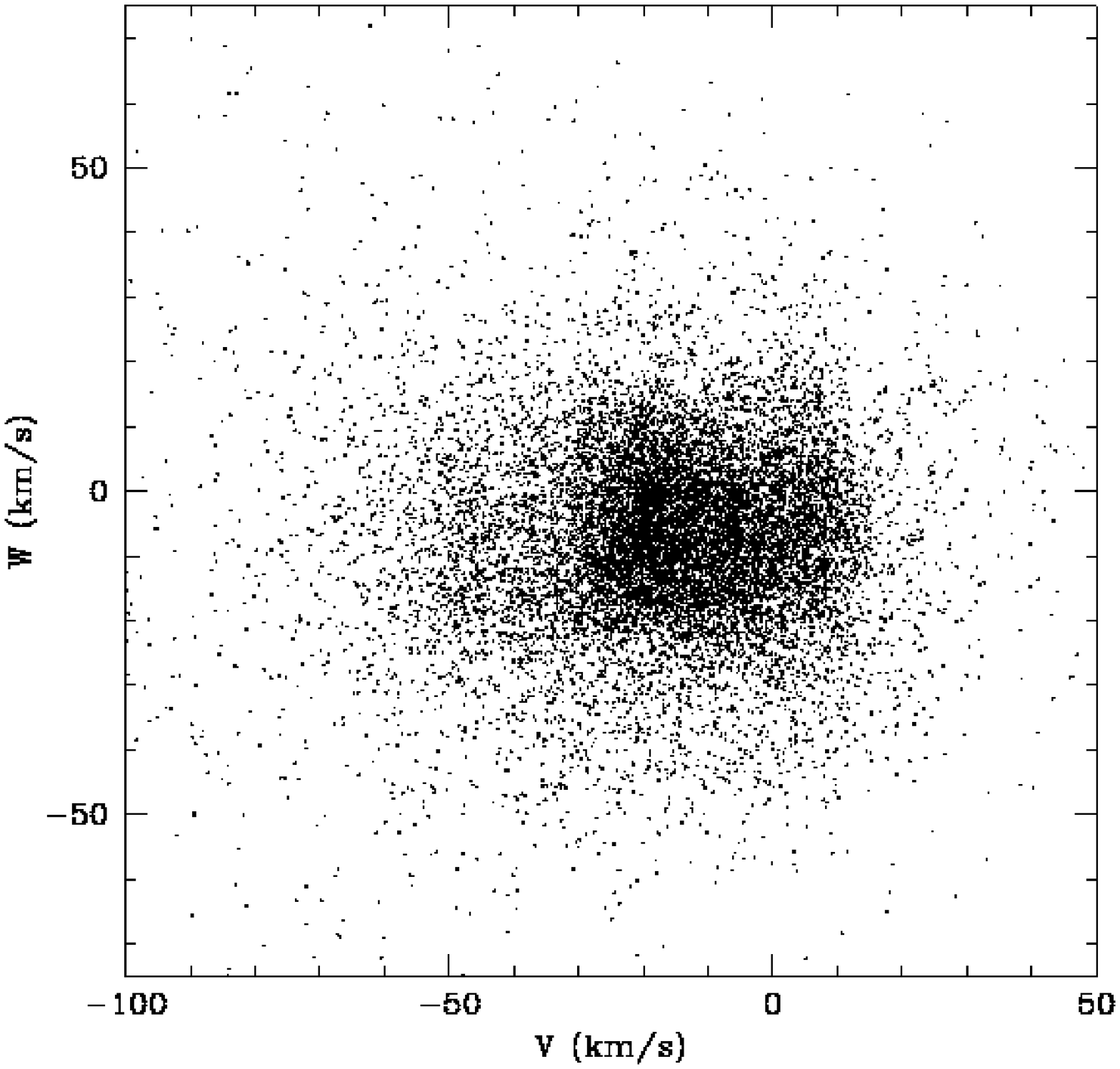}}
\caption{{\it U-V, U-W}, and {\it V-W} diagrams for all stars in the sample.}
\label{uvw}
\end{figure}

Fig. \ref{uvw} displays ({\it U, V,} and {\it W}) for all stars in the sample
with a measured radial velocity. The {\it U-W} and {\it V-W} diagrams show a
smooth distribution, the Hyades being the only clearly discernible structure
(cf. Fig. \ref{vr} and Sect. \ref{sample}). The {\it U-V} diagram, on the other
hand, shows abundant structure in addition to the Hyades, with four curved or
tilted bands of stars aligned along approximately constant $V$-velocities. Note
that the preponderance of stars from the two nearest spiral arms in our local
sample leads to the well-defined limits of these structures seen both in the
{\it U-V} diagram and in the {\it V-W} diagram (at $V$ = -30 and +20 km
s$^{-1}$). 

These structures, discussed e.g. by Dehnen (\cite{dehnen98}), Skuljan et al.
(\cite{skuljan99}), have velocities resembling classic moving groups or stellar
streams and have been named (top to bottom): The Sirius-UMa, Coma Berenices (or
local), Hyades-Pleiades, and $\zeta$ Herculis branches. The location of the
$\zeta$ Herculis branch in the {\it U,V} plane raises the interesting prospect
that kinematically selected samples of local thick-disk stars (e.g. Bensby et
al. \cite{bensby03}) might contain an admixture of somewhat younger, more
metal-rich thin-disk stars. 

Because these structures do not consist of coeval stars (see Fig. \ref{avr}),
they are not simply the remnants of broken-up systems such as classic moving
groups, but could be produced by transient spiral arm structures (De Simone et
al. \cite{desimon04}) or result from kinematic focusing by non-axisymmetric
structures of the Galaxy such as the bar (e.g. Fux \cite{fux01} and references
therein). Our data will allow more refined searches for detailed dynamical
substructure resulting, e.g., from past merger events (e.g., Chiba \& Beers
\cite{chiba00}, Helmi et al. \cite{helmi03}) or from the dynamical effect of
the bar.

\subsection{Galactic orbits}\label{orbits}

From the present positions and space motion vectors of the stars, we can
integrate their orbits back in time for several Galactic revolutions and
estimate their average orbital parameters. Correlating such Galactic orbits
with the chemical characteristics and ages of well-defined groups of stars may
yield insight of great interest into their origin; see e.g. Edvardsson et al.
(\cite{edv93}) and many later papers.

Before using the observed space motions for this task, they must be
transformed to the local standard of rest. For this, we have adopted a Solar
motion of (10.0, 5.2, 7.2) km~s$^{-1}$ (Dehnen \& Binney \cite{dehnbin98}). 
For the orbit integrations we used the potential of Flynn et al.
(\cite{flynnjsl96}), adopting a solar Galactocentric distance of 8 kpc. The
key properties of this model are: Circular rotation speed 220 km~s$^{-1}$, 
disk surface density 52 $M_{\odot}~pc^{-2}$, and disk volume density 0.10
$M_{\odot}~pc^{-3}$, in agreement with recent observational values
(Reid et al. \cite{reid99}, Backer \& Sramek \cite{BaSra99}, Flynn \& Fuchs
\cite{flynnfu94}, Holmberg \& Flynn \cite{holmb00}).

\begin{figure}[htbp] 
\resizebox{\hsize}{!}{\includegraphics[angle=0]{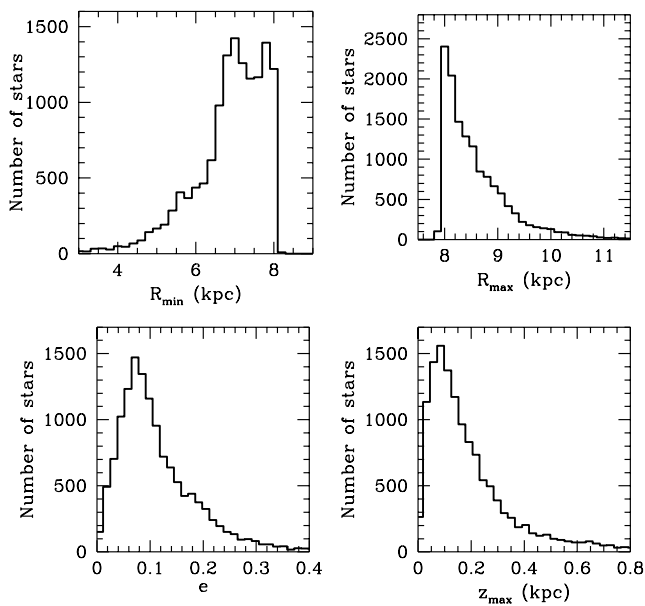}}
\caption{Key parameters of the Galactic orbits for the whole sample.} 
\label{rer} 
\end{figure}

In the catalogue we give the present radial and vertical positions ($R_{gal}$
and $z$) of each star as well as the computed mean peri- and apogalactic
orbital distances $R_{min}$ and $R_{max}$, the orbital eccentricity $e$, and
the maximum distance from the Galactic plane, $z_{max}$. Fig.~\ref{rer}
summarises the distribution of these parameters. 

The detailed shape of especially the $R_{min}$ distribution is strongly
dependent on the exact value of the Solar V-velocity applied, and the peaks can
be related to the enhanced density of stars in the streams seen in
Fig.~\ref{uvw}. Sirius-UMa causes the peak at $\sim$ 8kpc in the $R_{min}$
distribution, Hyades-Pleiades the one at $\sim$ 7kpc, and the small bump at
$\sim$ 5.5 kpc can be traced to the $\zeta$ Herculis stream. Tests have shown
that an increase of the Solar $V$-velocity by a few km/s changes the
double-peaked structure into a smooth increase, due to the changed orbit
distribution across the circular velocity.

\section{Statistical properties of the sample}\label{completeness}

Few if any complete and unbiased samples of stars with complete, homogeneous
astrophysical data exist in Galactic astronomy. Indeed, the conclusions of an
observational study may well depend more on the way the stars are selected
than on the actual data obtained. Thus, for any application of the data 
presented here it is crucial to be aware of the manner in which the stars
have been selected and observed, and how their astrophysical parameters have
been derived.

A chief concern at the start of the project was to avoid any kinematic bias
in the selection of stars. Therefore, stars for the photometric surveys were
originally selected based on the HD visual magnitudes. These are known to
have appreciable errors, so some stars brighter than the limits described in
Sect. \ref{survey} will certainly have been missed, while many of the newly
measured magnitudes were fainter than those limits. The wide range of HD
spectral types included in the surveys should ensure that few if any
metal-poor stars were missed from the final (much smaller) sample of F and G
dwarfs defined from the measured $uvby\beta$ indices. The cool limit is
somewhat less rigorously defined, as old metal-rich dwarf stars might be
mistaken for giants because of their strong CN bands, possibly even in the
MK spectral classes used to extend the red limit of the sample to K2~V south
of $\delta$ = -26$\degr$. 

As a practical factor, many of the radial-velocity observations were made from
preliminary versions of the photometric catalogues, so the final selection
criteria did not necessarily coincide precisely with the way the observations
were made. Stars that had already been observed could be excluded in the final
sample if needed, but new stars could not necessarily be added. This is the
case for a fraction of the unevolved F stars; they had been omitted initially
because no reliable ages can be determined for them, and not all were included
in the Hipparcos Input Catalogue which was later used to complete the data.
Also, several hundred stars yielded no radial-velocity determination because of
fast rotation, bright hot companions, or for other reasons. 

Inevitably, when many tens of thousands of observations are performed manually
on hundreds of nights, occasional misidentifications and other errors occur.
The thousands of double stars of all possible varieties provide rich
opportunities for ambiguity and inconsistency in the attribution of individual
observations. Great effort has been devoted to detecting and eliminating errors
of all conceivable kinds, but not all cases can be resolved and rejected
observations could not always be repeated.

Finally, we caution again that the astrophysical parameters for any star in
the catalogue flagged as a binary are likely to be inaccurate and potentially
misleading.

\subsection{Binary stars}\label{binaries}

Binary stars are abundant amongst field stars, especially in samples limited by
apparent magnitude because binary stars are on average brighter than single
stars. If unrecognised, the binary stars will appear as a substantial, but
unknown fraction of the sample for which the derived ages, metallicities,
distances, and space motions will all be wrong. Therefore, while corrected
values cannot always be derived, identifying such cases is important, and much
effort has been spent to that end. 

Our sample contains numerous double stars with separations ranging all the way
from invisible to fully resolved in both photometric and radial-velocity
observations. Similarly, brightness ratios range from unity to several
magnitudes, with associated observational difficulties. As far as possible, all
such cases have been recorded at the telescope (see comments in the photometry
papers), and stars with visual companions are flagged in the catalogue and the
available information given. As noted above, many radial-velocity observations
of such visual companions have been made;  these will be made available
separately at the web site of Observatoire de Gen{\`e}ve.

The accuracy of our new radial velocities varies with the rotational line
broadening, but the large majority are of excellent quality (cf. Fig.
\ref{vrstat}). Our average of more than four observations per star should
allow good binary detection statistics, although the actual number varies
considerably from star to star and many stars have only two observations.
Accordingly, some long-period and/or low-amplitude binary stars will likely
remain in the sample, but a large fraction of the systems with periods below
1000 days should have been revealed by our data (see below). Many
double-lined systems also manifested themselves by a double cross-correlation
peak already at the telescope. All available spectroscopic information on
duplicity for the stars in our sample is also given in the catalogue.

The complete sample of 16,682 stars contains 3,537 (21\%) visual double stars
and 3,223 (19\%) spectroscopic binaries of all kinds. The total number of
binary stars of any type is 5,622 (34\%), as some visual binary components are
also spectroscopic binaries. This fraction corresponds well to the total
frequency (32\%) of binary systems of all types with periods less than $10^5$
days found for G dwarfs by Duquennoy \& Mayor (\cite{duqmayor91}), after
careful correction for detection incompleteness. Our sample thus comprises
11,060 stars that are not known to be double; of these, 7,817 have measured
radial velocities consistent with their being true single stars.

A priori, we would expect a larger fraction of binary systems in our
magnitude-limited sample than in the volume-limited sample of Duquennoy \&
Mayor (\cite{duqmayor91}) because binaries are, on average, brighter than
single stars. However, restricting the statistics to the volume-limited
subsample within 40 pc, we still find a total binary fraction of 32\%,
suggesting that the great majority of binaries in the sample have indeed been
detected.

Finally, we warn that the mass ratios for 510 double-lined binaries derived
from our (generally sparse) radial-velocity observations and listed in Table 2
should not be used to derived mass ratio distributions or any other statistical
properties for binary stars in general. This sample is heavily biased towards
binary stars with near-equal components and periods of a few days, which show
significant line separation without excessive rotation (i.e. larger than
$\sim$30 km s$^{-1}$).

\subsection{Magnitude completeness}\label{magcompl}

\begin{figure}[htbp] 
\begin{center}
\resizebox{\hsize}{!}{\includegraphics[angle=0]{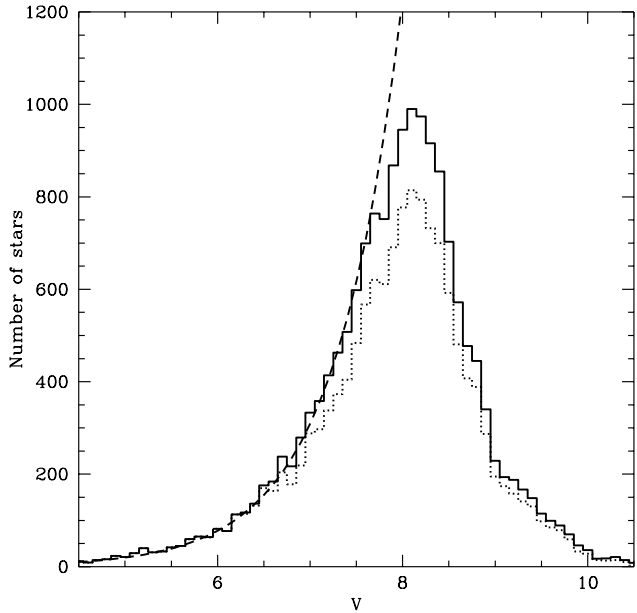}}
\end{center}
\caption{Distribution of observed $V$ magnitudes for the full sample (full
line), and for the stars with measured radial velocities (dotted line). The
dashed curve shows the expected relation for a uniform, volume complete 
sample.}
\label{mv} 
\end{figure}

The distribution of the photoelectric $V$ magnitudes for all stars in the
sample as well as for those with complete astrophysical data are shown in
Fig.~\ref{mv}. From this diagram, the magnitude-completeness of the sample
can be estimated by comparing with the distribution expected for a uniform
volume density of stars. The full sample begins to depart from completeness
near $V=7.6$. 

\begin{figure}[htbp] 
\resizebox{\hsize}{!}{\includegraphics[angle=0]{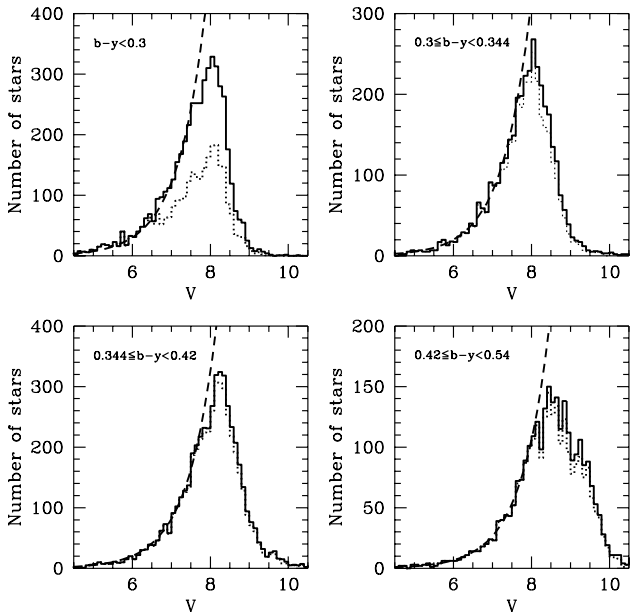}}
\caption{Distribution of $V$ magnitudes by interval of {\it b-y} colour.
Markings as in Fig.~\ref{mv}.}
\label{smallmv} 
\end{figure}

The original surveys were designed with variable magnitude limits in order to
better approximate a volume-complete sample. The magnitude limit thus depends
on colour, as shown in Fig. \ref{smallmv}. Note that a substantial fraction of
the earliest F dwarfs ($0.200 < b-y < 0.300$) lack radial-velocity data due to
fast rotation, while the cooler stars are essentially complete in this regard. 

The magnitude completeness as a function of colour may be characterized by
two numbers: $V_{lim}$, the magnitude at which incompleteness sets in, and
$V_{cut}$, the magnitude beyond which a negligible fraction of the stars were
measured. Estimates of these completeness limits are given in the table below. 

\begin{center}
\begin{tabular}{r c l r r} 
&& & $V_{lim}$ & $V_{cut}$ \\ \hline 
$ 0.205 \le$&$ b-y$&$ < 0.300$ & 7.7 & 8.9 \\ 
$ 0.300 \le$&$ b-y$&$ < 0.344$ & 7.8 & 8.9 \\ 
$ 0.344 \le$&$ b-y$&$ < 0.420$ & 7.8 & 9.3 \\ 
$ 0.420 \le$&$ b-y$&$ < 0.540$ & 8.2 & 9.9 \\ \hline 
\end{tabular} 
\end{center}

We note again that the coolest dwarfs are included only for $\delta < 
-26 \degr$. These 1,277 stars are flagged in Table 1, so the sample can be
cleanly divided into one subsample that is homogeneous over the sky to a
constant colour limit, and another that is similarly homogenous to a redder
colour limit, but covers only the southernmost 28\% of the sky.

\begin{figure}[htbp] 
\resizebox{\hsize}{!}{\includegraphics[angle=0]{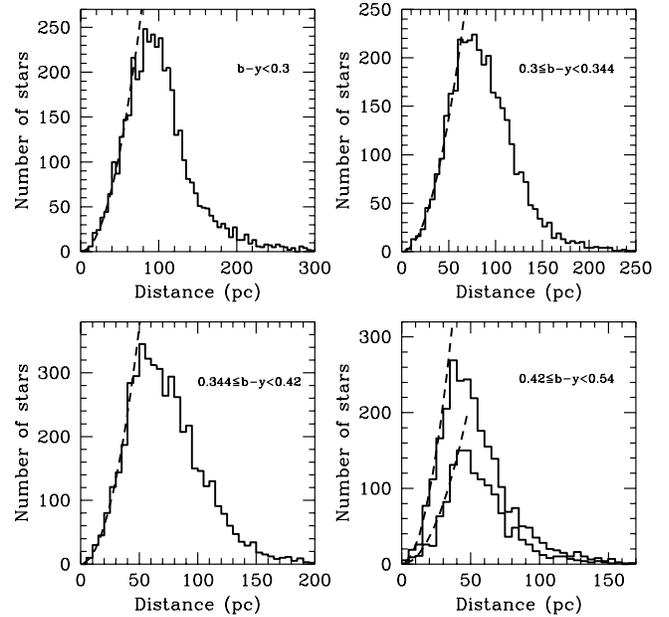}}
\caption{Distance distributions by interval in {\em b-y} colour (full line).
The dashed curves show the expected relation for a uniform, volume complete
sample. In the last panel, the lower histogram shows the distribution of stars
south of $\delta$ = -26$\degr$.} 

\label{distlimits} 
\end{figure} 

\subsection{Volume completeness}

The degree of volume completeness of the sample can be estimated from the
computed distances of the stars. The programme stars have a wide range of
absolute magnitudes, but are drawn from the apparent-magnitude limited
photometric surveys, so the distance to which the sample is volume complete
will vary considerably through the sample, primarily -- but by no means
exclusively -- with the {\it b-y} colour index. 

Fig.~\ref{distlimits} shows the distribution in distance of the four colour
intervals defined in Fig.~\ref{smallmv}. As expected, the earliest and
intrinsically brightest F stars are complete to the largest distance,
$\sim$70 pc (but note that many of these have no measured radial velocities).
The G5 dwarfs as well as the later-type stars south of $\delta$ = -26$\degr$
appear to be complete to $\sim$40 pc, as intended. We recall that 1,449 stars
have no distance listed in the catalogue because of inadequate data (see Sect.
\ref{distance}).

In the end, the volume within the limiting distance of 40 pc contains only 
1,685 stars with complete data, out of the 16,682 stars in the full sample (or
the over 30,000 stars in the original $uvby$ surveys!). However, as distances 
are known for nearly all the stars, the absolute-magnitude bias is quantifiable
and can be allowed for in computing the relative frequencies of the different
types of star.

\subsection{Completeness of masses}

Stellar mass varies much more slowly over the HR diagram than age. Accordingly,
the M-functions used to derive masses and their errors for our stars (Sect.
\ref{mass}) are much better behaved than the corresponding G-functions for the
ages, and useful mass estimates can be derived also for stars to which no
meaningful age can be assigned. Accordingly, masses have been derived for all
14,381 stars with determinations of $\log T_{eff}$, $M_V$, and [Fe/H] and
located inside the region in the HR diagram covered by the theoretical
isochrones (see Fig. \ref{massdist}).

\begin{figure}[htbp] 
\begin{center}
\resizebox{\hsize}{!}{\includegraphics[angle=0]{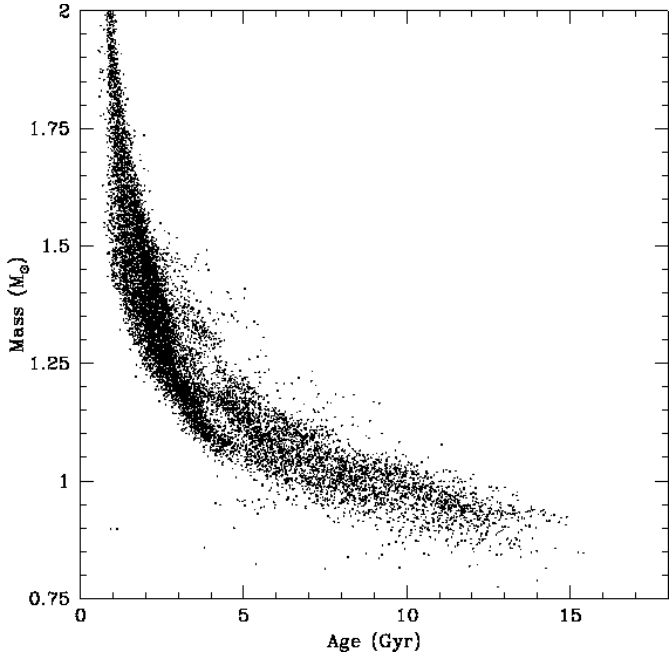}}
\end{center}
\caption{Age vs. mass for stars with well-defined ages. The apparent
correlation of age and mass is an artifact caused by the over-representation of
evolved stars and the fact that reliable ages cannot be determined for
unevolved stars.} 
\label{ama}
\end{figure}

\subsection{Completeness of ages}

In Sect. \ref{ageerr} we discussed how we have determined errors for the ages
derived for the individual stars. The degree of completeness of the age data
depends, then, on how large errors one is willing to accept. For the total
sample, publishable ages as defined in Sect. \ref{ageerr} are available for
13,636 stars or 82\% of the full sample (9,158 or 83\% of the single stars),
of which 11,445 (and 7,566 single stars -- 69\% of the total in both cases)
have ``well-defined'' ages by the definition in Sect. \ref{ageerr}. If the
relative error limits are set to 50\% or 25\%, the numbers are 9,428 and
5,472 (57\% and 33\%), respectively, for all stars, and 6,144 and 3,528 (or
56\% and 32\%) for the single stars. 

The fraction of stars for which reliable ages can be derived varies strongly
over the HR diagram. For the evolved F stars the situation is relatively
favourable. For stars on or near the ZAMS, only an upper limit to the age
can, in effect, be determined. And for the still-unevolved low-mass stars no
usable ages can be determined at all; in our sample, essentially no star
below 0.90 $M_{\odot}$ or fainter than $M_V = 4.5$ yields a meaningful
age.

These effects are evident in Fig. \ref{ama}, which shows the correlation
between the ages and masses derived for the stars in the sample. Three
effects are obvious from the figure: {\it (i):} The upper envelope of masses
for a given age, set by the evolution of the stars onto the giant branch (but
blurred by metallicity effects); {\it (ii):} the lack of stars younger than
$\sim 1$ Gyr, a direct result of our blue colour cut-off at {\it b-y} = 0.205; 
and {\it (iii):} the lack, at each age, of the low-mass, unevolved stars for 
which well-defined ages cannot be determined. The few scattered stars below the
main relation are probably undetected giants or other peculiar stars.

\section{Discussion}\label{discussion}

The data base presented here provides a basis for a thorough re-evaluation of
some of the global properties of the Solar neighbourhood and the Galactic
disk. Several of these analyses will require detailed simulations of the
predictions of various competing models, subjecting the simulated stellar
populations to the same selection criteria as used to establish our samples,
and comparing the results with the real data set. 

Such simulations will be the subject of future papers, but are beyond the
scope of the present discussion. Here, we will simply review a few of the
'classic' diagnostic relations for the Galactic disk, i.e, the metallicity
distribution for long-lived stars, the age-metallicity relation for the Solar
neighbourhood, the radial metallicity gradient, and the age-velocity and
metallicity-velocity relations.

In each case, two features are essential: ({\it i}) the lack of kinematic
selection bias in our sample; and ({\it ii}) the new radial-velocity data which
allow to identify stars that have not taken part in the evolution of the local
disk, but just happen to pass through it at this time; remove contamination by
unrecognised binary stars; complete the velocity information for the sample,
and estimate scale-height corrections. At the same time, the very strong
absolute-magnitude bias in our sample and limitation of the coolest stars to
the cap south of $-26\degr$ declination must be kept firmly in mind and
corrected for as appropriate. 

As a preliminary, we warn that the true uncertainty of the ages of especially
the oldest stars must be kept in mind when interpreting the following diagrams
(note in this connection that the final refinements in our age computation
methods have resulted in significant changes in the ages of the oldest stars
since our preliminary discussion in Holmberg et al. \cite{holmbetal03}). 

As shown in Fig. \ref{agedist}, restricting the sample to stars with
increasingly better-determined ages preferentially eliminates the oldest stars
and makes it increasingly difficult to determine the maximum age of stars in
the disk. Conclusions based on single points in, e.g. Figs. \ref{amrmag} or
\ref{avr}, will be risky at best. E.g., we note that the statement by Sandage
et al. (\cite{sandage03}), that ``the age of the field stars in the solar
neighbourhood is found to be $7.9\pm0.7$ Gyr'', is based on a few stars assumed
(but not known) to be super-metal-rich and ignores the existence of binary
stars on the main sequence. At the same time, the authors recognise that the
open cluster \object{NGC 6791}, with [Fe/H] = +0.3-0.4, is about 10 Gyr old, a
warning against quick conclusions drawn from a single diagram.

\subsection{Metallicity distribution (the ``G dwarf problem'')}\label{gdwarf}

Relative to the predictions of a closed-box model with constant initial mass
function (see e.g. Pagel \cite{pagel97}), the observed metallicity distribution
function for unevolved low-mass dwarfs shows a significant lack of low-mass
metal-poor stars (van den Bergh \cite{vdbergh62}, Schmidt \cite{schmidt63}).
These should have been formed along with the high-mass stars that produced the
heavy-element content of later stellar generations, but have not been found in
the predicted numbers in previous surveys (``the G-dwarf problem''). While
closed-box models are now obsolete, the metallicity distribution of long-lived
stars remains a fundamental constraint on any of its successors.

The G-dwarf problem could in principle have two different explanations: {\it
(i):} The 'missing' metal-poor dwarfs do in fact exist, but have not been found
by previous searches; a population of slow-moving metal-poor stars could exist
which would be missed in proper-motion surveys, but be included in our sample.
Or {\it (ii):} the Galactic disk did not evolve according to simple closed-box
models.

The G-dwarf problem was rediscussed exhaustively by J{\o}rgensen
(\cite{bjarner00}), based on an early version of the present data set. We refer
the interested reader to that paper for the full discussion, since no material
changes have been made to the underlying data since that study. 

\begin{figure}[htbp] 
\begin{center}
\resizebox{\hsize}{!}{\includegraphics[angle=0]{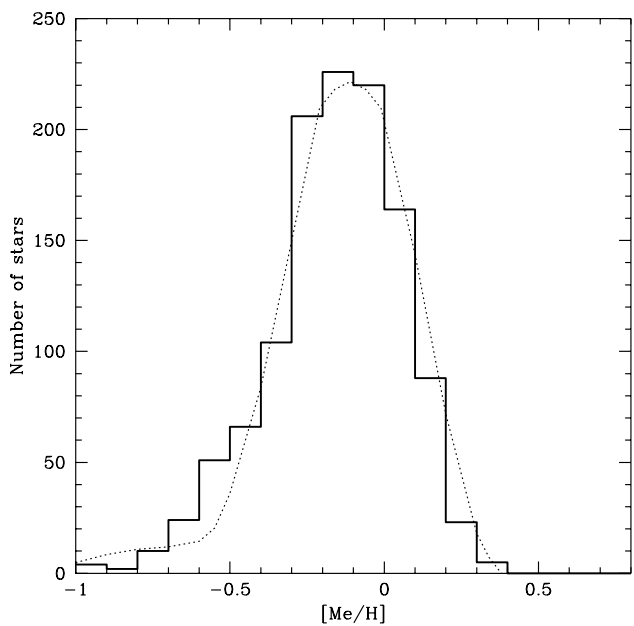}}
\end{center}
\caption{
Distribution of metallicities for the volume complete sample of single stars
(full histogram). For comparison the dotted curve shows the reconstructed
distribution for G dwarfs from J{\o}rgensen (\cite{bjarner00}), which is
corrected for scale height effects and measurement errors.}
\label{gprob}
\end{figure}

Here we just recall the salient conclusion, i.e. that the solution to the
G-dwarf problem is {\it not} a previously undiscovered metal-poor population.
On the contrary, when drawn from a volume-complete and kinematically unbiased
sample, the true fraction of metal-poor dwarfs is in fact only about half as
large as previously believed, and the agreement with closed-box models even
worse than before. Better physical models, not better data, are now the most
urgent requirement. Fig. \ref{gprob} shows the metallicity distribution for the
volume complete part of our survey.

\subsection{Age-metallicity relations}\label{amrsect}

Since the pioneering studies of Mayor (\cite{mayor74}) and Twarog
(\cite{twarog80}), the relationship between average age and metallicity in the
solar neighbourhood has been a subject of continuing debate, summarised in the
recent study by Feltzing et al. (\cite{feltz01}). The debate concerns both the
overall shape of the mean relation, whether there is scatter in the relation
over and above that due to observational errors, and if so, what might be the
cause of this scatter. 

The accurate spectroscopic metallicities of Edvardsson et al. (\cite{edv93})
established beyond reasonable doubt that real scatter exists in the relation,
far above that caused by observational errors. Little if any variation of mean
metallicity with age was found, except for the very oldest stars ($> 10$ Gyr)
where a downturn was observed in agreement with previous studies. However, as
pointed out by Edvardsson et al. (\cite{edv93}) themselves, the restriction of
their sample to F-type dwarfs automatically excluded any old, metal-rich stars
if such existed, a fact that has been overlooked in several later discussions.
The very recent similar analysis by Reddy et al. (\cite{reddy03}) of a larger
number of elements in 181 F and G dwarfs reached substantially the same
conclusions as Edvardsson et al. (\cite{edv93}). 

\begin{figure}[htbp] 
\begin{center}
\resizebox{\hsize}{!}{\includegraphics[angle=0]{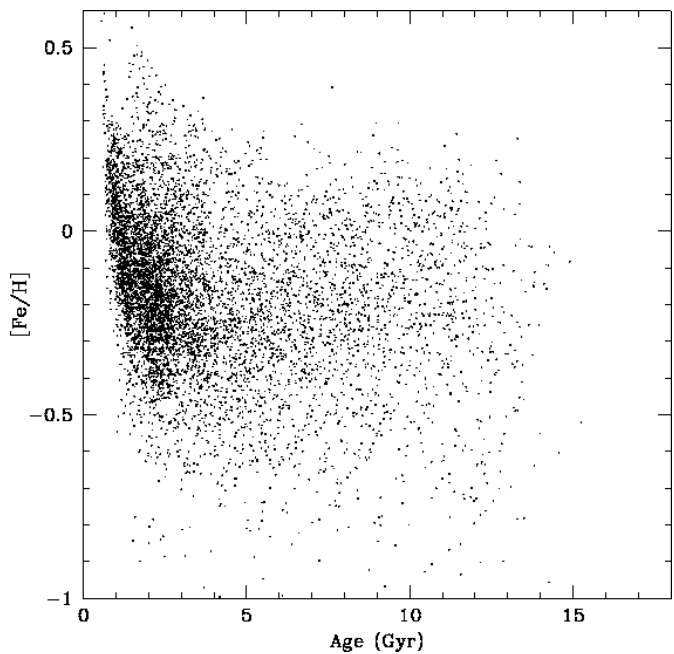}}
\end{center}
\caption{Age - metallicity diagram for 7,566 single stars with ``well-defined''
ages in the magnitude-limited sample. Note that individual age errors may still
exceed 50\% (cf. Fig. \ref{relagerror}).}
\label{amrmag}
\end{figure}

\begin{figure}[htbp] 
\begin{center}
\resizebox{\hsize}{!}{\includegraphics[angle=0]{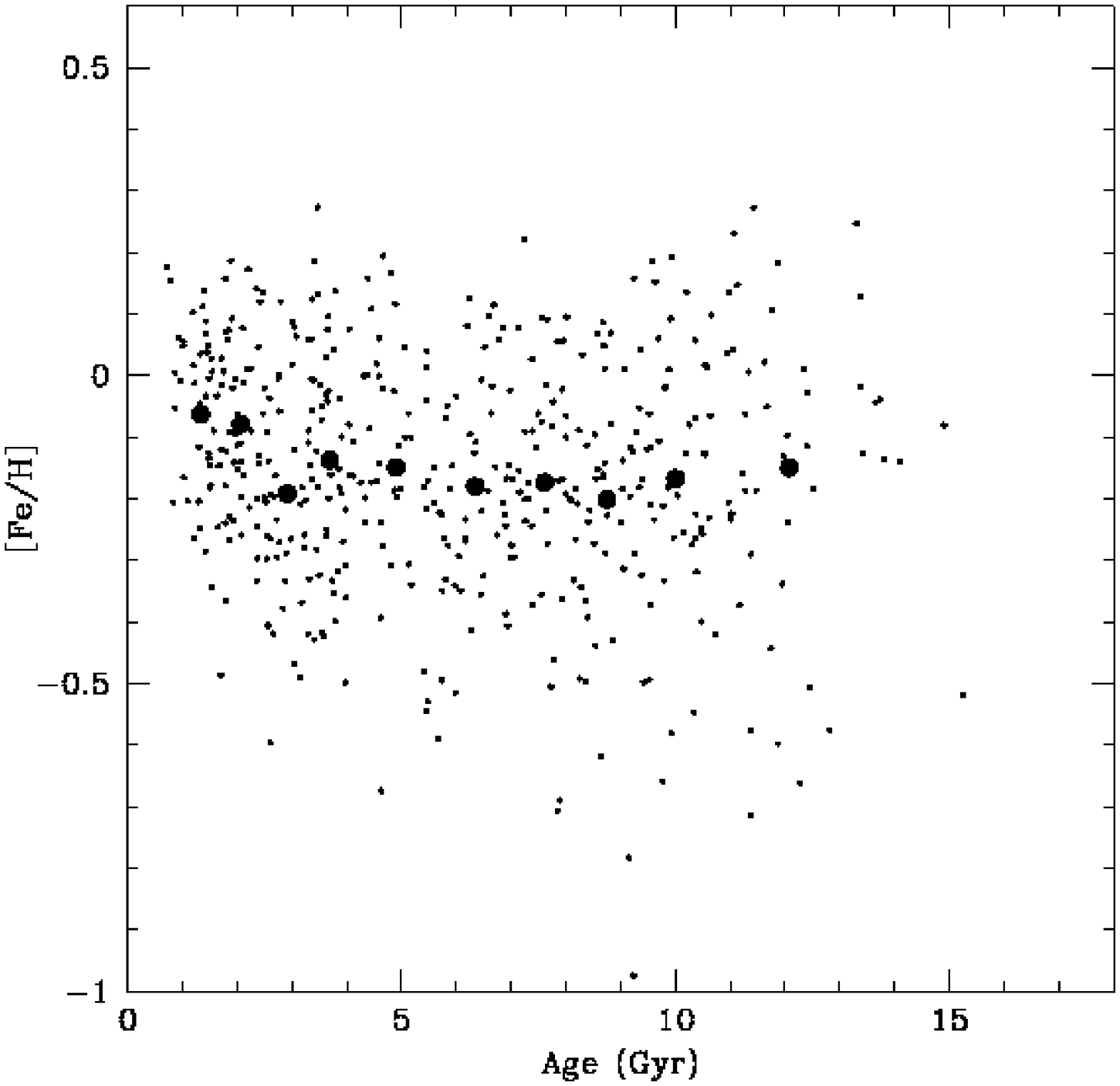}}
\resizebox{\hsize}{!}{\includegraphics[angle=0]{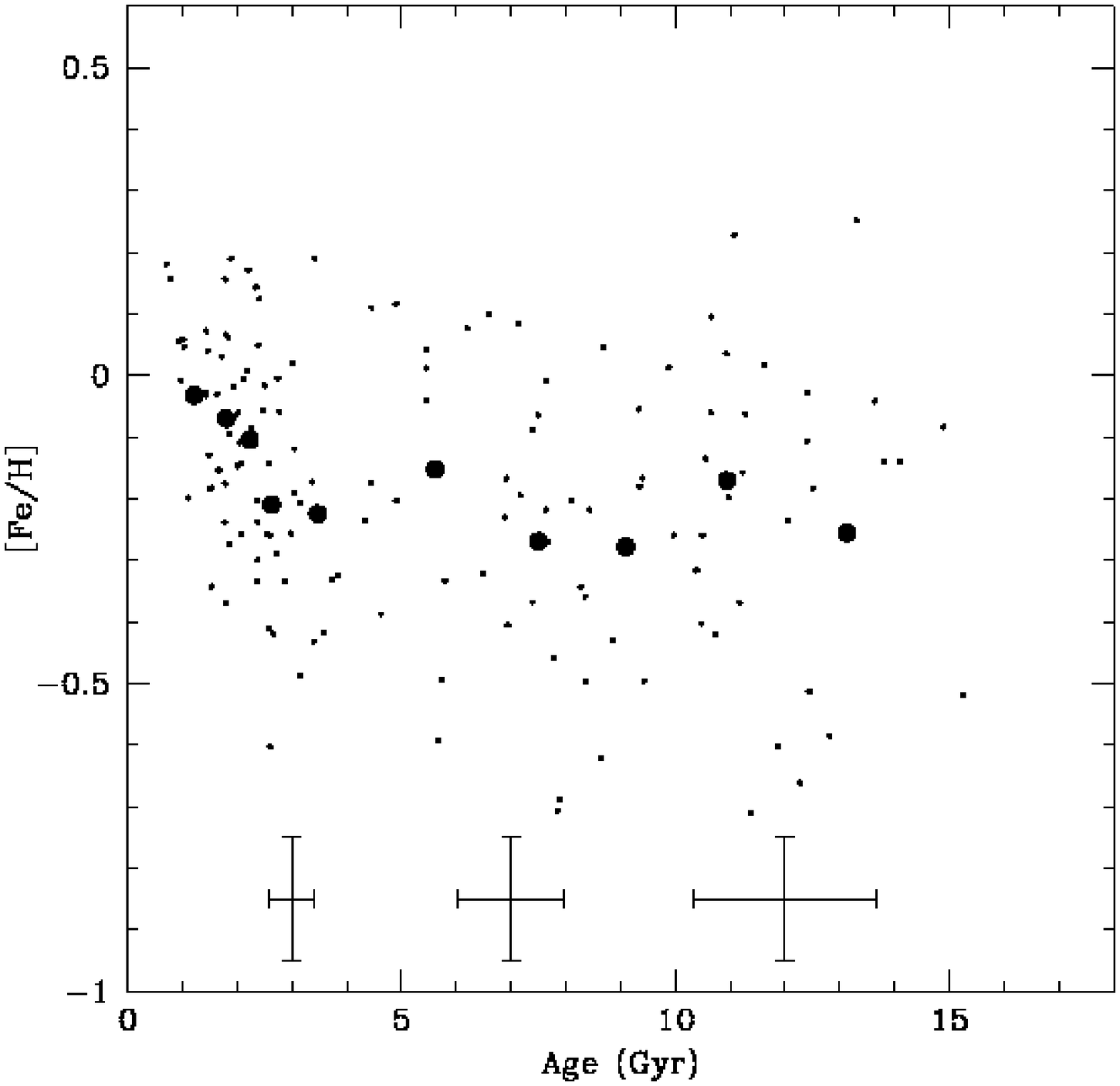}}
\end{center}
\caption{Age - metallicity diagrams for single stars in the volume-limited
subsample within 40 pc; large dots show mean ages and metallicities in 10 bins
with equal numbers of stars. {\it Top:} The 462 stars with ``well-defined''
ages (no limit on actual errors). {\it bottom:} The 142 stars with ages better
than 25\%; average error bars (14\%) in three age bins are shown.}
\label{amrvol}
\end{figure}

The selection criteria underlying the present sample were specifically designed
to include the old, metal-rich stars that would have been missed by Edvardsson
et al. (\cite{edv93}) even if they existed. At the same time, it was realised
that these are just the stars for which the derived ages will be the most
uncertain, as demonstrated implicitly by Feltzing et al. (\cite{feltz01}). In
this context we note that Reddy et al. (\cite{reddy03}) find again a more
clear-cut rise of mean metallicity with age than the two studies just
mentioned, primarily because the Reddy et al. sample contains few old, metal
rich stars. It is unclear whether this is caused by their selection procedure,
and it is therefore of interest to re-examine the age-metallicity diagram
constructed from the stars with well-defined ages in the unbiased sample
presented here (see Figs. \ref{amrmag} and \ref{amrvol}). 

The most obvious features of Fig. \ref{amrmag} remain direct results of our
selection criteria: The absence of stars near the lower left edge of the
diagram is caused by our blue colour cutoff. The predominance of young,
metal-rich stars is due to their intrinsic brightness and the correspondingly
large volume they sample; the few young, apparently super-metal-rich stars are
probably at least in part distant giant stars for which the interstellar
reddening has been overestimated (see below and Burstein \cite{burst03}).

Apart from these features, little variation in mean metallicity is seen, except
possibly for the very oldest stars which in general have kinematics
characteristic of the thick disk. Even some of these have solar-like
metallicities, however, and we note that Bensby et al. (\cite{bensby03})
recently derived a metallicity distribution for the thick disk that extends to
stars with [Fe/H] $\geq 0$ and Sun-like [$\alpha$/Fe] abundance ratios; those
stars are, however, also relatively young, so their thick-disk pedigree may
remain open to question. The ``$\zeta$ Herculis branch'' of disk stars in the
{\it U-V} diagram (Dehnen \cite{dehnen98}, Skuljan et al. \cite{skuljan99})
could be a source of such stars. 

When interpreting Fig. \ref{amrmag} (and Fig. \ref{avr}), the substantial
errors of even the ``well-defined'' ages should always be kept in mind; the
presence of stars appearing to be as old as 14 Gyr is easily explained by
observational errors. Uncertainties in the temperature scales of the observed
stars and theoretical isochrones (Sect. \ref{tempadj}) remain a potential
source of systematic error, and numerical details of the age computation may
introduce spurious features in diagrams such as Figs. \ref{amrmag} and
\ref{avr}, which can appear dramatic without the elaborate precautions
described in Sect. \ref{agedet}, but may remain in more subtle form. 

In order to avoid the strong absolute-magnitude bias in Fig. \ref{amrmag}, we
plot in Fig. \ref{amrvol} the same data for the volume-limited subsample within
40 pc. Despite the drastic reduction in number of stars, the lack of young
metal-poor stars produced by our blue colour cutoff remains well visible and is
responsible for the upturn of the mean relation for the youngest ages. The
disappearance of the young ``super-metal-rich'' stars supports their
interpretation as an artifact of the de-reddening procedure for distant giant
stars. The lack of any overall metallicity variation in the thin disk is even
more pronounced than before. Finally, the scatter in [Fe/H] at all ages again
greatly exceeds the observational error of $\sim$0.1 dex. Limiting the sample
to the best-determined ages leads to the same conclusions.

This picture of the age-metallicity distribution for field stars agrees well 
with the most recent studies of open clusters by Friel et al. (\cite{friel02}) 
and Chen et al. (\cite{chenl03}). These studies show the same constant mean 
metallicity and large scatter at all ages for the clusters as our study does
for the field stars. The discussion by Sandage et al. (\cite{sandage03}) also
focuses on the existence of old, metal-rich subgiants in their sample (the
10-Gyr-old metal-rich cluster NGC 6791 being another example) -- as in fact
pointed out already by Str{\"o}mgren (\cite{stromgren63}). Clearly, more
realistic models of the true complexities of star formation and chemical
enrichment of the interstellar medium are required.

We recall that we have taken special pains to verify that our age computation 
technique will not distort the overall trends in the resulting age-metallicity 
diagrams (cf. Sect. \ref{agecheck}). Individual ages shown in Figs.
\ref{amrmag} and \ref{amrvol} (top) may still be uncertain by 50\% or more,
however (cf. Fig. \ref{relagedist}), which must be taken into account in any
discussion of these diagrams. 

``Cleaner'' subsamples of stars can be selected from the catalogue (e.g. Fig.
\ref{amrvol}, bottom), but introduce further strong selection effects; cf. the
top and bottom panels of Figs. \ref{amrvol} and \ref{avr}. Evaluating the true
interplay of cosmic scatter in the chemical evolution of the disk with
observational errors and uncertainties in the age determinations will require
detailed numerical simulations, which must include models of the statistical
biases discussed above and by Pont \& Eyer (\cite{pont04}). Such detailed
analyses are beyond the scope of the present paper.

\subsection{Radial metallicity gradients in the disk}

\begin{figure}[hbtp] 
\begin{center}
\resizebox{\hsize}{!}{\includegraphics[angle=0]{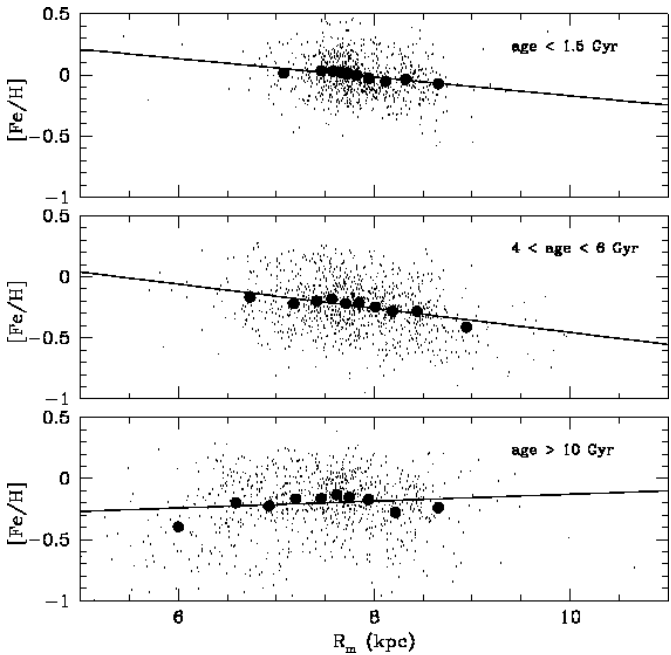}}
\end{center}
\caption{Radial metallicity gradient for single stars in three age ranges.
$R_m$ is the mean radius of the stellar orbits.}
\label{radmet}
\end{figure}

In addition to the evolution of metallicity with age, our sample can be used to
study the radial metallicity gradient in the Galaxy. From the $R_{min}$ and
$R_{max}$ of the stellar orbit, the mean orbital radius $R_{m}$ can be
calculated. Fig.~\ref{radmet} shows the radial metallicity gradient for three
groups of stars of different ages. The slopes of the fitted lines are
-0.076$\pm$0.014, -0.099$\pm$0.011, and +0.028$\pm$0.036 dex/kpc. 

This can be compared to other studies using FG dwarfs and giants by Mayor
(\cite{mayor76}), Cepheids by Andrievsky et al. (\cite{andrievsky02}),
planetary nebulae by Maciel et al. (\cite{maciel03}), and open clusters by
Friel et al. (\cite{friel02}) and Chen et al. (\cite{chenl03}). In general they
find the radial gradient to evolve over time from values between -0.02 and
-0.06 for the youngest stars to between -0.08 and -0.12 for the older stars.
This is compatible with our two younger age groups which show a mild steepening
with age of the radial gradient. The oldest stars in our sample, on the other
hand, show no radial gradient at all. This is an indication that these stars do
not follow the general evolution of the younger stars in the disk, but are of a
different origin, perhaps from the thick disk.

The radial metallicity gradient of $\sim$-0.09 dex/kpc seen for stars younger
than 10 Gyr can be used in an attempt to correct the age-metallicity
distribution for the effects of radial migration in the disk (cf. Wielen et al.
\cite{wfd96}). This correction has no discernible effect on the distribution,
and the dispersion around the mean for the volume-limited sample is unchanged:
Computing the mean and standard deviation of the metallicities in the age range
2-12 Gyr in Fig. \ref{amrvol} (top), we find $<$[Fe/H]$>$ = -0.16, $\sigma$ =
0.20 dex with no correction, $<$[Fe/H]$>$ = -0.19, $\sigma$ = 0.20 dex after
applying the correction for the radial metallicity gradient. 

Stars migrating into the sample from orbits centered elsewhere in the disk
obviously account for only a minute part of the scatter in the age-metallicity
diagram, and both dispersions are clearly much larger than can be explained by
errors in the metallicities, as also concluded by Edvardsson et al.
(\cite{edv93}) and Reddy et al. (\cite{reddy03}) from detailed spectroscopy.
Our removal of binaries from the sample and careful study of the age
uncertainties also guarantee that poorly-determined ages do not affect this
conclusion (and the main trends of Figs. \ref{amrvol} are insensitive to even
substantial horizontal redistribution of the points in the range 2-12 Gyr).

\subsection{Age-velocity relations}\label{avrsect}

\begin{figure}[htbp] 
\begin{center}
\resizebox{\hsize}{!}{\includegraphics[angle=0]{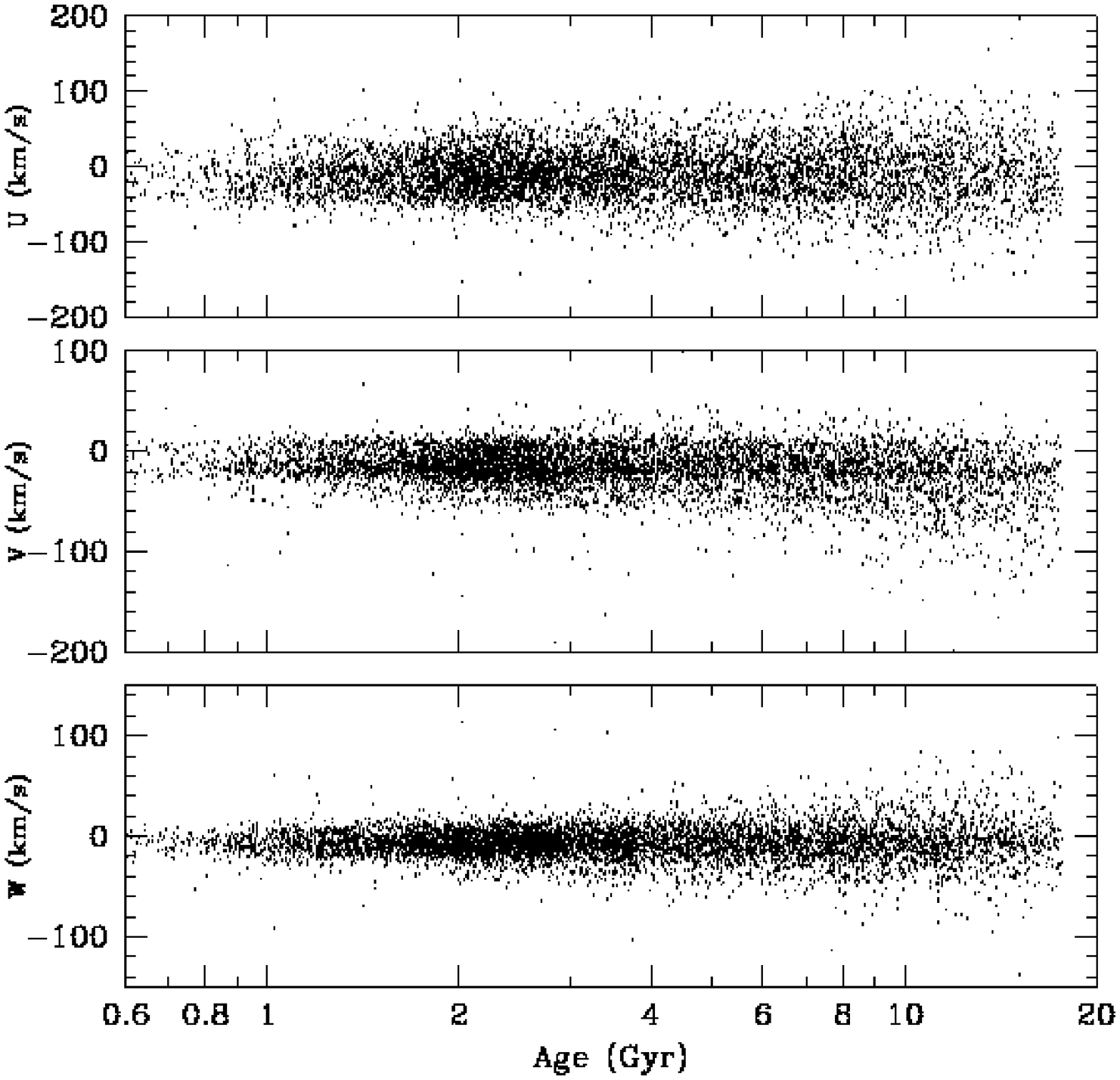}}
\resizebox{\hsize}{!}{\includegraphics[angle=0]{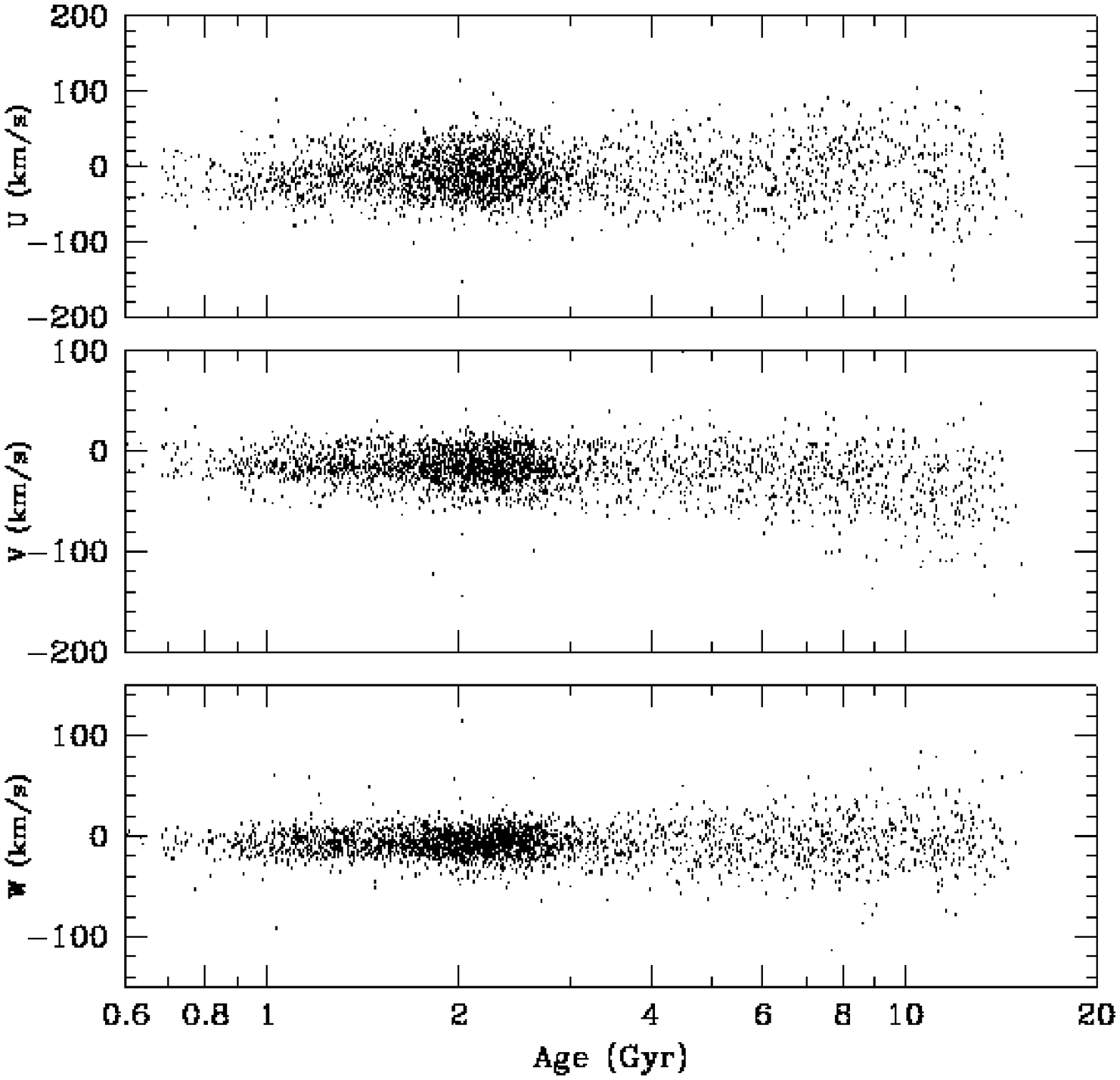}}
\end{center}
\caption{{\it U, V} and $W$ velocities as functions of age for all single stars
with complete data. {\it Top panels:} All 7,237 stars with ``well-defined''
ages and all velocity data; {\it bottom panels:} The 2,852 stars with ages
better than 25\%.}
\label{avr} 
\end{figure}

The observed space velocity components {\it U, V, W} for all single stars in
the sample are shown as functions of age in Fig. \ref{avr}, with weak as well
as strong limits on the accuracy of the ages. The population of bright, early
F-type stars is prominent, but otherwise the diagrams reflect the slow increase
of the random velocities with age which is attributed to heating of the disk by
massive objects such as spiral arms or giant molecular clouds. The rate of
change and maximum velocity dispersion reached in the thin disk are of key
importance for the interpretation in terms of the local dynamics of the disk
itself. 

\begin{figure}[htbp] 
\begin{center}
\resizebox{\hsize}{!}{\includegraphics[angle=0]{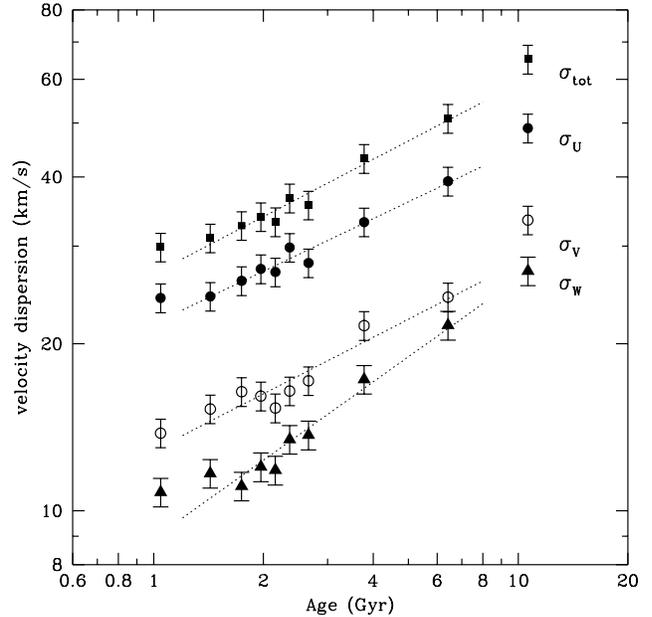}}
\end{center}
\caption{Velocity dispersions for single stars with relative age errors 
$<$25\% as functions of age (Fig. \ref{avr}, bottom). From top to bottom, the
total, {\it U, V,} and {\it W} velocity dispersion are plotted in 10 bins with
equal numbers of stars. The lines show fitted power laws; see text for full
details. The youngest and oldest age bins have been excluded from the fits to
avoid biases due to unrelaxed young structures and thick disk stars,
respectively.}
\label{vdisp} 
\end{figure}

In order to quantify these relations and also search for a kinematic signature
of the thick disk, we have computed the velocity dispersions of the single
stars with the best-determined ages. Fig. \ref{vdisp} shows the resulting
relations and the power laws fitted to them, excluding the youngest and oldest
bins. The resulting exponents are 0.31, 0.34, 0.47, and 0.34 for {\it U, V, W,}
and the total velocity dispersion, respectively, with an uncertainty of 0.05,
in close agreement with Holmberg (\cite{holmb01}) and Binney et al.
(\cite{binney00}). Aside from the low exponent for the total velocity
dispersion, the evolution of the velocities over time is characterised by a
small increase in the ratio of $\sigma_{V}/\sigma_{U}$ and a larger increase in
the ratio $\sigma_{W}/\sigma_{U}$. The consistency of these recent studies,
which all use isochrone ages but otherwise different samples and methods, is in
sharp contrast to studies using ages derived from chromospheric emission, which
give values of either 0.26$\pm0.01$ or 0.59$\pm0.04$ (H\"anninen \& Flynn
\cite{hanninen02}).

Fig. \ref{avr} shows no distinct jump in the velocity dispersion for the oldest
stars as derived by Quillen \& Garnett (\cite{qg01}) from the data by
Edvardsson et al. (\cite{edv93}). Our result, derived from a sample $\sim75$
times larger and with far more refined procedures for the age determination,
should clearly be more reliable. 

According to the classic Oort relation, $\sigma_{V}/\sigma_{U}$ should be
constant and equal to 0.5 for a flat rotation curve (we find a value of
$\sim0.63$). M\"uhlbauer \& Dehnen (\cite{muhlbauer03}) however, show that when
the true velocity dispersion and non-axisymmetric disturbances of the disk are
taken into account, large variations can occur. The smaller exponents for the
in-disk heating ($\sigma_{U},\sigma_{V}$) compared to the out-of-the-disk
heating ($\sigma_{W}$) gives further constraints to be fulfilled by models
trying to explain the observed kinematic heating of the disk. At least four
mechanisms have been proposed to explain the heating: fast perturbers from the
halo, such as $10^{6} M_{\odot}$ black holes; slow perturbers in the disk, such
as giant molecular clouds; large scale perturbations of the disk caused by
spiral arms (De Simone et al. \cite{desimon04}) or the bar (e.g. Fux
\cite{fux01}); and finally heating caused by infalling satellite galaxies. 

Massive black holes are improbable candidates due to other observational
constraints, apart from the fact that their heating index of 0.5 is too high.
Infalling satellites should result in a single dramatic heating of the disk,
such as the creation of the thick disk. Stable spiral arm patterns increase the
random motions of stars within the plane but not perpendicular to it, and also
become inefficient heaters when the epicyclic radius of a star gets larger than
the length scale of the spiral pattern. Molecular clouds heat stars in all
three directions, but in isolation they are inefficient heaters with a total
exponent of only 0.21 and a vertical exponent of 0.26 in the simulations of
H\"anninen \& Flynn (\cite{hanninen02}). Clearly, further simulations including
realistic descriptions of all heating components are needed, and indeed the
recent work on the effects of stochastic, transient spiral wave structures by
De Simone et al. (\cite{desimon04}) is able to produce exponents in the
observed range.

Wielen et al. (\cite{wfd96}) discussed the diffusion of stellar Galactic orbits
as a possible explanation of the scatter from an otherwise single-valued
age-metallicity relation, combined with a radial metallicity gradient in the
disk. From our redetermination of the age-velocity relations based on a much
larger sample and better data than available to them, we find a true dispersion
of velocities and orbits which is insufficient to support that interpretation
(see also Nordstr{\"o}m et al. \cite{paris99}). One might suspect the velocity
data used by Wielen et al. (\cite{wfd96}) to be contaminated by thick-disk or
halo stars, or by undetected binary orbital motion. 

Finally, we note that the continued heating of the disk throughout its lifetime
shown in Fig. \ref{vdisp} is inconsistent with the results of Quillen \&
Garnett (\cite{qg01}) who found the heating of the thin disk to saturate after
$\sim2$ Gyr, with an abrupt increase in velocity dispersion when the thick disk
appeared at $\sim10$ Gyr. Presumably, this is due to their much smaller sample
of stars (from Edvardsson et al. \cite{edv93}) and consequently larger
statistical errors than ours, as well as possible temperature or colour shifts
of the evolution models used to determine ages for cool dwarfs (see discussion
in Sect. \ref{selectmod}). 

Fig. \ref{vdisp} does show a modest increase in velocity dispersion for the
oldest age bin, suggestive of the appearance of the thick disk. The key new
result is, however, that the disk heating does {\it not} saturate at an early
stage. Continued heating of the disk would appear to make it more difficult to
trace old moving groups and other fossils of past merger events in the present
Galactic disk (see Freeman \& Bland-Hawthorn \cite{freeman02}).

\subsection{Metallicity-velocity relations}

Given the lack of correlation between age and metallicity demonstrated in 
Sect. \ref{amrsect}, plots of space motions vs. metallicity are not merely
trivial transformations of the age-velocity relations in Sect. \ref{avrsect},
but convey independent information. 

\begin{figure}[htbp] 
\resizebox{8cm}{!}{\includegraphics[angle=0]{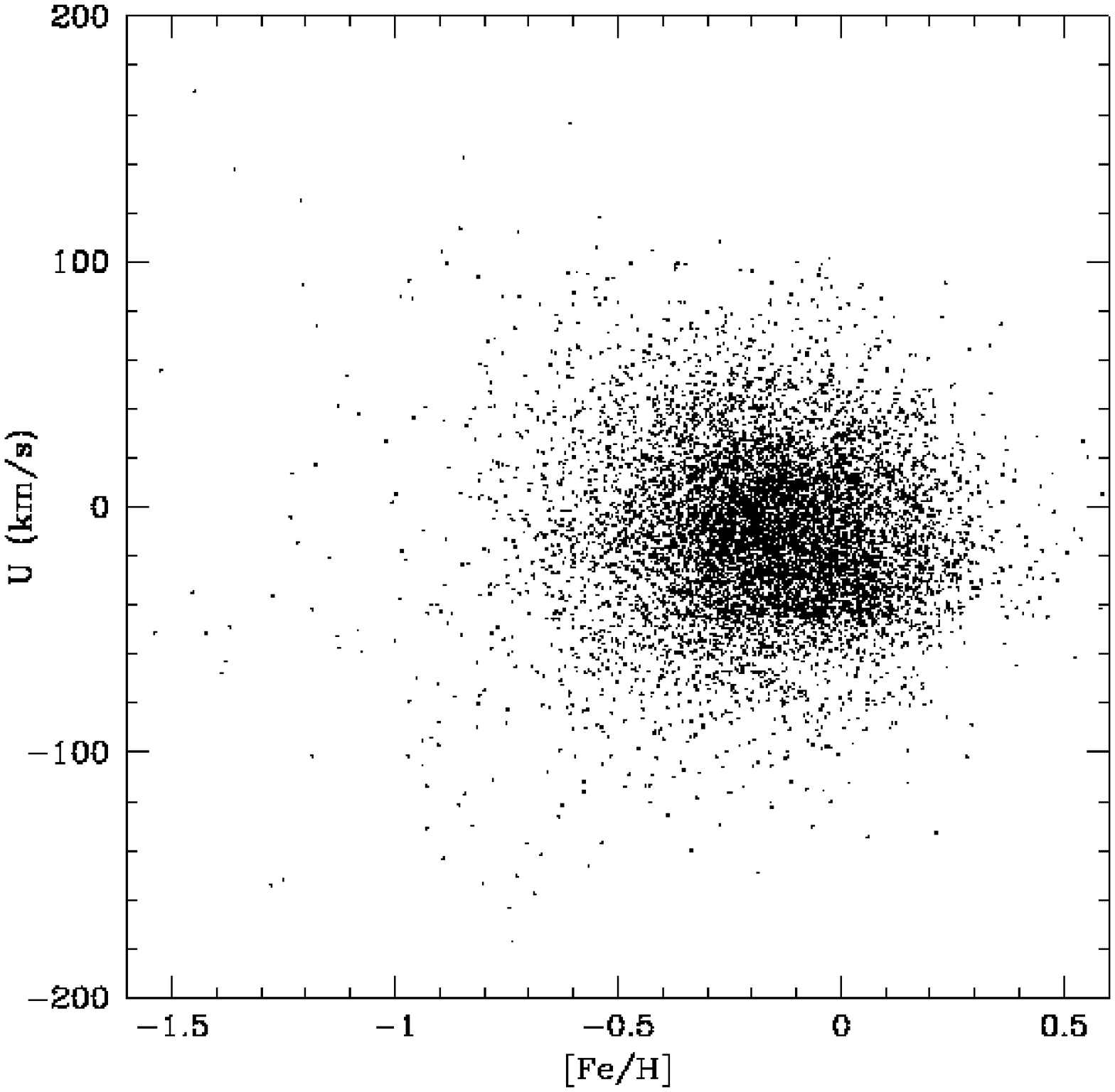}}
\resizebox{8cm}{!}{\includegraphics[angle=0]{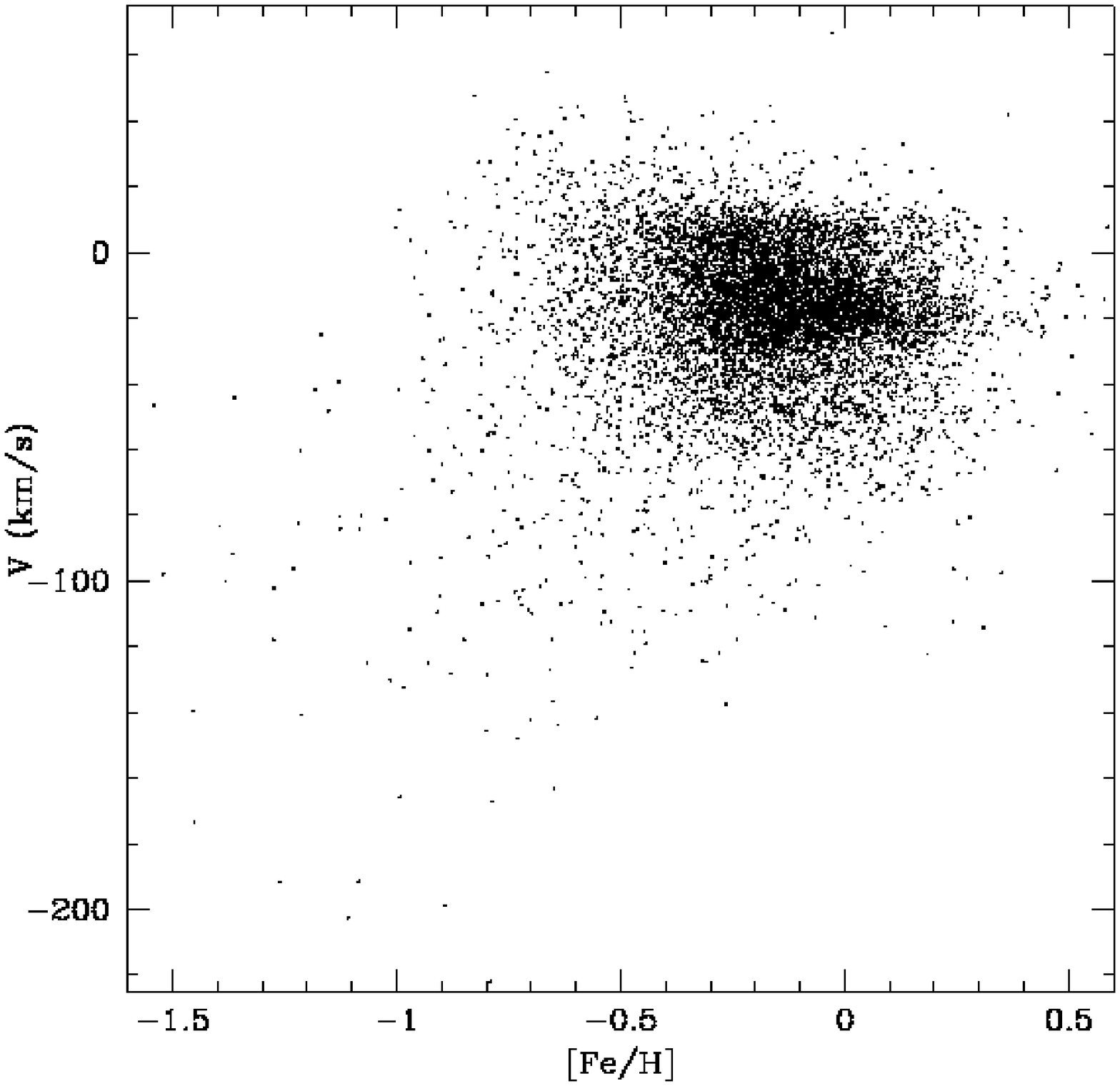}}
\resizebox{8cm}{!}{\includegraphics[angle=0]{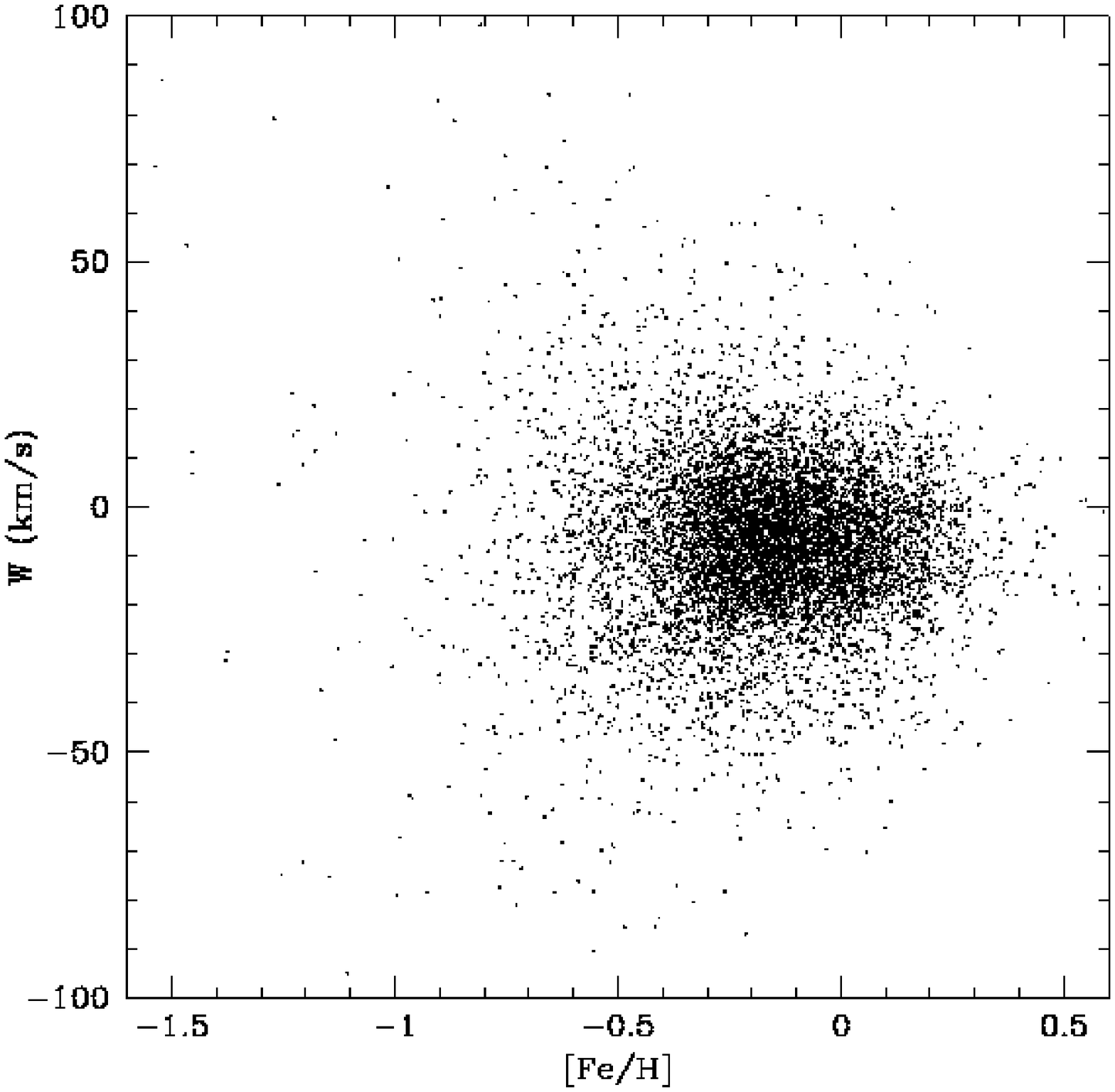}}
\caption{{\it U, V}, and {\it W} as functions of metallicity for all single
stars in the sample.}
\label{fehuvw}
\end{figure}

Fig. \ref{fehuvw} shows these diagrams, which display the expected similarity
with both Figs. \ref{uvw} and \ref{avr} but reveal new, significant features.
The [Fe/H] - $V$ and [Fe/H] - $U$ diagrams are especially interesting, as they
show that the structures seen in Fig. \ref{uvw} cover a wide range in [Fe/H] as
well as in age (Fig. \ref{avr}). This is quite unlike the behaviour expected
for stars in classical moving groups, which are presumed to have been born
together, but resembles the kinematic structures produced by an inhomogeneous
Galactic potential, e.g. by stochastic spiral waves (De Simone et al.
\cite{desimon04}) or the bar (e.g. Fux \cite{fux01}).

In particular, the group of stars near $V = -50$ km s$^{-1}$ and $U = -50$ km
s$^{-1}$ and with [Fe/H] between $\sim -0.7$ and $\sim +0.2$ suggest the
existence of a population of kinematically focused stars with a range of
thin-disk metallicities in addition to the {\it bona fide} thick-disk stars.
This possibility is of interest in studies of the detailed characteristics of
kinematically defined samples of thick-disk and thin-disk stars (see Bensby et
al. \cite{bensby03} for a recent example).

\section{Conclusions and outlook}\label{conclusion}

The results presented here are the culmination of many years of effort by many
people. Indeed, this is in essence the project for which not only the
Str{\"o}mgren $uvby$ photometric system (Str{\"o}mgren \cite{stromgren63}), but
also the CORAVEL radial-velocity scanners (Mayor \cite{mayor85}) were developed
(and an important part of the science case for the Hipparcos satellite). We
trust that the material presented here will remain useful to workers in the
field for years to come.

The properties of the data set and the results of our first analysis of it
are described in detail above; these descriptions will not be repeated here.
Instead, it is interesting to reflect on the change of scene in Galactic
research that has taken place in the three decades since the inception of
this project (cf. the recent review by Freeman \& Bland-Hawthorn
\cite{freeman02}, ``The New Galaxy''). 

At the time, the fundamental diagnostics of the evolution of the Galactic
disk were the relations discussed in Sect. \ref{discussion}, and
deriving isochrone ages was considered a fairly straightforward matter. What
was needed was a much larger and better-defined data base and an improvement
in accuracy to remove what was considered to be merely(?) observational
scatter in the relations. 

Having completed the observational effort and the first analysis of the data,
we find ourselves with much better techniques for determining isochrone ages,
but also a rather more conservative view of their accuracy than before.
Ironically, the similar improvements in observational accuracy (Edvardsson et
al. \cite{edv93} being the standard example) have only underscored the
disagreement of the mean relations with the predictions of simple Galactic
evolution models and the significance of the scatter around the relations in
terms of the real mechanisms of the evolution of the disk. 

Our age-metallicity diagrams (Fig. \ref{amrmag}-\ref{amrvol}) should lay to
rest the discussion whether closed-box models or single-valued predictions are
adequate representations of the evolution of the Galactic disk. Models that
allow for the coexistence of metal-poor and metal-rich stars throughout most of
the life of the thin disk will be needed for further progress. Similarly, our
new age-velocity relations will place much stronger constraints on models of
the dynamical heating of disk stars; few if any of the traditional candidate
mechanisms seem able to match the data. The new results will also be of
importance for assessing the survival rates of star clusters and merger
remnants in the disk and perhaps allow to detect the dynamical effects of the
bar of the Milky Way (Fux \cite{fux01}). Finally, a reanalysis of the ``G-dwarf
problem'' from a much larger and more rigorously selected data base points to a
similar conclusion: The Galaxy is a far more complicated and interesting
subject than ever before. The present work should lay the foundation for
learning more about it.

Looking ahead a decade from now, the ESA cornerstone mission GAIA (Perryman et
al. \cite{gaia01}) will provide the next quantum leap in our knowledge of the
Galaxy. Obtaining the complementary photometry and radial velocities needed to
fully exploit the astrometric data from Hipparcos was left to such independent
ground-based efforts as the present programme (see also Udry et al.
\cite{udryetal97}). In contrast, GAIA will obtain these observations with the
same satellite payload as the astrometric data, although with much larger
radial-velocity errors than achieved here (cf. Fig. \ref{vrstat}). The present
material should remain useful until the results from GAIA appear, not only for
studying our Galaxy, but also in the efforts to optimise the observing and data
reduction strategies for GAIA and for such precursor programmes as the RAVE
survey (Steinmetz \cite{steinmetz}).

\begin{acknowledgements}
This project was made possible by the very large amounts of observing time and
travel support granted over many years by ESO, through the Danish Board for
Astronomical Research, and by the Fonds National Suisse pour la Recherche
Scientifique (which also funded the development of the CORAVELs). Several
observers from our own institutes, ESO, and Observatoire de Marseille, France,
made many CORAVEL observations which are included in the data set presented
here; and Dr. David W. Latham and colleagues were indispensable collaborators
at the Center for Astrophysics. BRJ thanks Prof. L. Lindegren for many valuable
discussions on isochrone interpolations and statistical issues. We gratefully
acknowledge these essential contributions as well as the substantial financial
support received from the Fonds National Suisse pour la Recherche Scientifique
(to MM, FP, SU, and NM), the Carlsberg Foundation (to BN, JA, and JH), the
Danish Natural Science Research Council (to BN, JA, and EHO), the Smithsonian
Institution (to BN and JA), the Swedish Research Council and the Nordic Academy
for Advanced Study (to BN), and the Nordic Optical Telescope Scientific
Association (to JH and JA).
\end{acknowledgements}


\newpage

\noindent{\bf Table 1.} Table 1 will only be available in electronic form at
the CDS. The two first pages of the table, listing the first 100 stars, are
given at the end of this paper as sample of its content
and format.\\ 

\noindent{\bf Table 2.} Table 2 contains the mass ratios determined for 
511 double-lined binary systems; it will be available as for Table 1.\\

\clearpage 

\noindent{\bf Description of the columns in the catalogue\\
(Table 1; in electronic form only)}\label{catalog}
\medskip

\noindent {\bf Left-hand pages (Fields 1-26)}\\[2mm]
\noindent {\bf (1)} HIP : Number in the Hipparcos catalogue (ESA
\cite{hipp97}), 
when available. \\[1mm]
{\bf (2)} Name : Other designation (HD, BD, CD or CP number, in order of
preference). A second number following a slash indicates that a double star
with separate HD numbers has been observed together.\\[1mm]
{\bf (3)} Comp : If the star is a member of a multiple system, the
component(s) included in the photometry are identified here.\\[1mm]
{\bf (4)} $f_{b}$ : This flag identifies confirmed and suspected binaries.
The information can come from one or several sources such as photometry,
radial velocity or astrometry.\\[1mm]
{\bf (5)} $f_{s}$ : This flag identifies stars in the additional sample of 
cool dwarfs south of declination $-26\degr$ (see sect. \ref{sample} and
\ref{magcompl}).\\[1mm]
{\bf (6-7)} RA \& Dec : Equatorial coordinates of the star (J2000 equinox and
epoch), from the Tycho-2 catalogue. \\[1mm]
{\bf (8-9)} $l, b$ : Galactic coordinates of the star.  \\[1mm]
{\bf (10)} $V$ : Johnson $V$ magnitude from the published $uvby$
photometry.\\[1mm]
{\bf (11)} {\em b-y} : The {\em b-y} colour in the Str{\"o}mgren $uvby$
system.\\[1mm]
{\bf (12)} $\beta$ : The $H_{\beta}$ index in the Str\"omgren system.\\[1mm]
{\bf (13)} {\em E(b-y)} : Colour excess, from the calibrations of Olsen
(\cite{eho88}).\\[1mm]
{\bf (14)} log$T_{eff}$ : Logarithm of the effective temperature, from the
$uvby$ indices and the calibration by Alonso et al. (\cite{alonso96}).\\[1mm]
{\bf (15)} [Fe/H] : Metallicity of the star as determined from the
$uvby\beta$ indices and the calibrations by Schuster \& Nissen
(\cite{schuni89}), Edvardsson et al. (\cite{edv93}) or the new calibration
defined in Sect. \ref{FeH}.\\[1mm]
{\bf (16)} $d$ : Distance of the star, in parsecs (pc). Sect. \ref{distance}
explains why some stars have no entry in this column. \\[1mm]
{\bf (17)} $M_V$ : Absolute magnitude, from the apparent magnitude, distance,
and colour excess.\\[1mm]
{\bf (18)} $\delta M_V$ : The magnitude difference between the star and the
ZAMS.\\[1mm]
{\bf (19)} $f_r$ : Source for the distance (see text). H = Hipparcos
parallax; F,G = F or G-star photometric distance.\\[1mm]
{\bf (20)} $f_g$ : Suspected giant flag. An asterisk (*) indicates a
disagreement between the photometric determination and the Hipparcos parallax
at the $3\sigma$ level, suggesting that the star is a giant not detected from
the photometry.\\[1mm]
{\bf (21-23)} age, $\sigma_{age}^{low}$, $\sigma_{age}^{high}$ : The age
determined for the star, if any, with upper and lower $1\sigma$ confidence
limits, all in Gigayears.\\[1mm]
{\bf (24-26)} mass, $\sigma_{mass}^{low}$, $\sigma_{mass}^{high}$ : The mass
determined for the star, with upper and lower $1\sigma$ confidence limits,
all in Solar masses.\\[5mm]

\noindent {\bf Right-hand pages (Fields 27-51)}\\[2mm]
\noindent {\bf (27)} Name : Repeated from field 2.\\[1mm]
{\bf (28)} $V_{r}$ : Mean radial velocity of the star. For double-lined
binaries, the computed systemic velocity is given if so indicated by the flag
$f_{d}$ in field 34.\\[1mm]
{\bf (29)} $m.e.$ : Mean error of the mean radial velocity. See text for
details. \\[1mm]
{\bf (30)} $\sigma_{V_{r}}$ : Standard deviation of the individual
radial-velocity measurements if $N >$ 1.\\[1mm]
{\bf (31)} $N$ : The number of individual radial-velocity measurements.\\[1mm]
{\bf (32)} $\Delta$T : Time-span, in days, from the first to the last
radial-velocity observation.\\[1mm]
{\bf (33)} $P(\chi^{2})$ : Probability that the observed scatter of the
radial velocities is due to random observational errors only.\\[1mm]
{\bf (34)} $f_{d}$ : This flag indicates that the star is a spectroscopic
binary with a systemic radial velocity given in field 28.\\[1mm]
{\bf (35)} $f_{v}$ : Source of the radial velocity: C = Coravel; A = CfA; L =
literature.\\[1mm]
{\bf (36)} $v$sin$i$ : Rotational velocity of the star, in km~s$^{-1}$.\\[1mm]
{\bf (37-39)} $\mu_{\alpha*}$ $\mu_{\delta}$ $\sigma_{\mu}$ : Proper motion
in right ascension and declination and standard error of the combined motion,
in milliarcsec/year. Mostly from the Tycho-2 catalogue.\\[1mm]
{\bf (40-41)} $\pi$ $\sigma_{\pi}$ : Hipparcos parallax and its standard
error, in milliarcsec.\\[1mm]
{\bf (42-44)} {\em U V W} : Heliocentric space velocity components for the
star (U positive towards the Galactic centre), in km~s$^{-1}$.\\[1mm]
{\bf (45-46)} $R_{gal}$, $z$ : Galactic position of the star, in kpc.\\[1mm]
{\bf (47-50)} $R_{min}$ $R_{max}$ $e$ $z_{max}$ : Perigalactic and
apogalactic distance, eccentricity, and maximum distance from the galactic
plane of the computed galactic orbit, in kpc.\\[1mm]
{\bf (51) $f_{n}$} : General note (see below).\\
a: Double star with $\Delta m <$ 5 mag\\
b: Double star with $\Delta m <$ 5 mag from Hipparcos\\ 
c: Variable star\\
d: Simbad note\\
e: See Olsen (\cite{eho83})\\ 
f: See Olsen (\cite{eho79})\\ 
g: See Olsen (\cite{eho80})\\ 
h: See Olsen (\cite{eho93})\\ 
i: See Olsen (\cite{eho94a})\\ 
j: See Olsen (\cite{eho94b})\\ 
k: See Abt et al. (\cite{abt79})\\ 
l: See Abt (\cite{abt84})\\ 
m: See Abt (\cite{abt86})\\ 
n: See Gray \& Garrison (\cite{graygar89})\\ 
o: See Gray (\cite{gray89})\\ 
p: See Henry et al. (\cite{henry96})\\ 

\clearpage 
\newpage

\begin{table*}
\caption[]{Sample left-hand page of the catalogue (Fields 1-25 for the first 
100 stars).}
\tiny
\begin{tabular}{|r@{\hspace{1mm}}l@{\hspace{0mm}}l@{\hspace{0mm}}l@{\hspace*{-0mm}}l@{\hspace{0mm}}|@{\hspace{2mm}}r@{\hspace{1mm}}r@{\hspace{2mm}}r@{\hspace{1mm}}r|@{\hspace{1mm}}r@{\hspace{1mm}}r@
{\hspace{1mm}}r@{\hspace{1.5mm}}r@{\hspace{1.0mm}}|r@{\hspace{1mm}}r@{\hspace{1mm}}r@{\hspace{1mm}}r@{\hspace{1mm}}r@{\hspace{1mm}}r@{\hspace{0mm}}r@{\hspace{1mm}}|@{\hspace{1mm}}r@{\hspace{1mm}}r@
{\hspace{1mm}}r@{\hspace{2mm}}r@{\hspace{1mm}}r@{\hspace{1mm}}r|}
\hline
 HIP&Name&Comp&$f_{b}$&$f_{s}$&RA J2000{\hspace*{0mm}}&Dec J2000{\hspace*{-1mm}}&l&b&$V$&b-y&$\beta$&E(b-y)&$logT_{e}$&[Fe/H]&d&$M_{v}$&
$\delta M_{v}$&$f_{r}$&$f_{g}$&age&$\sigma_{age}^{low}$&$\sigma_{age}^{high}$&mass&$\sigma_{mass}^{low}$&$\sigma_{mass}^{high}$\\
&&&&&h{\hspace{1.5mm}}m{\hspace{2.5mm}}s{\hspace*{2mm}}&$^{o}${\hspace{2.5mm}}$^{\prime}${\hspace{2mm}}$^{\prime\prime}$&$^{o}$&$^{o}$&mag&mag&&mag&&&pc&mag&mag&&&Gy&Gy&Gy&$M_{\odot}$&$M_{\odot}$&
$M_{\odot}$\\
1&2&3&4&5&6{\hspace*{5.5mm}}&7{\hspace*{3.5mm}}&8&9&10&11&12&13&14&15&16&17&18&19&20&21&22&23&24&25&26\\
\hline
   437&HD 15      &    &*& &00 05 17.8&+48 28 37& 115& -14& 8.304&0.592&     &      &     &     &   &     &     &G&*&    &    &    &    &    &    \\
   431&HD 16      &    & & &00 05 12.4&+36 18 13& 113& -26& 8.092&0.311&     &      &3.810& 0.10&   &     &     & & &    &    &    &    &    &    \\
   420&HD 23      &    & & &00 05 07.4&-52 09 06& 319& -64& 7.552&0.366&2.626& 0.015&3.770&-0.19& 42& 4.44& 0.49&H& & 7.6& 3.6&10.9&0.99&0.94&1.06\\
   425&HD 24      &    &*& &00 05 09.7&-62 50 42& 312& -53& 8.146&0.377&2.607& 0.013&3.763&-0.31& 70& 3.91& 1.29&H& & 9.3& 7.8&11.0&1.01&0.97&1.04\\
      &HD 25      &    & & &00 05 22.3&+49 46 11& 115& -12& 7.590&0.256&2.675&-0.005&3.828&-0.38& 87& 2.89& 0.91&F& & 2.0& 1.6& 2.4&1.31&1.24&1.39\\
   447&HD 26      &    &*& &00 05 22.2&+08 47 16& 104& -52& 8.238&0.645&2.546&      &     &     &   &     &     &G&*&    &    &    &    &    &    \\
   461&HD 39      &AB  &*& &00 05 29.0&+34 06 20& 112& -28& 7.852&0.337&2.628& 0.014&3.784&-0.55& 91& 3.06& 1.85&H& & 4.7& 4.0& 5.6&1.14&1.09&1.20\\
      &HD 59      &    & & &00 05 33.5&+46 39 46& 115& -15& 8.595&0.363&2.613& 0.007&3.772&-0.24& 87& 3.90& 1.03&F& & 7.4& 5.3& 9.8&1.02&0.97&1.09\\
   462&HD 63      &    &*& &00 05 31.1&-09 37 02&  89& -69& 7.132&0.298&2.663& 0.008&3.806&-0.20& 51& 3.62& 0.50&H& & 2.9& 1.7& 3.8&1.21&1.15&1.26\\
   459&HD 67      &    & &*&00 05 28.4&-61 13 33& 313& -55& 8.822&0.424&     &      &3.743&-0.14& 54& 5.17& 0.33&H& &    &    &    &0.87&0.83&0.96\\
   475&HD 70      &    & & &00 05 41.6&+58 18 47& 117&  -4& 8.221&0.395&2.582& 0.010&3.752&-0.49& 48& 4.83& 0.74&H& &14.9& 7.4&    &0.84&0.81&0.90\\
   482&HD 85      &AB  &*& &00 05 44.4&+17 50 25& 108& -44& 7.754&0.275&2.661&-0.014&3.820&-0.08& 87& 3.06& 0.64&F& & 1.9& 1.3& 2.3&1.35&1.28&1.43\\
   493&HD 101     &    & & &00 05 54.7&+18 14 06& 108& -43& 7.456&0.373&2.598& 0.001&3.765&-0.32& 38& 4.55& 0.61&H& & 9.4& 5.4&14.2&0.94&0.87&1.00\\
   490&HD 105     &    & & &00 05 52.5&-41 45 11& 333& -73& 7.509&0.373&2.627& 0.019&3.766&-0.21& 40& 4.49& 0.55&H& & 8.6& 4.8&12.6&0.97&0.92&1.03\\
      &HD 117     &ABC &*& &00 05 57.0&-30 19 41&  12& -80& 9.047&0.385&2.644& 0.077&3.800&-0.26& 98& 3.76& 0.55&F& & 3.3& 2.0& 4.7&1.14&1.07&1.21\\
   518&HD 123     &AB  &*& &00 06 15.8&+58 26 12& 117&  -4& 5.978&0.421&     &      &3.746& 0.04& 20& 4.44& 0.85&H& &12.8& 8.5&15.1&0.98&0.92&1.02\\
   510&HD 126     &    &*& &00 06 08.0&+09 42 53& 105& -52& 7.803&0.312&2.647& 0.002&3.798&-0.21& 90& 3.03& 1.28&H& & 2.7& 2.3& 3.1&1.31&1.23&1.38\\
   522&HD 142     &AB  &*& &00 06 19.1&-49 04 30& 322& -66& 5.710&0.330&2.640& 0.006&3.791&-0.09& 26& 3.67& 0.68&H& & 3.6& 2.8& 4.3&1.20&1.12&1.25\\
   530&HD 153     &    & & &00 06 26.0&+42 45 09& 114& -19& 8.357&0.388&2.601& 0.010&3.763&-0.16&123& 2.90& 2.16&H& & 3.8& 3.1& 4.9&1.30&1.18&1.36\\
   529&HD 156     &    & & &00 06 24.9&-18 02 17&  72& -76& 7.311&0.248&2.699& 0.007&3.844& 0.23&131& 1.72& 1.27&F& & 1.1& 0.9& 1.3&1.87&1.75&2.01\\
   519&HD 160     &AB  &*& &00 06 16.8&-64 14 25& 311& -52& 7.801&0.291&2.689& 0.036&3.840& 0.25&160& 1.63& 1.41&F& & 1.1& 0.8& 1.3&1.92&1.79&2.04\\
   544&HD 166     &A   & & &00 06 36.7&+29 01 17& 111& -33& 6.093&0.460&2.576&      &3.727&-0.07& 14& 5.41& 0.36&H& &16.6& 3.0&    &0.85&0.83&0.91\\
   547&HD 189     &    & & &00 06 39.5&-24 37 15&  44& -80& 8.560&0.347&2.633& 0.018&3.783&-0.25&129& 3.01& 1.67&F& & 4.3& 2.8& 5.4&1.17&1.12&1.29\\
   556&HD 200     &    & & &00 06 46.9&-04 21 00&  96& -65& 8.222&0.357&2.605& 0.000&3.774&-0.45&110& 3.02& 2.04&H& & 4.9& 3.7& 6.2&1.14&1.06&1.22\\
   560&HD 203     &    & & &00 06 50.0&-23 06 27&  52& -79& 6.190&0.256&2.689& 0.003&3.830&-0.15& 39& 3.23& 0.32&H& & 1.5& 0.5& 2.2&1.34&1.29&1.40\\
   597&HD 219     &    & & &00 07 13.7&+73 12 37& 120&  11& 7.875&0.257&     &      &3.835&-0.17&   &     &     & & &    &    &    &    &    &    \\
      &HD 220     &    & & &00 07 04.7&+67 16 27& 119&   5& 8.686&0.320&2.639& 0.027&3.806&-0.58&133& 2.95& 1.49&F& & 2.8& 2.3& 4.5&1.18&1.11&1.29\\
   578&HD 222     &    & & &00 07 02.0&+22 50 40& 110& -39& 8.290&0.241&2.692&-0.007&3.840&-0.08&119& 2.91& 0.39&H& & 1.3& 0.6& 1.7&1.43&1.37&1.50\\
   584&HD 233     &    &*& &00 07 07.6&-15 51 00&  78& -75& 8.680&0.331&2.638& 0.018&3.787&-0.44& 86& 4.01& 0.74&F& & 5.2& 2.5& 7.6&1.04&0.97&1.10\\
   603&HD 251     &    &*& &00 07 18.3&+07 42 12& 104& -54& 7.637&0.309&2.636&-0.004&3.802&-0.32&108& 2.47& 1.84&F& & 2.2& 1.9& 2.6&1.42&1.32&1.51\\
   606&HD 268     &    & & &00 07 22.5&-25 21 23&  41& -80& 7.064&0.306&2.643& 0.003&3.800&-0.37& 49& 3.63& 0.77&H& & 3.5& 2.7& 4.3&1.15&1.08&1.20\\
   596&HD 276     &    & & &00 07 13.4&-76 43 51& 306& -40& 7.556&0.271&2.677& 0.006&3.821&-0.29& 84& 2.93& 0.93&H& & 2.2& 1.9& 2.5&1.35&1.30&1.40\\
   612&HD 285     &    & & &00 07 28.1&-56 00 41& 315& -60& 7.489&0.297&2.649& 0.002&3.809&-0.29&133& 1.86& 2.27&H& & 1.6& 1.4& 1.8&1.61&1.53&1.71\\
   630&HD 291     &    & & &00 07 40.4&+39 02 05& 114& -23& 8.019&0.297&2.660& 0.009&3.808&-0.24&104& 2.93& 1.18&H& & 2.4& 2.1& 2.8&1.33&1.26&1.40\\
   634&HD 292     &AB  &*& &00 07 43.1&+37 11 07& 113& -25& 7.436&0.317&     &      &3.800&-0.12&139& 1.72& 2.46&H& & 1.5& 1.3& 1.8&1.70&1.57&1.81\\
   641&HD 299     &    & & &00 07 52.0&+55 34 37& 117&  -7& 7.831&0.388&     &      &3.762&-0.04& 46& 4.52& 0.46&H& & 8.1& 3.6&11.8&0.99&0.94&1.06\\
   624&HD 307     &    & & &00 07 37.4&-45 07 10& 326& -70& 8.198&0.379&2.609& 0.015&3.768&-0.17&122& 2.77& 2.19&H& & 3.6& 2.8& 4.5&1.32&1.25&1.39\\
      &HD 308     &    & &*&00 07 37.8&-51 57 17& 318& -64& 9.402&0.438&     &      &3.737&-0.24& 78& 4.94& 0.76&G& &    &11.7&    &0.86&0.83&0.91\\
   618&HD 309     &AB  &*& &00 07 34.7&-60 38 30& 312& -56& 8.379&0.382&2.621& 0.030&3.781&-0.04& 87& 3.55& 0.99&F& & 3.6& 2.7& 5.9&1.15&1.09&1.25\\
   656&HD 330     &    & & &00 08 04.6&+53 47 46& 116&  -9& 8.151&0.384&2.597& 0.005&3.762&-0.39&102& 3.11& 2.18&H& & 5.1& 3.9& 6.5&1.12&1.05&1.24\\
   650&HD 333     &    & & &00 08 00.7&+29 50 15& 112& -32& 7.538&0.261&2.697& 0.022&3.849& 0.13&189& 1.06& 1.93&F& & 0.8& 0.7& 0.9&2.09&1.97&2.26\\
   649&HD 334     &    & & &00 08 00.2&-07 32 40&  93& -68& 7.848&0.317&2.638& 0.004&3.797&-0.27& 93& 2.99& 1.39&H& & 2.7& 2.3& 4.2&1.30&1.18&1.38\\
   669&HD 361     &    & & &00 08 16.3&-14 49 28&  82& -74& 7.045&0.388&2.598& 0.004&3.758&-0.20& 28& 4.84& 0.38&H& & 9.2& 0.7&16.1&0.92&0.85&1.01\\
   697&HD 372     &    &*& &00 08 35.3&+53 15 14& 116&  -9& 7.752&0.281&     &      &3.817&-0.24&114& 2.47& 1.44&H& & 1.9& 1.7& 2.2&1.46&1.38&1.55\\
   691&HD 373     &    & & &00 08 30.1&+52 20 11& 116& -10& 7.782&0.241&2.688&-0.004&3.835&-0.40& 69& 3.57& 0.11&H& &    &    & 1.8&1.24&1.19&1.29\\
   689&HD 375     &    & & &00 08 28.4&+34 56 04& 113& -27& 7.409&0.377&     &      &3.771& 0.05& 79& 2.93& 1.77&H& & 2.7& 2.4& 4.0&1.44&1.33&1.49\\
   682&HD 377     &    & & &00 08 25.7&+06 37 00& 104& -55& 7.581&0.391&2.603& 0.001&3.760&-0.02& 40& 4.58& 0.43&H& & 8.2& 3.4&12.2&0.99&0.93&1.06\\
      &HD 382     &    & & &00 08 25.4&-21 24 31&  61& -79& 8.346&0.512&     &      &3.702&-0.42& 27& 6.19& 0.31&G& &    &    &    &0.75&0.72&0.79\\
   688&HD 392     &    & & &00 08 27.6&-24 05 37&  48& -80& 7.599&0.321&2.643& 0.008&3.799&-0.12&132& 2.00& 2.20&H& & 1.7& 1.4& 2.0&1.61&1.52&1.71\\
   699&HD 400     &    & & &00 08 40.9&+36 37 37& 113& -25& 6.190&0.332&2.615&-0.011&3.787&-0.35& 33& 3.60& 1.08&H& & 6.1& 3.4& 7.1&1.09&1.04&1.15\\
   723&HD 404     &A   & & &00 08 57.2&+66 27 23& 119&   4& 8.622&0.516&     &      &3.708& 0.31& 32& 6.10&-0.14&H& &    &    &    &0.88&0.83&0.90\\
      &HD 410     &    & &*&00 08 35.1&-66 25 01& 310& -50& 9.425&0.426&     &      &3.745& 0.06& 90& 4.65& 0.64&G& &11.2& 6.3&15.6&0.95&0.89&1.02\\
   706&HD 427     &    & & &00 08 46.1&-34 47 39& 350& -78& 7.882&0.299&2.636&-0.016&3.808&-0.12&111& 2.66& 1.34&F& & 2.1& 1.8& 2.5&1.42&1.34&1.51\\
   746&HD 432     &A   & & &00 09 10.6&+59 08 59& 118&  -3& 2.270&0.216&2.709&-0.005&3.862& 0.18& 17& 1.16& 1.60&H& & 0.7& 0.6& 0.8&2.09&2.07&2.24\\
   722&HD 435     &    & & &00 08 56.5&+12 48 15& 107& -49& 8.477&0.353&2.637& 0.017&3.779&-0.11& 78& 4.03& 0.62&H& & 5.2& 2.9& 7.7&1.08&1.02&1.15\\
   709&HD 439     &    & & &00 08 48.4&-47 03 38& 323& -68& 7.817&0.304&2.670& 0.020&3.820& 0.13&116& 2.40& 1.11&H& & 1.5& 1.2& 1.7&1.60&1.52&1.70\\
   740&HD 447     &    & & &00 09 04.3&+19 55 28& 110& -42& 7.114&0.301&2.645&-0.006&3.810&-0.08&115& 1.81& 2.11&H& & 1.4& 1.2& 1.6&1.74&1.65&1.82\\
   726&HD 457     &    & & &00 08 59.6&-39 44 13& 335& -75& 7.725&0.391&2.617& 0.018&3.765& 0.11& 55& 4.03& 0.76&H& & 6.6& 3.6& 8.7&1.12&1.06&1.19\\
   747&HD 460     &A   &*& &00 09 10.2&+44 43 18& 115& -18& 8.782&0.306&2.640& 0.001&3.800&-0.44&129& 3.23& 1.23&F& & 3.1& 2.5& 5.0&1.16&1.09&1.24\\
   732&HD 466     &    & & &00 09 02.6&-35 05 30& 349& -78& 7.779&0.268&2.686& 0.012&3.827&-0.05&154& 1.84& 1.68&H& & 1.3& 1.1& 1.5&1.71&1.60&1.82\\
   730&HD 469     &AB  &*& &00 09 02.3&-54 00 06& 316& -62& 6.332&0.460&     &      &     &     &143& 0.56&     &H& &    &    &    &    &    &    \\
   754&HD 471     &A   &*& &00 09 15.7&+25 16 55& 111& -37& 7.789&0.421&2.590&      &3.745&-0.16& 45& 4.51& 0.96&H& &14.4&10.6&    &0.90&0.86&0.96\\
   759&HD 483     &    &*& &00 09 19.4&+17 32 02& 109& -44& 7.065&0.404&2.601& 0.009&3.756& 0.04& 52& 3.49& 1.57&H& & 5.6& 5.0& 6.4&1.18&1.14&1.22\\
   768&HD 489     &AB  &*& &00 09 28.1&+19 06 56& 109& -43& 7.953&0.421&2.591&      &3.747& 0.01& 72& 3.65& 1.64&H& & 7.3& 5.7& 9.0&1.11&1.04&1.18\\
   761&HD 493     &AB  &*& &00 09 21.0&-27 59 16&  25& -81& 5.420&0.273&2.668&-0.003&3.826&-0.05& 68& 1.24& 2.30&H& & 1.0& 0.9& 1.1&1.97&1.87&2.03\\
   762&HD 494     &    &*&*&00 09 23.3&-41 02 40& 332& -74& 9.118&0.492&     &      &3.715& 0.08& 84& 4.50& 1.42&H& &17.3&13.2&    &0.93&0.90&1.00\\
   795&HD 531     &AB  &*& &00 09 51.2&+08 27 11& 106& -53& 8.650&0.439&2.583&      &3.737&-0.04& 53& 5.03& 0.51&G& &15.6& 4.6&    &0.89&0.85&0.95\\
   786&HD 536     &AB  &*& &00 09 41.5&-37 18 53& 341& -77& 8.355&0.287&2.669& 0.008&3.814&-0.17&123& 2.91& 1.00&F& & 2.2& 1.8& 2.6&1.34&1.26&1.43\\
   794&HD 546     &AB  &*&*&00 09 50.0&-33 47 20& 354& -79& 8.832&0.498&     &      &3.709&-0.11& 58& 5.00& 1.16&H& &    &    &    &    &    &    \\
   791&HD 547     &    & & &00 09 48.5&-40 53 34& 332& -74& 8.578&0.408&2.590& 0.013&3.750&-0.34& 72& 4.29& 1.22&H& &13.8&11.0&16.7&0.91&0.85&0.96\\
   801&HD 564     &    & & &00 09 52.8&-50 16 04& 319& -66& 8.323&0.381&2.584&-0.005&3.761&-0.33& 53& 4.69& 0.57&H& &10.6& 4.5&16.0&0.91&0.84&0.99\\
      &HD 570     &AB  &*& &00 10 24.7&+58 31 25& 118&  -4& 8.217&0.207&     &      &3.868& 0.26&   &     &     & & &    &    &    &    &    &    \\
   815&HD 578     &    & & &00 10 02.5&-63 17 48& 311& -53& 8.134&0.290&2.661& 0.001&3.812&-0.19&104& 3.05& 0.92&H& & 2.4& 1.9& 2.8&1.33&1.26&1.39\\
   848&HD 583     &    &*& &00 10 24.0&+58 29 22& 118&  -4& 7.738&0.297&     &      &3.807&-0.32& 96& 2.82& 1.38&H& & 2.4& 2.1& 2.7&1.35&1.28&1.41\\
   851&HD 604     &    & & &00 10 26.3&-14 27 08&  84& -74& 8.353&0.331&2.610&-0.029&3.790&-0.24&131& 2.77& 1.75&F& & 2.7& 2.2& 4.1&1.34&1.19&1.45\\
   819&HD 610     &    & & &00 10 04.4&-79 00 06& 305& -38& 7.878&0.294&2.663& 0.006&3.810&-0.17& 90& 3.10& 0.90&H& & 2.4& 2.0& 2.8&1.33&1.27&1.38\\
   856&HD 615     &    & & &00 10 29.5&+15 13 54& 109& -46& 8.245&0.306&2.650&-0.004&3.804&-0.02& 78& 3.80& 0.19&H& & 1.9&    & 3.4&1.22&1.16&1.28\\
   867&HD 631     &    & & &00 10 39.1&+12 49 12& 108& -49& 8.475&0.366&2.611& 0.003&3.771&-0.18& 64& 4.44& 0.46&H& & 7.1& 2.6&10.6&1.00&0.95&1.07\\
   870&HD 633     &    & & &00 10 40.7&+02 03 19& 103& -59& 7.514&0.313&2.643&-0.004&3.799&-0.08& 55& 3.81& 0.35&H& & 2.9& 0.8& 4.2&1.19&1.14&1.25\\
   880&HD 639     &    &*& &00 10 46.8&+37 28 26& 114& -25& 7.764&0.272&     &      &3.824&-0.16&114& 2.49& 1.19&H& & 1.8& 1.5& 2.0&1.49&1.42&1.57\\
   889&HD 652     &A   & & &00 10 56.0&+48 06 37& 116& -14& 8.429&0.354&2.621& 0.004&3.777&-0.16& 61& 4.52& 0.22&H& & 4.0&    & 8.8&1.03&0.96&1.09\\
   872&HD 659     &    &*& &00 10 42.4&-54 24 01& 315& -62& 8.548&0.456&     &      &3.728&-0.25&   &     &     &G&*&    &    &    &    &    &    \\
   865&HD 661/2   &    &*& &00 10 38.5&-73 13 27& 307& -44& 6.657&0.233&2.719& 0.014&3.849& 0.20& 66& 2.55& 0.38&H& & 0.9& 0.6& 1.3&1.67&1.60&1.72\\
   910&HD 693     &    & & &00 11 15.8&-15 28 04&  82& -75& 4.910&0.326&2.617&-0.008&3.789&-0.49& 19& 3.53& 1.22&H& & 5.7& 4.8& 6.9&1.07&1.03&1.14\\
      &HD 694     &    & & &00 11 18.0&-21 14 58&  64& -79& 8.339&0.341&2.644& 0.018&3.786&-0.12& 84& 3.72& 0.78&F& & 3.7& 2.7& 5.9&1.13&1.06&1.20\\
   920&HD 700     &    & & &00 11 24.4&+23 49 05& 111& -38& 8.239&0.362&2.621& 0.011&3.771&-0.25& 54& 4.59& 0.37&H& & 7.3& 0.3&11.6&0.97&0.91&1.05\\
   914&HD 705     &    &*&*&00 11 18.0&-32 48 01& 357& -80& 8.601&0.430&     &      &3.738&-0.25& 66& 4.50& 1.19&H& &17.5&14.2&    &0.88&0.86&0.92\\
   931&HD 717     &AB  &*& &00 11 35.2&-03 04 40& 100& -64& 7.580&0.345&2.638& 0.018&3.783&-0.21& 66& 3.48& 1.17&F& & 5.2& 3.1& 6.7&1.12&1.06&1.20\\
   937&HD 732     &    & & &00 11 36.4&-23 37 56&  52& -80& 7.945&0.355&2.639& 0.022&3.796& 0.20&107& 2.70& 1.28&F& & 2.2& 1.8& 2.7&1.49&1.40&1.59\\
   924&HD 734     &A   & &*&00 11 30.8&-49 37 45& 319& -66& 9.143&0.448&     &      &3.728&-0.33& 48& 5.74& 0.20&H& &    &    &    &0.81&0.77&0.86\\
   924&HD 734     &B   &*&*&00 11 30.2&-49 37 41& 319& -66&10.987&0.655&     &      &     &     & 48& 7.59&     &H& &    &    &    &    &    &    \\
   952&HD 737     &    &*& &00 11 45.7&+27 30 36& 112& -35& 8.449&0.303&2.642&-0.004&3.804&-0.25& 92& 3.64& 0.57&H& & 3.1& 2.0& 4.1&1.17&1.10&1.24\\
   947&HD 738     &    &*& &00 11 40.8&-09 42 10&  93& -70& 8.530&0.324&2.645& 0.014&3.794&-0.24&129& 2.98& 1.44&F& & 2.8& 2.3& 4.4&1.27&1.13&1.38\\
   950&HD 739     &    & & &00 11 44.0&-35 07 59& 347& -78& 5.250&0.290&2.653&-0.010&3.811&-0.13& 22& 3.55& 0.39&H& & 2.4& 1.0& 3.2&1.25&1.21&1.30\\
   929&HD 741     &    & &*&00 11 33.0&-58 54 35& 312& -57& 8.355&0.398&     &      &3.755&-0.26& 59& 4.51& 0.83&H& &13.2& 8.8&    &0.91&0.85&0.97\\
   956&HD 744     &AB  &*& &00 11 50.1&+28 25 24& 112& -34& 7.510&0.322&2.622&-0.008&3.791&-0.45& 78& 3.05& 1.62&H& & 4.4& 2.9& 5.3&1.16&1.10&1.25\\
   944&HD 749     &    &*& &00 11 38.0&-49 39 21& 319& -66& 7.903&0.672&     &      &     &     &   &     &     &G&*&    &    &    &    &    &    \\
   934&HD 750     &    & &*&00 11 35.7&-57 28 21& 313& -59& 9.013&0.492&     &      &3.707&-0.35& 36& 6.21& 0.15&H& &    &    &    &0.76&0.73&0.81\\
   971&HD 755     &    &*& &00 12 01.4&+28 36 23& 113& -33& 7.194&0.253&2.690& 0.006&3.836&-0.08& 92& 2.38& 0.99&H& & 1.4& 1.2& 1.7&1.56&1.50&1.63\\
   964&HD 768     &    & & &00 11 54.7&-22 49 03&  56& -80& 7.934&0.238&2.696&-0.004&3.840&-0.15& 92& 3.12& 0.24&H& & 1.2&    & 1.8&1.37&1.32&1.44\\
 \hline
 \end{tabular}
 \end{table*}


\addtocounter{table}{-1}

\begin{table*}
\caption[]{Sample right-hand page of the catalogue (Fields 26-50 for the first 
100 stars).}
\tiny
\begin{tabular}{|l|@{\hspace{1mm}}r@{\hspace{1mm}}r@{\hspace{1mm}}r@{\hspace{1.5mm}}r@{\hspace{1mm}}r@{\hspace{2mm}}r@{\hspace{1mm}}r@{\hspace{0mm}}r@{\hspace{0mm}}r|r@{\hspace{0mm}}r@{\hspace{0mm}}
r@{\hspace{2mm}}r@{\hspace{1mm}}r@{\hspace{1mm}}|@{\hspace{0mm}}r@{\hspace{1mm}}r@{\hspace{1mm}}r|@{\hspace{2mm}}r@{\hspace{1mm}}r@{\hspace{1mm}}r@{\hspace{1mm}}r@{\hspace{2mm}}r@{\hspace{1mm}}r@
{\hspace{1mm}}|r|}
\hline
 Name&$V_{r}$&m.e&$\sigma_{V_{r}}$&N&$\Delta T$&$P(\chi^{2})$&$f_{d}$&$f_{v}$&$v$sin$i$&$\mu_{\alpha *}$&
$\mu_{\delta}$&$\sigma_{\mu}$&$\pi$&$\sigma_{\pi}$&U&V&W&$R_{gal}$&$z$&$R_{min}$&$R_{max}$&$e$&$z_{max}$&$f_{n}$\\
 &km/s&km/s&km/s&&d&&&&km/s&&{\hspace{-2mm}}mas/yr&&mas&mas&km/s&km/s&km/s&kpc&kpc&kpc&kpc&&kpc&\\
 27&28&29&30&31&32&33&34&35&36&37&38&39&40&41&42&43&44&45&46&47&48&49&50&51\\   
\hline
HD 15      & -22.6& 2.4&  3.4&  2& 381&0.000& &C&  2&   -4&  -10& 2&  3.2& 1.0&     &     &     &      &      &       &       &     &       &a   \\
HD 16      & -10.7& 1.7&  1.4&  2&1848&0.556& &C&    &   28&   -7& 2&  2.5& 0.9&     &     &     &      &      &       &       &     &       &    \\
HD 23      &  34.3& 0.2&  0.3&  3&1767&0.309& &C&  4& -111& -131& 2& 23.9& 0.9&  40& -22& -16& 7.986&-0.037&   6.39&   8.88& 0.16&   0.15&    \\
HD 24      & -26.9& 0.2& 19.0&  5&1806&0.000&*&C&  9&   64&   20& 2& 14.2& 0.8& -31&   7&  15& 7.972&-0.057&   7.74&   9.09& 0.08&   0.38&a   \\
HD 25      &  -2.2& 1.5&     &   &    &     & &L&    &  -30&  -68& 2&     &    &  19&   1& -24& 8.036&-0.019&   7.54&   9.03& 0.09&   0.28&    \\
HD 26      &-215.1& 0.2&  1.6&103&5879&0.000& &C&  5&  261&  -59& 2&  1.7& 1.3&     &     &     &      &      &       &       &     &       &a   \\
HD 39      & -26.8& 0.8&  1.4&  3& 661&0.006& &C& 14&  -18&   26& 2& 11.0& 0.9&  11& -15&  24& 8.030&-0.042&   7.24&   8.37& 0.07&   0.55&a   \\
HD 59      & -34.2& 0.3&  0.2&  2&  94&0.626& &C&  4&  159&   40& 2&     &    & -48& -57&  13& 8.035&-0.023&   5.10&   8.31& 0.24&   0.33&    \\
HD 63      &   0.2& 2.1&  3.0&  2&1507&0.000& &C& 24&   88&  -72& 1& 19.8& 0.9& -10& -24& -10& 8.000&-0.047&   6.91&   8.00& 0.07&   0.06&a   \\
HD 67      &   6.3& 0.3&  0.1&  2&1085&0.986& &C&  4&   84&  -78& 2& 18.6& 0.9& -10& -28&   2& 7.979&-0.044&   6.65&   7.98& 0.09&   0.14&    \\
HD 70      & -28.4& 0.2&  0.4&  3& 727&0.287& &C&  3&   65& -166& 2& 21.0& 1.1&   7& -31& -37& 8.022&-0.003&   6.49&   8.13& 0.11&   0.52&    \\
HD 85      & -12.0& 2.4&     &  1&   0&     & &C&    &   60&  -18& 2&  5.9& 1.7& -16& -24&  -1& 8.019&-0.060&   6.91&   8.03& 0.08&   0.10&a   \\
HD 101     & -45.6& 0.1&  0.2&  7&2452&0.924& &C&  2& -149& -151& 1& 26.2& 0.8&  46& -34&  17& 8.009&-0.026&   5.86&   8.84& 0.20&   0.41&b   \\
HD 105     &   1.6& 0.3&  0.5&  4&3363&0.176& &C& 15&   98&  -76& 1& 24.9& 0.9& -10& -21&  -2& 7.989&-0.038&   7.02&   7.99& 0.06&   0.09&d   \\
HD 117     &  15.0& 0.3&  0.2&  3& 361&0.842& &C& 11&   -6&  -10& 3&     &    &   7&  -2& -14& 7.983&-0.096&   7.68&   8.54& 0.05&   0.15&a   \\
HD 123     &  -8.0& 5.0&  1.3& 10&2676&0.000&*&C&  5&  271&   30& 2& 49.3& 1.0& -20& -19&  -1& 8.009&-0.001&   7.13&   8.07& 0.06&   0.09&a   \\
HD 126     &  11.9& 1.7&  2.4&  2& 829&0.276& &C&    &   20&  -13& 2& 11.1& 1.0&  -7&   0& -14& 8.014&-0.071&   8.00&   8.34& 0.02&   0.13&b   \\
HD 142     &   5.3& 0.3&  0.1&  2&1528&0.800& &C& 11&  575&  -37& 1& 39.0& 0.6& -58& -37& -15& 7.992&-0.023&   5.82&   8.58& 0.19&   0.12&    \\
HD 153     & -31.8& 0.3&  0.2&  2& 133&0.645& &C&  5&   78&   -1& 2&  8.1& 0.9& -27& -48&   2& 8.048&-0.041&   5.61&   8.12& 0.18&   0.14&    \\
HD 156     &  13.6& 2.4&     &   &    &     & &L&    &   16&    1& 2&  6.7& 0.9&  -8&  -1& -15& 7.990&-0.127&   7.98&   8.30& 0.02&   0.17&    \\
HD 160     &  17.2& 1.8&     &   &    &     & &L&    &   28&   23& 2&  5.4& 1.7& -16&  -4& -28& 7.936&-0.126&   7.82&   8.16& 0.02&   0.35&a   \\
HD 166     &  -6.9& 0.1&  0.3& 21&5410&0.162& &C&  4&  380& -178& 1& 73.0& 0.8& -15& -22& -10& 8.004&-0.007&   7.01&   8.02& 0.07&   0.04&c   \\
HD 189     &   7.6& 0.3&  0.1&  2& 244&0.846& &C&  9&  -25&  -68& 2&  5.6& 1.1&  34& -28&  -8& 7.983&-0.127&   6.21&   8.61& 0.16&   0.13&    \\
HD 200     &  -0.2& 0.2&  0.2&  2& 360&0.635& &C&  5&   59&  -88& 1&  9.1& 1.1&  -4& -50& -23& 8.005&-0.099&   5.56&   8.01& 0.18&   0.27&    \\
HD 203     &   6.5& 3.5&     &   &    &     & &L&    &   97&  -47& 1& 25.6& 0.8& -11& -15& -10& 7.995&-0.038&   7.40&   8.00& 0.04&   0.06&d   \\
HD 219     & -20.0& 3.7&     &   &    &     & &L&    &   43&  -11& 2&  4.2& 0.7&     &     &     &      &      &       &       &     &       &    \\
HD 220     &      &    &     &   &    &     & & &    &  -32&   14& 2&     &    &     &     &     & 8.064& 0.011&       &       &     &       &    \\
HD 222     &      &    &     &   &    &     & & &    &   13&  -16& 1&  8.4& 0.9&     &     &     & 8.031&-0.075&       &       &     &       &c   \\
HD 233     & -12.2& 0.4&  0.5&  2& 360&0.178& &C& 11&   70&  -23& 2&  7.9& 2.3& -21& -24&   5& 7.995&-0.083&   6.83&   8.05& 0.08&   0.20&b   \\
HD 251     &  17.2& 1.1&  1.9&  3& 365&0.364& &C& 45&   -5&   -6& 2&  5.9& 1.7&   1&   9& -15& 8.016&-0.087&   7.95&   9.05& 0.07&   0.16&b   \\
HD 268     &   8.4& 0.2&  0.1&  2& 361&0.797& &C&  9&  213& -133& 2& 20.6& 0.9& -27& -48& -18& 7.994&-0.048&   5.58&   8.06& 0.18&   0.17&    \\
HD 276     &  23.0& 0.3&  0.5&  2& 381&0.216& &C&  7&   78&   -5& 2& 11.9& 0.6& -16& -30& -19& 7.962&-0.054&   6.55&   7.98& 0.10&   0.18&    \\
HD 285     &  -0.9& 0.5&  0.6&  2& 358&0.151& &C& 13&   74&   24& 2&  7.5& 0.8& -46&  -9& -14& 7.953&-0.116&   7.05&   8.78& 0.11&   0.16&    \\
HD 291     & -17.6& 0.3&  0.1&  2& 363&0.946& &C&  6&  -68&  -58& 1&  9.6& 0.9&  45&  -7& -13& 8.039&-0.041&   6.82&   9.48& 0.16&   0.11&    \\
HD 292     &   0.7& 0.3&  0.2&  2& 363&0.664& &C& 10&   11&  -16& 2&  7.2& 0.9&  -3&  -6& -11& 8.050&-0.058&   7.86&   8.21& 0.02&   0.08&b   \\
HD 299     &   1.4& 0.2&  0.2&  3&1765&0.820& &C&  8&  201&    8& 2& 21.8& 0.8& -39& -19&  -6& 8.021&-0.005&   6.87&   8.44& 0.10&   0.02&c   \\
HD 307     &  14.6& 0.2&  0.2&  4&1477&0.580& &C&  4&   29&  -98& 2&  8.2& 1.0&  12& -59&   0& 7.965&-0.115&   5.03&   8.06& 0.23&   0.16&    \\
HD 308     & -33.4& 0.3&  0.1&  2& 360&0.870& &C&  1&  114&   38& 2&     &    & -53&   1&  18& 7.974&-0.070&   7.24&   9.41& 0.13&   0.47&d   \\
HD 309     &  56.2& 0.7&  1.0&  2&1215&0.007& &C&  7&   34&  -53& 2&  8.9& 1.5&  16& -46& -38& 7.967&-0.072&   5.67&   8.13& 0.18&   0.56&a   \\
HD 330     & -42.5& 0.3&  0.6&  3& 714&0.076& &C&  4&   73&  -63& 2&  9.8& 1.0&  -5& -56& -29& 8.045&-0.015&   5.32&   8.05& 0.20&   0.34&    \\
HD 333     &      &    &     &   &    &     & & &    &    1&    6& 2&  3.8& 0.9&     &     &     & 8.059&-0.100&       &       &     &       &    \\
HD 334     & -25.4& 0.3&  0.5&  2& 607&0.260& &C& 10&  -29&   29& 2& 10.7& 1.0&   6&   6&  30& 8.002&-0.087&   7.85&   9.06& 0.07&   0.73&    \\
HD 361     &   7.7& 0.2&  0.4&  6&3621&0.154& &C&  3&  -14&  -10& 2& 36.2& 1.0&   2&   2&  -7& 7.999&-0.027&   7.86&   8.62& 0.05&   0.03&    \\
HD 372     & -18.1& 3.2&  4.5&  2& 132&0.001& &C& 35&   85&   -3& 1&  8.8& 0.9& -31& -37&  -6& 8.050&-0.018&   6.12&   8.19& 0.14&   0.02&a   \\
HD 373     &  -6.2& 0.6&  0.8&  3&1147&0.543& &A& 35&   80&   24& 2& 14.4& 1.0& -22& -17&   5& 8.030&-0.012&   7.24&   8.13& 0.06&   0.17&g   \\
HD 375     &  -5.7& 0.2&  0.1&  2& 363&0.815& &C&  5&  114&   10& 2& 12.7& 0.9& -36& -23&  -1& 8.027&-0.036&   6.74&   8.34& 0.11&   0.10&b   \\
HD 377     &   1.3& 0.3&  0.3&  2&1437&0.475& &C& 13&   85&   -3& 2& 25.1& 1.0& -14&  -7&  -4& 8.006&-0.033&   7.84&   8.06& 0.01&   0.06&c   \\
HD 382     &   1.2& 0.2&  0.2&  2&1305&0.452& &C&  2&  137&   35& 2&     &    & -17&  -4&  -4& 7.997&-0.026&   7.86&   8.22& 0.02&   0.06&    \\
HD 392     &  -3.0& 2.0&     &   &    &     & &L&    &   46&    8& 2&  7.6& 0.9& -28&  -9&  -2& 7.984&-0.130&   7.44&   8.33& 0.06&   0.16&    \\
HD 400     & -15.2& 0.1&  0.3& 20&4858&0.831& &C&  6& -115& -124& 2& 30.3& 0.7&  28& -10&  -8& 8.012&-0.014&   7.03&   8.84& 0.11&   0.02&    \\
HD 404     & -60.8& 0.4&     &  1&   0&     & &C&  2&  178&    2& 1& 31.3& 2.1&   6& -66&  -8& 8.015& 0.002&   4.79&   8.06& 0.25&   0.02&c   \\
HD 410     & -36.2& 0.3&  0.4&  2& 362&0.283& &C&  3&  149&   79& 2&     &    & -80&  13&  -3& 7.963&-0.069&   6.96&  10.81& 0.22&   0.12&    \\
HD 427     & -20.9& 0.6&  1.0&  3&1161&0.310& &C& 27&    6&    7& 2&  5.0& 0.9&  -8&   2&  20& 7.977&-0.109&   7.97&   8.52& 0.03&   0.48&    \\
HD 432     &  12.6& 1.6&     &   &    &     & &L&    &  523& -180& 1& 59.9& 0.6& -39&  -8& -22& 8.008&-0.001&   7.28&   8.69& 0.09&   0.22&c   \\
HD 435     &   8.7& 0.3&  0.2&  2&1904&0.531& &C&  7&   38&  -38& 2& 12.9& 1.0&  -8& -10& -18& 8.015&-0.058&   7.74&   8.03& 0.02&   0.16&    \\
HD 439     &   5.9& 2.6&  3.6&  2& 923&0.044& &C& 55&  100&  -45& 1&  8.6& 0.9& -37& -48&  -7& 7.966&-0.108&   5.52&   8.13& 0.19&   0.11&    \\
HD 447     &  11.3& 1.1&  0.2&  2& 437&0.910& &C& 50&   11&   21& 1&  8.7& 0.8& -13&  11&   0& 8.029&-0.077&   8.02&   9.14& 0.06&   0.14&    \\
HD 457     & -19.4& 0.2&  0.1&  3&1477&0.856& &C&  3&  112&  -16& 2& 18.2& 0.9& -28& -16&  15& 7.987&-0.053&   7.20&   8.21& 0.07&   0.36&    \\
HD 460     & -13.9& 0.6&  1.2&  4& 439&0.016& &C&  6&   53&  -33& 1&  4.8& 2.0& -17& -31& -20& 8.052&-0.039&   6.54&   8.07& 0.10&   0.20&    \\
HD 466     & -16.3& 0.9&  0.9&  4&1116&0.758& &A& 60&   37&  -12& 2&  6.5& 0.8& -23& -20&  13& 7.968&-0.150&   7.05&   8.06& 0.07&   0.36&    \\
HD 469     &  -1.1& 0.2&  0.2&  2&1050&0.585& &C&  3&   47&   13& 1&  7.0& 0.7& -31&  -7&  -8& 7.952&-0.126&   7.43&   8.43& 0.06&   0.13&a   \\
HD 471     &   5.0&10.9& 15.5&  2& 793&0.000& &C&  1&  173& -149& 1& 22.1& 2.3& -21& -29& -34& 8.013&-0.027&   6.66&   8.06& 0.10&   0.45&a   \\
HD 483     & -31.4& 0.1& 20.4& 36& 753&0.000&*&C&  4&    3& -131& 1& 19.3& 0.8&  21& -40&  -1& 8.012&-0.036&   5.89&   8.26& 0.17&   0.10&a   \\
HD 489     & -31.0& 0.2&  0.2&  3& 739&0.554& &C&  4&  269&   29& 1& 13.8& 1.3& -77& -59&  13& 8.018&-0.049&   4.80&   8.81& 0.30&   0.35&a   \\
HD 493     &   7.7& 1.6&     &   &    &     & &L&    &   78&  -16& 6& 14.6& 1.3& -18& -16& -12& 7.990&-0.068&   7.30&   8.05& 0.05&   0.09&a   \\
HD 494     &   6.0& 5.0& 67.2&  2& 803&0.000&*&C&  0&  137& -137& 2& 11.9& 1.3& -23& -74&  -2& 7.979&-0.081&   4.43&   8.01& 0.29&   0.11&a   \\
HD 531     &  13.9& 0.4&     &  1&   0&     & &C&  9&   54&  -10& 3& 14.2& 4.2& -13&   0& -15& 8.009&-0.042&   7.99&   8.37& 0.02&   0.12&a   \\
HD 536     &  -1.4& 0.9&  0.8&  2& 370&0.533& &C& 30&  -53&  -24& 2&  6.9& 1.4&  33&   2&   8& 7.973&-0.120&   7.24&   9.42& 0.13&   0.28&a   \\
HD 546     &  11.3& 0.2&  0.1&  2& 795&0.737& &C&  6&  -62& -150& 2& 17.1& 1.9&  36& -29&  -5& 7.989&-0.057&   6.18&   8.65& 0.17&   0.07&a   \\
HD 547     &  14.3& 0.2&  0.3&  3&1476&0.340& &C&  3&  138& -118& 1& 13.9& 1.1& -20& -59& -13& 7.982&-0.069&   5.10&   8.00& 0.22&   0.11&    \\
HD 564     &  11.1& 0.2&  0.3&  5&3659&0.448& &C&  3&  -96&  -19& 2& 18.8& 1.0&  26&   4&  -4& 7.983&-0.048&   7.41&   9.32& 0.11&   0.07&    \\
HD 570     &      &    &     &   &    &     & & &    &   -5&   -9& 2&     &    &     &     &     &      &      &       &       &     &       &a   \\
HD 578     &  -2.9& 0.4&  0.1&  2& 769&0.906& &C& 14&  -47&  -39& 2&  9.6& 0.8&  25&  -2&  17& 7.959&-0.083&   7.28&   9.01& 0.11&   0.43&    \\
HD 583     & -20.1& 1.4&  1.9&  2& 132&0.001& &C& 13&  -99&  -10& 2& 10.4& 0.8&  49&   3&   4& 8.044&-0.007&   7.01&  10.07& 0.18&   0.19&a   \\
HD 604     &   3.5& 0.5&  0.6&  2& 607&0.201& &C& 10&   69&  -52& 2&  6.2& 1.0& -21& -46& -18& 7.996&-0.126&   5.72&   8.03& 0.17&   0.20&    \\
HD 610     &  11.2& 0.4&  0.6&  2& 312&0.188& &C& 14&   13&   -4& 2& 11.1& 0.7&   0& -11&  -7& 7.959&-0.055&   7.49&   8.09& 0.04&   0.06&    \\
HD 615     &  -4.6& 0.4&  0.1&  3&1790&0.953& &C& 16&  -70&  -63& 1& 12.9& 1.1&  34&  -5&  -8& 8.017&-0.056&   7.09&   9.20& 0.13&   0.06&    \\
HD 631     &  32.3& 0.2&  0.3&  4&1162&0.618& &C&  6&   38&  -17& 1& 15.6& 1.2& -14&  12& -29& 8.013&-0.048&   8.00&   9.19& 0.07&   0.38&    \\
HD 633     &  17.0& 0.3&  0.1&  2& 734&0.955& &C&  6&  -13&   17& 1& 18.2& 0.9&  -1&  13& -12& 8.006&-0.047&   7.97&   9.29& 0.08&   0.09&    \\
HD 639     & -11.8& 1.8&  3.1&  3& 368&0.000& &C& 13&   50&   16& 2&  8.8& 0.8& -22& -20&   8& 8.042&-0.047&   7.10&   8.13& 0.07&   0.24&a   \\
HD 652     &   1.2& 0.6&  0.8&  2& 618&0.050& &C&  6&  169&    0& 2& 16.5& 0.9& -43& -22&  -8& 8.026&-0.015&   6.70&   8.49& 0.12&   0.02&    \\
HD 659     &  -1.8& 0.6&  1.3&  5&3659&0.000& &C&  3&    8&   18& 2&  4.4& 1.1&     &     &     &      &      &       &       &     &       &a   \\
HD 661/2   & -14.0& 5.0&     &  1&   0&     &*&C& 10&  123&   20& 2& 15.1& 0.7& -41&  -6&  -1& 7.971&-0.046&   7.26&   8.75& 0.09&   0.11&a   \\
HD 693     &  14.9& 0.1&  0.3&228&6678&0.961& &C&  5&  -82& -271& 1& 52.9& 0.8&  19& -13& -19& 7.999&-0.018&   7.10&   8.53& 0.09&   0.17&    \\
HD 694     &      &    &     &   &    &     & & &    &   -6&  -42& 2&     &    &     &     &     & 7.993&-0.082&       &       &     &       &    \\
HD 700     &   4.1& 0.3&  0.1&  2& 346&0.894& &C&  4&  -34&   32& 1& 18.6& 0.9&   3&  11&   5& 8.015&-0.033&   7.93&   9.22& 0.08&   0.20&    \\
HD 705     &  14.0& 5.0& 45.5&  2&1714&0.000&*&C&  5& -116&   -8& 2& 15.1& 1.1&  35&  15&  -8& 7.988&-0.065&   7.43&  10.15& 0.15&   0.08&a   \\
HD 717     & -19.8& 0.3&  0.4&  2&1132&0.217& &C&  7&  150&  -12& 2&  9.9& 1.5& -37& -33&   9& 8.005&-0.059&   6.24&   8.24& 0.14&   0.26&a   \\
HD 732     &  -2.1& 0.3&  0.2&  2& 791&0.604& &C& 14&   55&    1& 2&  7.6& 1.1& -25& -13&  -2& 7.989&-0.105&   7.35&   8.18& 0.05&   0.13&    \\
HD 734     &   4.9& 0.5&  0.8&  2& 362&0.050& &C&  5&  -38&  -38& 2& 20.9& 2.2&  12&  -4&   0& 7.985&-0.044&   7.50&   8.58& 0.07&   0.11&    \\
HD 734     &   6.7& 0.5&     &  1&   0&     & &C&  4&  -38&  -38& 2& 20.9& 2.2&  13&  -5&  -2& 7.985&-0.044&   7.48&   8.58& 0.07&   0.09&    \\
HD 737     &  -6.8& 1.6&  2.3&  2&1420&0.000& &C& 10&    5&  -27& 2& 10.9& 1.1&   5& -12&  -6& 8.029&-0.052&   7.45&   8.24& 0.05&   0.05&a   \\
HD 738     &  -6.8& 0.4&  0.2&  2&1006&0.626& &C& 16&  101&  -11& 2&  6.9& 1.6& -50& -37&  -5& 8.002&-0.121&   5.93&   8.44& 0.18&   0.13&b   \\
HD 739     & -21.1& 0.3&     &  1&   0&     & &C&  5&  169&  115& 1& 45.8& 0.7& -25&   3&  16& 7.996&-0.021&   7.83&   8.77& 0.06&   0.40&d   \\
HD 741     &  -4.8& 0.2&  0.4&  4&2273&0.354& &C&  1&  273&   -5& 3& 17.0& 0.8& -67& -35&  -7& 7.979&-0.050&   5.78&   8.82& 0.21&   0.05&    \\
HD 744     &   0.6& 0.3&  0.4&  2& 357&0.416& &C&  7&  -96&   36& 2& 12.8& 1.4&  26&  23&  16& 8.025&-0.043&   7.71&  10.50& 0.15&   0.44&a   \\
HD 749     &  21.6& 1.6&  2.8&  3&2520&0.000& &C&  1&   28&  -58& 1&  7.1& 1.1&     &     &     &      &      &       &       &     &       &a   \\
HD 750     &  26.1& 0.3&  0.4&  2& 360&0.233& &C&  4&  136&  -68& 2& 27.5& 1.1&  -7& -30& -21& 7.987&-0.031&   6.56&   7.99& 0.10&   0.20&    \\
HD 755     & -38.8& 2.7&  5.4&  4&1089&0.000& &A& 70&  -14&  -92& 2& 10.9& 0.9&  33& -44& -10& 8.029&-0.051&   5.60&   8.46& 0.20&   0.07&a   \\
HD 768     &      &    &     &   &    &     & & &    &   46&  -39& 2& 10.9& 1.0&     &     &     & 7.991&-0.090&       &       &     &       &    \\
 \hline

\hline
\end{tabular}
\label{catalogue.tab}
\end{table*}

\clearpage


\begin{thebibliography}{}

\bibitem[1984]{abt84} 
Abt, H.A. 1984, ApJ 285, 247

\bibitem[1986]{abt86} 
Abt, H.A. 1986, ApJ 309, 260

\bibitem[1979]{abt79} 
Abt, H.A., Brodzik, D., \& Schaefer, B. 1979, PASP 91, 176

\bibitem[1996]{alonso96} 
Alonso, A., Arribas, S. \& Mart\'{i}nez-Roger, C.\ 1996, A\&A 313, 873

\bibitem[1983a]{an83a}
Andersen, J., \& Nordstr{\"o}m, B. 1983a, A\&AS 52, 479

\bibitem[1983b]{an83b}
Andersen, J., \& Nordstr{\"o}m, B. 1983b, A\&A 122, 23

\bibitem[1985]{jaetal85}
Andersen, J., Nordstr{\"o}m, B., Ardeberg, A., Benz, W., Imbert, M.,
Lindgren, H., Martin, N., Maurice, E., Mayor, M., \& Pr{\'e}vot, L. 1985,
A\&AS 59, 15

\bibitem[2002]{andrievsky02}
Andrievsky, S.M., Kovtyukh, V.V., Luck, R.E., Lépine, J.R.D., Bersier, D.,
Maciel, W.J., Barbuy, B., Klochkova, V.G., Panchuk, V.E., \& Karpischek, R.U.
2002, A\&A 381,32

\bibitem[1999]{BaSra99}
Backer, D.C., \& Sramek, R.A. 1999, ApJ 524, 805

\bibitem[1979]{baranne79}
Baranne, A., Mayor, M., \& Poncet, J.-L. 1979, Vistas in Astron. 23, 279

\bibitem[2000]{barbier00}
Barbier-Brossat, M., \& Figon, P. 2000, A\&AS 142, 217

\bibitem[2002]{barklem02}
Barklem, P.S., Stempels, H.C., Allende Prieto, C., Kochukhov, O.P., Piskunov,
N., \& O'Mara, B.J. 2002, A\&A 385, 951

\bibitem[2003]{bensby03}
Bensby, T., Feltzing, S., \& Lundstr{\"o}m, I. 2003, A\&A 410, 527

\bibitem[1980]{benzm80}
Benz, W., \& Mayor, M. 1980, A\&A 93, 235

\bibitem[1984]{benzm84}
Benz, W., \& Mayor, M. 1984, A\&A 138, 183

\bibitem[2000]{binney00}
Binney, J., Dehnen, W., \& Bertelli, G. 2000, MNRAS 318, 658

\bibitem[1991]{burkhart91}
Burkhart, C. \& Coupry, M.F. 1991, A\&A 249, 205

\bibitem[2003]{burst03}
Burstein, D. 2003, AJ 126, 1849

\bibitem[1918-24]{hdcat}
Cannon, A.J., \& Pickering, E.C. 1918-1924, Harvard Ann. 91-99 (HD catalog)

\bibitem[2003]{chenl03}
Chen, L., Hou, J.L., \& Wang, J.J. 2003, AJ 125, 1397

\bibitem[2000]{chenyq00}
Chen, Y.Q., Nissen, P.E., Zhao, G., Zhang, H.W., \& Benono, T. 2000, A\&AS 141,
491

\bibitem[2003]{chenyq03}
Chen, Y.Q., Zhao, G., Nissen, P.E., Bai, G.S., \& Qiu, H.M. 2003, ApJ 591, 925

\bibitem[2000]{chiba00}
Chiba, M., Beers, T.C. 2000, AJ 119, 2843

\bibitem[1975]{craw75}
Crawford, D.L. 1975, AJ 80, 955

\bibitem[1966]{crawbar66}
Crawford, D.L., Barnes, J.V., Faure, B.Q., Golson, J.C., \& Perry, C.L. 1966,
AJ 71, 709

\bibitem[1972]{crawbar72}
Crawford, D.L., Barnes, J.V., Gibson, J., Golson, J.C., Perry, C.L., \&
Crawford, M.L. 1972, A\&AS 5, 109

\bibitem[1970]{crawbar70}
Crawford, D.L., Barnes, J.V., \& Golson, J.C. 1970, AJ 75, 624

\bibitem[1971a]{crawbar71a}
Crawford, D.L., Barnes, J.V., \& Golson, J.C. 1971a, AJ 76, 621

\bibitem[1971b]{crawbar71b}
Crawford, D.L., Barnes, J.V., \& Golson, J.C. 1971b, AJ 76, 1058

\bibitem[1973]{crawbar73}
Crawford, D.L., Barnes, J.V., Golson, J.C., \& Hube, D.P. 1973, AJ 78, 738 

\bibitem[2004]{desimon04}
De Simone, R.A., Wu, X., \& Tremaine, S. 2004, MNRAS, subm. (astro-ph/030106)

\bibitem[1998]{dehnen98}
Dehnen, W. 1998, AJ 115, 2384

\bibitem[1998]{dehnbin98}
Dehnen, W., \& Binney, J.J. 1998, MNRAS 298, 387

\bibitem[1991]{duqmayor91}
Duquennoy, A., \& Mayor, M. 1991, A\&A 248, 485

\bibitem[1993]{edv93}
Edvardsson, B., Andersen, J., Gustafsson, B., Lambert, D.L., Nissen, P.E., \&
Tomkin, J. 1993, A\&A 275, 101 

\bibitem[1997]{hipp97}
ESA 1997, The Hipparcos and Tycho Catalogues, ESA-SP 1200

\bibitem[2003]{feltz03}
Feltzing, S., Bensby, T., \& Lundstr{\"o}m, I. 2003, A\&A 397, L1

\bibitem[2001]{feltz01}
Feltzing, S., Holmberg, J., \& Hurley, J.R. 2001, A\&A 377, 911

\bibitem[1994]{flynnfu94}
Flynn, C., \& Fuchs, B. 1994, MNRAS 270, 471

\bibitem[1997]{flynnmo97}
Flynn, C., \& Morell, O. 1997, MNRAS 286, 617

\bibitem[1996]{flynnjsl96}
Flynn, C., Sommer-Larsen, J., \& Christensen, P.R. 1996, MNRAS 281, 1027

\bibitem[2002]{freeman02}
Freeman, K.C., \& Bland-Hawthorn, J. 2002, ARA\&A 40, 487

\bibitem[2002]{friel02}
Friel, E.D., Janes, K.A., Tavarez, M., Scott, J., Katsanis, R., Lotz, J., 
Hong, L., \& Miller, N. 2002, AJ 124, 2693

\bibitem[1998]{fuhrm98}
Fuhrmann, K. 1998, A\&A 338, 161

\bibitem[2001]{fux01}
Fux, R. 2001, A\&A 373, 511

\bibitem[2000]{girardi00}
Girardi, L., Bressan, A., Bertelli, G., \& Chiosi, C. 2000, A\&AS 141, 371

\bibitem[2001]{girardi01}
Girardi, L., \& Salaris, M. 2001, MNRAS 323, 109

\bibitem[1994]{glaspey94}
Glaspey, J.W., Pritchet, C.J., \& Stetson, P.B. 1994, AJ 108, 271

\bibitem[1989]{gray89}
Gray, R.O. 1989, AJ 89, 1049

\bibitem[1989]{graygar89}
Gray, R.O., \& Garrison, R.F. 1989, ApJS 69, 301

\bibitem[1987]{grenon87}
Grenon, M. 1987, JA\&A 8, 123

\bibitem[1976]{gronbech76}
Gr{\o}nbech, B., Olsen, E.H. 1976, A\&AS 25, 213

\bibitem[1977]{gronbech77}
Gr{\o}nbech, B., Olsen, E.H. 1977, A\&AS 27, 443

\bibitem[2003]{helmi03}
Helmi, A., Navarro, J.F., Meza, A., Steinmetz, M., \& Eke, V. 2003, ApJ 592,
L25

\bibitem[1996]{henry96}
Henry, T.J., Soderblom, D.R., Donahue, R.A., \& Baliunas, S.L. 1996, AJ 111,
439 (Table 6)

\bibitem[1982]{bsc82}
Hoffleit, D., \& Jaschek, C. 1982, The Bright Star Catalog, Yale Univ. Obs.
(New Haven, Conn.)

\bibitem[2001]{holmb01}
Holmberg, J. 2001, PhD thesis, Lund University

\bibitem[2004]{holmb04}
Holmberg, J. 2004, A\&A, in prep.

\bibitem[2003]{holmbetal03}
Holmberg, J., Nordstr{\"o}m, B., J{\o}rgensen, B.R., \& Andersen, J. 2003, 
in ``The Evolution of Galaxies. III - From Simple Approaches to
Self-Consistent Models'', Eds. G. Hensler, G. Stasinska, S. Harfst, P.
Kroupa, \& C. Theis. Ap\&SS 284, 685

\bibitem[2000]{holmb00}
Holmberg, J., \& Flynn, C. 2000, MNRAS 313, 209

\bibitem[1978]{houk78}
Houk, N. 1978, Michigan Catalogue of Two-Dimensional Spectral Types for the
HD Stars, Vol. 2, Univ. of Michigan (Ann Arbor, Michigan)

\bibitem[1982]{houk82}
Houk, N. 1982, Michigan Catalogue of Two-Dimensional Spectral Types for the
HD Stars, Vol. 3, Univ. of Michigan (Ann Arbor, Michigan)

\bibitem[1975]{houk75}
Houk N., \& Cowley A.P. 1975, Michigan Catalogue of Two-Dimensional Spectral
Types for the HD Stars, Vol. 1, Univ. of Michigan (Ann Arbor, Michigan)

\bibitem[2002]{hanninen02}
H\"anninen, J., \& Flynn, C. 2002, MNRAS 337, 731

\bibitem[2000]{tycho2}
H{\o}g, E., Fabricius, C., Makarov, V.V., Urban, S., Corbin, T., Wycoff, G.,
Bastian, U., Schwekendiek, P., \& Wicenec, A. 2000, A\&A 355, L27

\bibitem[2000]{bjarner00}
J{\o}rgensen, B.R. 2000, A\&A 363, 947

\bibitem[2004]{bjarner04}
J{\o}rgensen, B.R., \& Lindegren, L. 2004, A\&A, in prep.

\bibitem[1989]{kawaler89}
Kawaler, S.D. 1989, ApJ 343, L65

\bibitem[1993]{kroupa93} 
Kroupa, P., Tout, C.A., Gilmore, G., 1993, MNRAS 262, 545

\bibitem[1999]{lachaume99}
Lachaume, R., Dominik, C., Lanz, T., \& Habing, H.J. 1999, A\&A 348, 897

\bibitem[1985]{dwl85}
Latham, D.W. 1985, in ``Stellar Radial Velocities'' (IAU Colloq. 88), eds.
A.G.D. Philip \& D.W. Latham, L. Davis Press, Schenectady, N.Y., p. 21

\bibitem[1996]{dmvir96}
Latham, D.W., Nordstr{\"o}m, B., Andersen, J., Bester, M., Torres, G.,
Stefanik, R.P., \& Thaller, M. 1996, A\&A 314, 864

\bibitem[2001]{lebreton01}
Lebreton, Y. 2001, ARA\&A 38, 35

\bibitem[2001]{lejeune01}
Lejeune, T., \& Schaerer, D. 2001, A\&A 366, 538

\bibitem[2003]{maciel03}
Maciel, W.J., Costa, R.D.D., \& Uchida, M.M.M. 2003, A\&A 397, 667

\bibitem[1974]{mayor74}
Mayor, M. 1974, A\&A 32, 321

\bibitem[1976]{mayor76}
Mayor, M. 1976, A\&A 48, 301

\bibitem[1985]{mayor85}
Mayor, M. 1985, in ``Stellar Radial Velocities'' (IAU Colloq. 88), eds.
A.G.D. Philip \& D.W. Latham, L. Davis Press, Schenectady, N.Y., p. 35

\bibitem[1998]{nami98}
Mowlavi, N., Schaerer, D., Meynet, G., Bernasconi, P.A., Charbonnel C., \&
Maeder, A. 1998, A\&AS 128, 471

\bibitem[2003]{muhlbauer03}
M\"uhlbauer, G., \& Dehnen, W. 2003, A\&A 401, 975 

\bibitem[1985]{na85}
Nordstr{\"o}m, B., \& Andersen, J. 1985, A\&AS 61, 53

\bibitem[1997a]{naa97}
Nordstr{\"o}m, B., Andersen, J., \& Andersen, M.I. 1997a, A\&A 322, 460

\bibitem[1994]{bnetal94}
Nordstr{\"o}m, B., Latham, D.W., Morse, J.A., Milone, A.A.E., Kurucz, R.L.,
Andersen, J., \& Stefanik, R.P. 1994, A\&A 287, 338

\bibitem[1996]{bnetal96}
Nordstr{\"o}m, B., Olsen, E.H., Andersen, J., Mayor, M., \& Pont, F. 1996, in
``Galactic Chemodynamics 4: The History of the Milky Way and its Satellite
System'', Eds. A. Burkert, D. Hartmann \& S. Majewski, ASP Conf. Ser. 112,
145

\bibitem[1997b]{bnetal97}
Nordstr{\"o}m, B., Stefanik, R.P., Latham, David W., \& Andersen, J. 1997b,
A\&AS 126, 21

\bibitem[1999]{paris99}
Nordstr{\"o}m, B., Andersen, J., Olsen, E.H., Fux, R., Mayor, M., \& Pont, F.
1999, in ``Galactic Evolution: Connecting the distant Universe with the local
fossil record'', Ed. M. Spite. Ap\&SS 265, 235

\bibitem[1979]{eho79}
Olsen, E.H. 1979, A\&AS 37, 367

\bibitem[1980]{eho80}
Olsen, E.H. 1980, A\&AS 39, 205

\bibitem[1983]{eho83}
Olsen, E.H. 1983, A\&AS 54, 55

\bibitem[1984]{eho84}
Olsen, E.H. 1984, A\&AS 57, 443

\bibitem[1988]{eho88}
Olsen, E.H. 1988, A\&A 189, 173

\bibitem[1993]{eho93}
Olsen, E.H. 1993, A\&AS 102, 89

\bibitem[1994a]{eho94a}
Olsen, E.H. 1994a, A\&AS 104, 429

\bibitem[1994b]{eho94b}
Olsen, E.H. 1994b, A\&AS 106, 257

\bibitem[1984]{ehoper84}
Olsen, E.H., \& Perry C.L. 1984, A\&AS 56, 229

\bibitem[1997]{pagel97}
Pagel, B.E.J. 1997, Nucleosynthesis and Chemical Evolution of Galaxies,
Cambridge Univ. Press, Cambridge, U.K.

\bibitem[2001]{gaia01}
Perryman, M.A.C., de Boer, K.S., Gilmore, G., et al. 2001, A\&A 369, 339

\bibitem[2004]{pont04}
Pont, F., \& Eyer, L. 2004, MNRAS, in press

\bibitem[2001]{qg01}
Quillen, A.C., \& Garnett, D. 2001, in ``Galaxy Disks and Disk Galaxies'', 
Eds. G. Jose, S.J. Funes \ E.M. Corsini. ASP Conf. Ser. 230, 87

\bibitem[2003]{reddy03}
Reddy, B.E., Tomkin, J. Lambert, D.L., \& Allende Prieto, C. 2003, 
MNRAS 340, 304

\bibitem[1999]{reid99}
Reid, M.J., Readhead, A.C.S., Vermeulen, R.C., \& Treuhaft, R. N. 1999, 
ApJ 524, 816

\bibitem[2000]{rocha00}
Rocha-Pinto, H.J., Maciel, W.J., Scalo, J., \& Flynn, C. 2000, A\&A 358, 850

\bibitem[2000]{salasn00}
Salasnich, B., Girardi, L., Weiss, A., \& Chiosi, C. 2000, A\&A 361, 1023

\bibitem[2003]{sandage03}
Sandage, A., Lubin, L.M., \& VandenBerg, D.A. 2003, PASP 115, 1187

\bibitem[2001]{santos01}
Santos, N.C., Israelian, G., \& Mayor, M. 2001, A\&A 373, 1019

\bibitem[1963]{schmidt63}
Schmidt, M. 1963, ApJ 137, 758

\bibitem[1989]{schuni89}
Schuster, W.J., \& Nissen, P.E. 1989, A\&A 221, 65

\bibitem[2002]{sellw02}
Sellwood, J.A., \& Binney, J.J. 2002, MNRAS 336, 785

\bibitem[1999]{skuljan99}
Skuljan, J., Hearnshaw, J.B., \& Cottrell, P.L. 1999, MNRAS 308, 731

\bibitem[1991]{soderb91}
Soderblom, D.R., Duncan, D.K., \& Johnson, D.R.H 1991, ApJ 375, 722

\bibitem[2003]{steinmetz}
Steinmetz, M. 2003, in Munari, U. (ed.) GAIA Spectroscopy, Science and 
Technology, ASP Conf. Ser., Vol. 298, p. 381

\bibitem[1963]{stromgren63}
Str{\"o}mgren, B. 1963, QJRAS 4, 8

\bibitem[1965]{stromper65}
Str{\"o}mgren, B., \& Perry, C. 1965, ``Photoelectric {\em uvby} Photometry
for 1217 Stars Brighter than $V$ = 6.5, mostly of spectral classes A, F and G
(second version)'', Inst. for Advanced Study, Princeton, New Jersey 

\bibitem[1987]{stromgren87} 
Str\"omgren, B. 1987, in Gilmore, G. \& Carswell, R.F. (eds.) The Galaxy.
Reidel, Dordrecht, p. 299

\bibitem[2003]{taylor03}
Taylor, B.J. 2003, A\&A 398, 731

\bibitem[2000]{thoren00}
Thor{\'e}n, P., \& Feltzing, S. 2000, A\&A 363, 692

\bibitem[1999]{tomkin99}
Tomkin, J., \& Lambert, D.L. 1999, ApJ 523, 234 

\bibitem[1980]{twarog80}
Twarog, B.A. 1980, ApJ 242, 242

\bibitem[2002]{twarog02}
Twarog, B.A., Anthony-Twarog, B.J., \& Tanner, D. 2002, AJ 123, 2715

\bibitem[1997]{udryetal97}
Udry, S., Mayor, M., Andersen, J., Crifo, F., Grenon, M., Imbert, M.,
Lindgren, H., Maurice, E., Nordstr{\"o}m, B., Pernier, B., Pr{\'e}vot, L.,
Traversa, G., \& Turon, C. 1997, in ``Hipparcos-Venice'97'', Ed. B. Battrick.
ESA-SP-402, p. 693

\bibitem[1962]{vdbergh62}
van den Bergh, S. 1962, AJ 67, 486

\bibitem[2000]{davb00}
VandenBerg, D.A. 2000, ApJ 532, 430

\bibitem[1996]{wfd96}
Wielen, R., Fuchs, D., \& Dettbarn, C.: 1996, A\&A 314, 438

\bibitem[1941]{olinw41}
Wilson, O.C. 1941, ApJ 93, 29


\end{thebibliography}
\end{document}